\documentclass{ustthesis}  % default square logo 

\usepackage{mathtools}
\usepackage{amsmath}
\usepackage{amssymb}
\usepackage{graphics}
\usepackage{physics}
\usepackage{hyperref}
\usepackage{mathrsfs}
\usepackage[b5paper]{geometry}
\usepackage{pdfpages}
\usepackage{enumitem}
\usepackage[normalem]{ulem}

\usepackage[english]{babel}
\usepackage[T1]{fontenc}

\usepackage[scaled=1.1]{newtxtext}
\DeclareFontEncoding{LS2}{}{\noaccents@}
\DeclareFontSubstitution{LS2}{stix}{m}{n}
\usepackage{titlesec} %%% to allow custom headings

\titleformat{\chapter}[display]
  {\normalfont\huge\bfseries}
  {\chaptertitlename\ \thechapter}{20pt}{\huge}

\usepackage[makeroom]{cancel}

\hypersetup{colorlinks=true,citecolor={blue},linkcolor={blue},urlcolor={blue}}

\newcommand{\mh}{\mathcal{H}}
\newcommand{\mhfd}{\mh_\mathrm{FD}}
\newcommand{\mhsd}{\mh_\mathrm{SD}}
\newcommand{\mhabf}{\mh_\mathrm{ABF}}

\newcommand{\vac}{\ket{\varnothing}}
\newcommand{\kgsfb}{\ket{\mathrm{GS}(P_{\mathrm{FB}})}}
\newcommand{\cls}{\ket{\mathrm{CLS}}}
\newcommand{\kgs}{\ket{\mathrm{GS}}}
\newcommand{\kpsi}{\ket{\psi}}
\newcommand{\kfb}{\ket{\mathrm{FB}}}
\newcommand{\eig}{\ket{\mathrm{EIG}}}

\newcommand{\mbet}{\langle\tau\rangle}
\newcommand{\mbeet}{\langle\tau(e)\rangle}

\newcommand{\updated}[1]{\textcolor{black}{#1}}

\definecolor{myblue}{rgb}{0,0,0.75}

\title{{Quantum Phase Transitions\\and Dynamics in\\Perturbed Flatbands}}
\author{Sanghoon Lee}             %your name
\college{Center for Theoretical Physics of Complex Systems, IBS School}  %your college

\degree{Doctor of Philosophy}     %the degree
\degreedate{December 2023}         %the degree date

\begin{document}

%set the number of sectioning levels that get number and appear in the contents
\setcounter{secnumdepth}{3}
\setcounter{tocdepth}{3}

%\includepdf[pages=-]{ust_1.pdf}
%\includepdf[pages=-]{ust_2.pdf}
%\includepdf[pages=-]{ust_3.pdf}
\maketitle                  % create a title page from the preamble info
\pagenumbering{gobble}

\begin{center}
    \textbf{\large Ph.D. Thesis}
    
    \vskip 80pt
    
    \textbf{\huge Quantum Phase Transitions and\\\vspace{0.3em}Dynamics in Perturbed Flatbands}
    
    \vskip 250 pt
    
    \textbf{\large Sanghoon Lee}
    
    \vskip 10 pt
    
    \textbf{\large Basic Science}
    
    \vskip 10 pt
    
    \textbf{\large UNIVERSITY OF SCIENCE AND TECHNOLOGY}
    
    \vskip 10 pt
    
    \textbf{\large December 2023}
    
    \vskip 10 pt

\end{center}

\begin{center}
    \textbf{\huge Quantum Phase Transitions and\\\vspace{0.3em}Dynamics in Perturbed Flatbands}
    
    \vskip 79 pt
    
    \textbf{\large Sanghoon Lee}
    
    \vskip 79 pt
    
    \textbf{\large A dissertation Submitted in Partial Fulfillment of Requirements for the Degree of Doctor of Philosophy}
    
    \vskip 79 pt
    
    \textbf{\large December 2023}
    
    \vskip 79 pt
    
    \textbf{\large UNIVERSITY OF SCIENCE AND TECHNOLOGY}
    
    \textbf{\large Major of: Basic Science}
    
    \vskip 9 pt

     \textbf{\large Supervisor: Alexei Andreanov}
     
    \textbf{\large Co-supervisor: Sergej Flach}

\end{center}

\vspace*{\fill}
\begin{center}
\includegraphics[width = 1.0\columnwidth]{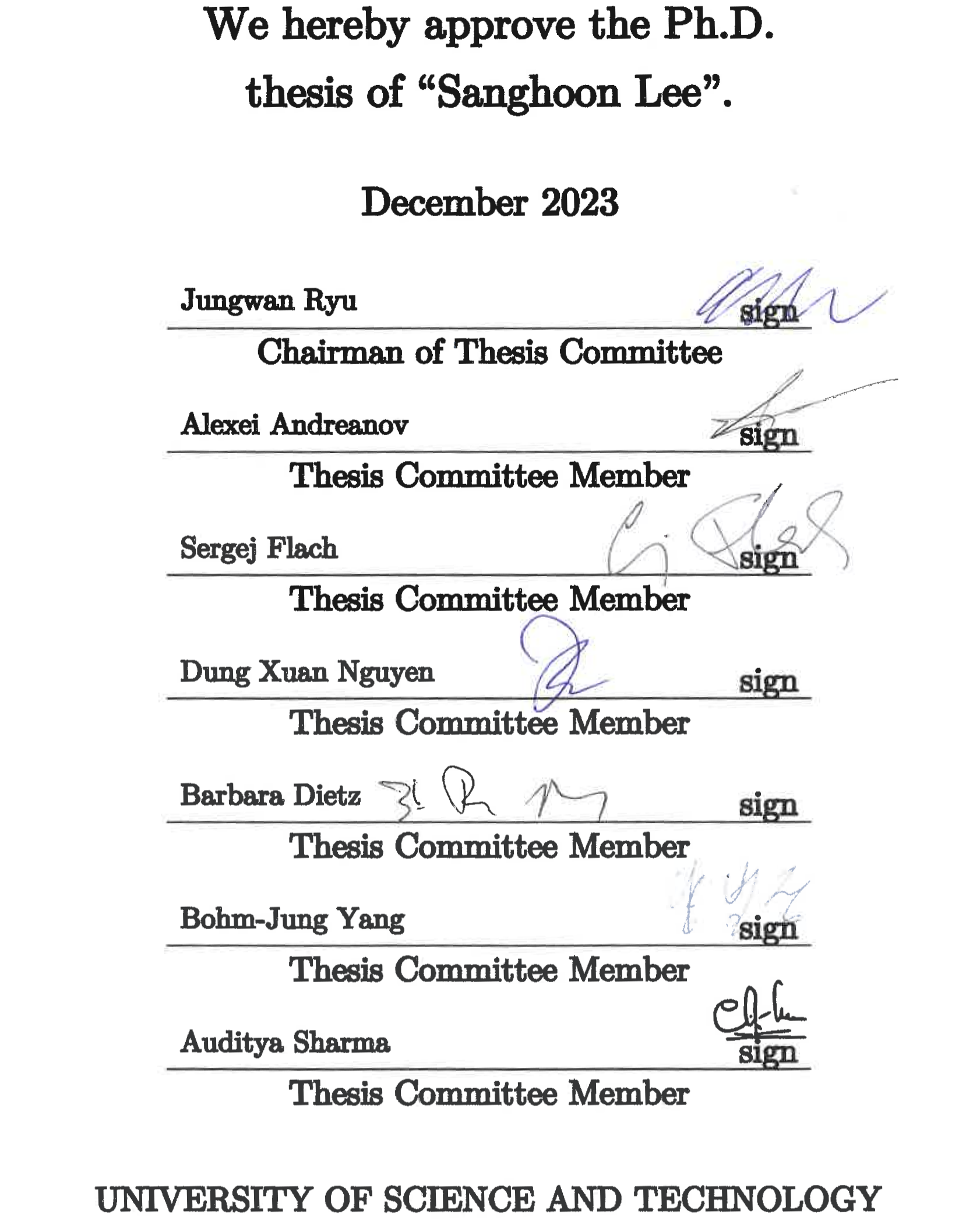}
\end{center}
\vspace*{\fill}
%\includepdf[pages={1}]{signature.pdf}
\begin{center}
    \textbf{ACKNOWLEDGEMENTS}
\end{center}

I want to thank my supervisors, Prof. Dr. Alexei Andreanov and Prof. Dr. Sergej Flach, for their amazing supervision and support during the Ph.D. program on the road to attaining the ability as a researcher. 
I am also very grateful to all colleagues and administration staff at IBS--PCS for the valuable discussion and memorable time at the center.
Last but not least, I would like to thank my family for their unconditional and endless support.   % include an acknowledgements.tex file
\begin{center}
    \textbf{ABSTRACT}
\end{center}

%\noindent\textcolor{blue}{Why I did}\\
In recent years, there has been a growing interest in flatband systems which exhibit macroscopic degeneracies.
These systems offer a valuable mathematical framework for the extreme sensitivity to perturbations and interactions.
This sensitivity unveils a wide variety of exotic and unconventional physical phenomena.
Moreover, the progress in their experimental realization contributes to the expanding landscape of exploration in this field.
%\noindent\textcolor{blue}{How was done and what was found}\\
This thesis aims to summarize all the works throughout the Ph.D. program.
Firstly, an in-depth exploration was conducted on the impact of weak quasiperiodic perturbations on one-dimensional two-band all-bands-flat lattices.
These \vphantom{lattices}\updated{tight-binding Hamiltonian} are diagonalized through a sequence of local unitary transformations.
By adjusting the quasiperiodic potential parameters, the key achievement involves finding a critical-to-insulator transition and identifying fractality edges in the flatband systems with quasiperiodic perturbations.
Next, the investigation delved into the effects of on-site interactions among hard-core bosons in one- and two-dimensional cross-stitch lattices.
One key finding is that groundstate energy primarily depends on compact localized states.
Moreover, the presence of barriers of compact localized states trap bosons, leading to the emergence of non-ergodic excitation and Hilbert space fragmentation.
\vphantom{Lastly, by using an electric circuit, a compact localized state of the one-dimensional diamond chain was successfully generated.}
\updated{Lastly, a compact localized eigenstate of the one-dimensional diamond chain using an electric circuit was successfully generated via local (linear and non-linear) driving.}
This achievement opens up a versatile circuit platform for the generation of flatbands and holds promise for potential applications in the field of quantum information.
I hope these collective efforts have expanded the frontiers of the field and made a meaningful contribution to the scientific community.          % include the abstract
\includepdf{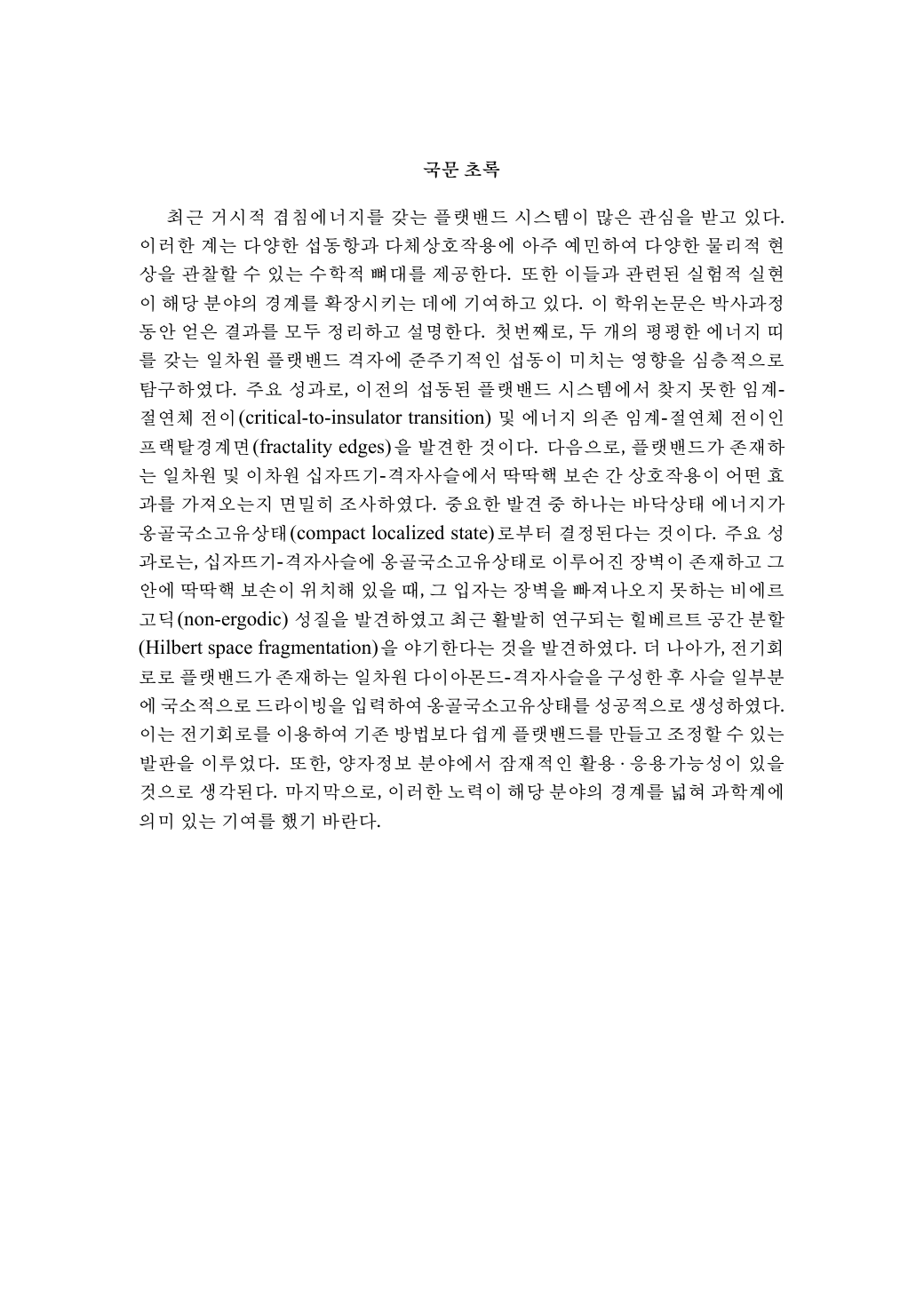}   
%\includepdf[pages=-]{ust_4.pdf}
%
%\include{remarks}
%
\begin{romanpages}          % start roman page numbering
\tableofcontents            % generate and include a table of contents
\listoffigures              % generate and include a list of figures
\end{romanpages}            % end roman page numbering

\pagenumbering{arabic}
%now include the files of latex for each of the chapters etc
\chapter{Introduction}\label{chpt1}

%\textcolor{blue}{Discovery of Anderson localization}\\
P. W. Anderson discovered in 1958 that when an electron is in a disordered system, the eigenstates can be spatially localized~\cite{anderson1958absence}.
By employing a tight-binding approach on a lattice with random on-site potentials, he showed that an electron ceases to diffuse away from its initial position depending on a disorder strength, resulting in the material converting from a metal to an insulator.
Such spatial localization is solely due to quantum interference between various electron paths, \(C_{n}\)s, with amplitudes, \(A_{n}\)s, respectively, from point \(\mathbf{r}_{0}\) back to \(\mathbf{r}_{0}\).
%\alexei{AA : 0 for destructive?}
For any two paths, average of interference terms goes to zero, because \(w_{\mathrm{int}}=2\operatorname{Re}(A_{1}^{*}A_{2})\) may be positive or negative.
Then, we remain only with \(w_{\mathrm{cl}} = |A_{1}|^{2} + |A_{2}|^{2}\).
If two paths are almost equal to each other, \(A_{1}\approx A_{2}\), and are related via time-reversal symmetry, constructive interference is obtained.
This explains the reduction in conductivity lead by increasing average return probability \(\langle w \rangle\)~\cite{doi:10.1142/7663},
\begin{gather}
    \langle w \rangle = \langle|A_{1} + A_{2}|^{2} \rangle = \langle w_{\mathrm{cl}} \rangle + \langle w_{\mathrm{int}} \rangle = \begin{dcases}
        \langle w_{\mathrm{cl}} \rangle \text{ }, &\text{destructive}\\
        2\langle w_{\mathrm{cl}} \rangle\text{ }, &\text{constructive}.
    \end{dcases}
\end{gather}

%\textcolor{blue}{Mathematical development of Anderson localization}\\
After Anderson initially proposed the idea of a disorder-induced localization, many mathematicians dived into the task of rigorously developing Anderson's intuition using a functional analysis.
One of the main results is the RAGE \updated{(acronym of Ruelle, Amrein, Georgescu, Enss)} theorem~\cite{ruelle1969remark,amrein1973characterization,enss1978asymptotic,hunziker2000quantum}.
It states that for any eigenstates \(\varphi\) corresponding to discrete eigenvalues (subspace \(\mathcal{H}_{\mathrm{pp}}\) of a given Hilbert space \(\mathcal{H}\) with pure point spectrum), their time evolved states remain bounded uniformly in time in the compact ball \(\{|x| < r\}\) with a radius \(r\).
In other words, all wave amplitudes must be zero outside of the compact ball \(\{|x| > r\}\),
\begin{gather}
    \lim_{r\to\infty}\sup_{t\in\mathbb{R}}\big\Vert \chi(|x| > r) e^{-iHt} \varphi \big\Vert = 0 \hspace{0.5em} \text{where }\chi(|x| > r) =
    \begin{dcases}
    0\text{ }, & |x| \leq r \\
    1\text{ }, & |x| > r
    \end{dcases}.
\end{gather}
Such bounded states are called to be a \emph{spectral localization} of eigenstates.
One may define Anderson localization by adding an exponential bound to the spectral localization so that a given state decays exponentially with a localization length \(\xi\) at \(\mathbf{r}\) compared to the center \(\mathbf{r}_{0}\) of the localization,
\begin{gather}
    |\varphi(\mathbf{r})| \leq C\exp(-|\mathbf{r} - \mathbf{r}_{0}|/\xi).
\end{gather}
Moreover, the absence of ballistic motion was proven for pure point spectrum for exponentially localized eigenstates in a disordered system~\cite{simon1990absence},
\begin{gather}
    \lim_{t\to\infty}\frac{\big\Vert \mathbf{r} e^{-iHt}\varphi \big\Vert^{2}}{t^{2}} = 0
\end{gather}
However, the argument above does not always hold for a non-random system with an arbitrary growth of \(\langle \mathbf{r}^{2} \rangle \), even though the spectrum is pure point with the exponentially localized states as discussed in Ref.~\cite{del1995localization,del1996operators}.
To have a more natural notion of localization without any transport, an additional condition of having finite \(\langle |\mathbf{r}|^{p} \rangle\) may be added to have a stronger form of Anderson localization, for any \(p\).
Such localization is called a \emph{strong (dynamical) localization}~\cite{damanik2001multi}.

%\textcolor{blue}{Other type of quantum localization?}\\
As it was discussed in the above paragraph, Anderson localization arises when wave functions encounter a random disorder in a system, leading to the prevention of long-range transport which implies the inability to propagate freely across the entire system.
Now, what we can ask is the following:
\vfill
\begin{center}
\emph{"Is there an alternative approach other than a disordered system\\to localize wavefunctions using the quantum mechanical effect?"}
\end{center}
\vfill

%========================================================================================
\clearpage
% =========================================================================== %
% =========================================================================== %
% SECTION : FLATBAND SYSTEMS
% =========================================================================== %
% =========================================================================== %
\section{Flatband systems and compact localization}\label{ch1:sec1:flatband}

Anderson localization originates from the enhanced return probability of a particle induced by a random disorder.
On the other hand, an alternative way of enhancing localization can be achieved without a random disorder in some lattices via destructive interference, known as compact localization.
Such localization is closely connected to \emph{flatbands}, which refer to constant energy bands in specific arrangements of translation-invariant tight-binding structures.
This section goes through the modern definition of a flatband system and a compact localization.

% =========================================================================== %
% SUBSECTION : Flatbands
% =========================================================================== %
\subsection{Flatbands}

%\textcolor{blue}{Modern definition of flatbands in lattice}\\
The modern understanding of a flatband system in a discrete lattice can be summarized as follows.
In a flatband system, there exists an energy band (or energy bands) that exhibits a completely flat energy dispersion relation, \(E_{\mathrm{FB}} = E(\mathbf{k})\), with short-range connectivities~\cite{sutherland1986localization,lieb1989two,mielke1992exactgs,creutz1999end,santos2004atomic} as well as some long--range connectivities~\cite{wang2019highly}.
Such energy band remains independent of the momentum, resulting in a vanishing group velocity \(\nabla_{\mathbf{k}}E = 0\) with effectively infinite mass.
This leads to macroscopic degeneracy at energy \(E_{\mathrm{FB}}\) and a strong suppression of transport.
Moreover, as we see soon through examples, the flatband features the eigenstates trapped over a strictly finite number of sites~\cite{sutherland1986localization,aoki1996hofstadter},
dubbed as the \emph{compact localized states} (CLS), due to destructive interference caused by network geometry.
This remarkable flatness gives rise to intriguing phenomena and opens up opportunities for unique physical results via the \emph{enhancement} of perturbation and interaction effects \updated{(see Sec.~\ref{ch1:sec3:perturbation})}, no matter how weak they are.

Flatband energy can also manifest in continuous systems, a concept distinct from the flatband systems discussed in this thesis. One such example is the Landau level~\cite{landau1930diamagnetismus}. It describes the movement of a non-interacting charged particle in a two-dimensional spatial domain in the presence of a magnetic field \(\mathbf{B}\) in \(z\)-axis.
In the Landau gauge, a vector potential \((-By, 0, 0)\) is a possible option.
With this specific gauge, the Hamiltonian can be expressed as follows %(\(\hat{p}_{z}\) is ignored).
which as a resemblance to the quantum harmonic oscillator, shifted in the coordinate space.
Then, the energy levels in the system are equivalent to those found in the conventional quantum harmonic oscillator,
\begin{gather}
    \mh = \frac{1}{2m} \left(\hat{p}_x - qB\hat{y}\right)^2 + \frac{\hat{p}_y^2}{2m} = \hslash \omega_{c}(\hat{n} + 1/2).
\end{gather}
Notably, the energy remains independent of the quantum number \(p_{x}\), leading to a macroscopic degeneracy.
Furthermore, in the high-energy regime, a holographic flatband system was also achieved~\cite{laia2011holographic}.

% =========================================================================== %
% SUBSECTION : Compact localized states
% =========================================================================== %
\subsection{Compact localized states}

\begin{figure}
\centering
  \includegraphics[width = 0.6\columnwidth]{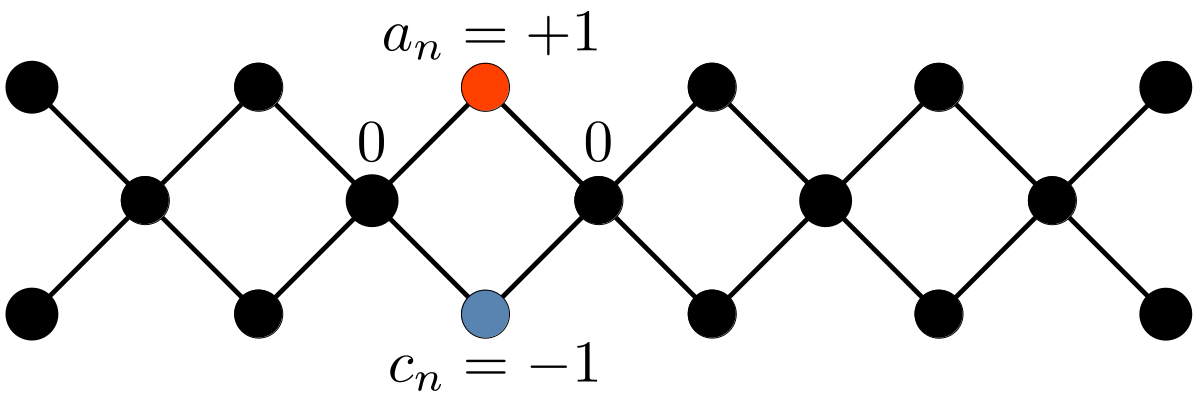}
  \caption[Compact localized state of a one-dimensional diamond chain]{
    A compact localized state of a one-dimensional diamond chain at flatband energy \(E_{\mathrm{FB}}\).
    Because of the lattice geometry, the wave function is trapped only at \(a_{n}\) and \(c_{n}\) and no leakage to \(b_{n}\) and \(b_{n+1}\).
    \(E_{\mathrm{FB}}\) has a macroscopic degeneracy and CLS can be placed at any unit cells.
    The linear combination of CLSs are again an eigenstate corresponding to \(E_{\mathrm{FB}}\).
  } \label{fig:diamondcls}
  \vspace{2.5em}
  \includegraphics[width = 0.45\columnwidth]{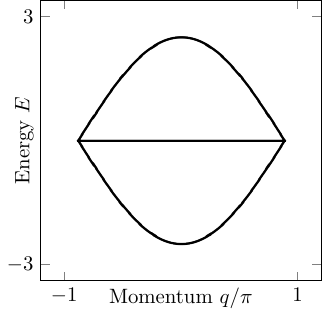}
  \caption[Energy bands of a one-dimensional diamond chain]{
    Energy bands of a one-dimensional diamond chain where \(E\) and \(q\) denoting the energy and crystal momentum, respectively.
    At \(E = E_{\mathrm{FB}} = 0\), we see a flatband energy.
    Other two bands are the dispersive bands.
  } \label{fig:diamond}
\end{figure}

% \textcolor{blue}{Definition of CLS}\\
CLS is an eigenstate of a particular tight-binding lattice that is limited to a finite system region and has no presence of amplitudes elsewhere.
These states exist in various systems.
For instance, let us consider the one-dimensional diamond chain (Fig.~\ref{fig:diamondcls}) without on-site potentials and with the same hopping strength,
\begin{gather}
    \mh = \sum_{n\in\mathbb{Z}} (\hat{b}_{n}^{\dagger} + \hat{b}_{n+1}^{\dagger})(\hat{a}_{n} + \hat{c}_{n}) + \mathrm{h.c}.
\end{gather}
\(\hat{a}_{n}\), \(\hat{b}_{n}\), and \(\hat{c}_{n}\) are annihilation operators at site \(a_{n}\), \(b_{n}\), and \(c_{n}\), respectively.
\updated{Fourier transforms of \(\hat{a}^{\dagger}_{n}\) and \(\hat{a}_{n}\) are defined as follows (same for \(\hat{b}^{\dagger}_{n},\hat{c}^{\dagger}_{n}\) and their conjugates transforms similarly)},
\begin{gather}
    \hat{a}^{\dagger}_{n} = \frac{1}{\sqrt{N}}\sum_{k\in\mathrm{BZ}}\exp(-ikn)\hat{A}^{\dagger}_{k} \quad \text{and} \quad \hat{a}_{n} = \frac{1}{\sqrt{N}}\sum_{k\in\mathrm{BZ}}\exp(ikn)\hat{A}_{k}
\end{gather}
\updated{\(\mathrm{BZ}\) stands for the first Brillouin zone.}
The Hamiltonian in a \(k\)-space, by means of Bloch's theorem, is the following,
\begin{gather}
    \mh = \sum_{k\in\mathrm{BZ}}\hat{\psi}_{k}^{\dagger} \mh(k) \hat{\psi}_{k}, \quad
    \mh(k) = \begin{bmatrix}
        0 & 1+e^{ik} & 0 \\
        1+e^{-ik} & 0 & 1+e^{ik} \\
        0 & 1+e^{ik} & 0
    \end{bmatrix}.
\end{gather}
We obtain three different energy bands: a flatband, \(E_{\mathrm{FB}} = 0\), and two dispersive bands, \(E_{\pm}(k) = \pm 2\sqrt{2}\cos(k/2)\).
The energy bands are drawn in Fig.~\ref{fig:diamond}.
\(E_{\mathrm{FB}}\) is macroscopically degenerated, and we have the CLSs, which make a complete orthogonal set in the system both in the \(k\)- and real-space \updated{(\(\vac\) is a vacuum state)},
\begin{gather}
    \vert \psi_{\mathrm{FB}}(k) \rangle = \frac{1}{\sqrt{2}} \left( \hat{a}^{\dagger}_{k} - c^{\dagger}_{k} \right)\vac  \\
    \vert \psi_{\mathrm{FB}}(n) \rangle = \frac{1}{\sqrt{N}}\sum_{k\in\mathrm{BZ}}e^{ikn}\vert \psi_{\mathrm{FB}}(k) \rangle = \frac{1}{\sqrt{2}} \left( \hat{a}^{\dagger}_{n} - c^{\dagger}_{n} \right)\vac
\end{gather}
In this setup, the compact localization relies on the system's geometry, as shown in Fig.~\ref{fig:diamondcls},
and each CLS is an anti-symmetric state occupying one unit cell, with zero amplitude on all other sites.
The CLSs are the eigenstates of the Hamiltonian \(\mh\), and they remain unchanged over time, except for a possible phase factor.
This counter-intuitive behavior, where the excitation remains localized, results from the \emph{destructive interference}.
The amplitudes for tunnelling or leaking of the CLS to the neighboring sites cancel each other out.
It is crucial to note that compact localization differs fundamentally from disorder-induced Anderson localization since CLSs can exist even in perfect translation-invariant systems.

% \textcolor{blue}{Why CLSs are interesting}\\
%\alexei{AA : Confusing!} \sout{The property of complete localization exhibited by CLSs makes them particularly intriguing for various reasons.}
\updated{Compact localized states (CLSs)} are intriguing for various reasons.
For instance, the temporal robustness of CLSs against perturbations is of significant interest due to their exact confinement within their localization region by destructive interference.
\updated{Such characteristics find} a potential application such as an information storage in quantum networks~\cite{rontgen2019quantum}.
Furthermore, CLSs have the potential to realize bound states within a scattering continuum~\cite{hsu2016bound, plotnik2011experimental}.
Alongside these prospects, CLSs offer different possible applications, such as in data transmission, where the interaction of light modes from different fibers can lead to crosstalk issues~\cite{vicencio2013diffraction, xia2016demonstration, rojas-rojas2017quantum}.

% =========================================================================== %
% SUBSECTION : Classification of flatband systems
% =========================================================================== %
\subsection{Construction of flatband systems}

In recent years, the existence of flatband systems has been experimentally demonstrated in a variety of photonic and dissipative condensate networks~\cite{rojas-rojas2017quantum, leykam2018perspective, vicencio2021photonic}.
%\alexei{AA : I don't agree and think the opposite!} \sout{In theoretical perspective, the generating of flatband Hamiltonians with specific properties, yet, remains a challenge.
%However, a significant amount of research has been conducted in recent decade, leading to a deeper understanding of the mechanisms involved in generating flatband systems.}
From a theoretical perspective, a significant amount of research has also been conducted in the recent decade, leading to a deeper understanding of the mechanisms involved in generating flatband systems.
Here, we summarize the current principles of generating flatband Hamiltonians.

% \textcolor{blue}{Inverse eigenvalue problem method}\\
The pioneering work of a flatband generator is based on local network properties and classifying flatband networks using orthogonal and non-orthogonal CLS.
The main method is to apply an inverse eigenvalue problem~\cite{maimaiti2017compact, maimaiti2019universal, maimaiti2021flat},
studied extensively for one-dimensional networks with the nearest unit cell interaction and a specific number of unit cells.
%The findings contribute to a better understanding of flatbands in networks with multiple bands and in higher dimensions.

% \textcolor{blue}{Non-singular and singular flatbands}\\
Flatband systems can also be classified into singular and non-singular flatband systems~\cite{rhim2019classification,hwang2021flat,hwang2021general,graf2021designing}.
Singular flatbands exhibit immovable discontinuities in their eigenstates, resulting from band touching with other dispersive bands.
\updated{Consequently, the compact localized states cannot form a complete set spanning the singular flatband, and we require an additional state.
This so-called non-contractible loop state is compact localized line that circles the entire system to have a complete set~\cite{rhim2019classification}.}
The non-contractible loop states lead to a novel bulk-boundary correspondence, where the presence of robust boundary modes is guaranteed by the singularity of the Bloch wave function and the non-contractible loop states.
Moreover, when the degeneracy at the band crossing point is lifted, the singular flatband becomes dispersive.
It can acquire a finite Chern number, offering a pathway for achieving nearly flat Chern bands.
In contrast, non-singular flatbands display no singularities in their Bloch wave functions.
They can be fully isolated from other bands while maintaining their perfect flatness.
It is worth noting that all one-dimensional flat bands belong to the non-singular class~\cite{rhim2019classification}.
Exploiting these topological characteristics, a general scheme has been developed to systematically construct flat-band model Hamiltonians, allowing for the design of singular or non-singular nature~\cite{hwang2021general}.

% \textcolor{blue}{Construction based on local symmetries}\\
Another way of constructing a flatband system is to use the concept of local and latent symmetries of a lattice structure and their role in explaining degeneracies in energy spectra~\cite{rontgen2018compact,morfonios2021flat,rontgen2021latent}.
Latent symmetries are local symmetries of an effective Hamiltonian derived through subsystem partitioning and isospectral reduction~\cite{smith2019hidden}.
This work offers new perspectives for analyzing accidental degeneracies in terms of latent symmetries, providing insights into the symmetries underlying physical systems.

% \textcolor{blue}{Systematic construction for non-Hermitian}\\
At the same time, a non-Hermitian system, which describes an open system with gain and loss, has also gained attention in recent decades.
Non-Hermitian systems with dispersionless energy bands were first studied using symmetries or specific geometries inspired by Hermitian models, both theoretically and experimentally~\cite{chern2015pt,qi2018defect, zyuzin2018flat,leykam2017flat,ramezani2017non,tobias2019experimental,zhang2020nonhermitian,zhang2020flatband, talkington2022dissipation}.
Later, the first systematic approach to construct non-Hermitian flat bands using one-dimensional two-band tight-binding networks was studied in Ref.~\cite{maimaiti2021nonhermitian}.
This method extends the techniques used for constructing Hermitian flat bands and demonstrates that the non-Hermitian case allows for fine-tuned flatbands unprotected by symmetry.
Furthermore, it can potentially reveal a greater variety of flatband types compared to the Hermitian case.

%\clearpage
% =========================================================================== %
% =========================================================================== %
% SECTION : EXPERIMENTAL REALIZATIONS
% =========================================================================== %
% =========================================================================== %
\section{Experimental realizations}\label{ch1:sec2:experiment}
%\sergej{SF : I am missing review of experimental work on FBs, Ideally somewhere at the end of the introductory part!}
%\shlee{SL : I include here.}

\noindent The precise arrangement of lattice geometry is crucial for fine-tuned flatband lattices.
As a result, finding actual materials featuring macroscopically degenerate flatbands in nature is difficult.
Such rarity presents a challenge for many researchers in the experimental field.
Advancements in fabrication technologies have facilitated the creation and examination of artificial lattices containing flatbands in various physical systems.
%The endeavor to observe flatbands in laboratory settings was initiated shortly after the formulation of theoretical models elucidating flatband ferromagnetism.
This section provides an overview of experimental methods employed in artificial flatband systems within diverse frameworks.
For details concerning the experimental realizations of artificial flatband lattices, readers are encouraged to consult the papers by Leykam et al.~\cite{leykam2018perspective, leykam2018artificial}.

\emph{Flatbands in electronic systems}: In 1998, Vidal et al.~\cite{vidal1998aharonov} discovered a completely flat spectrum in periodic electronic networks induced by a specific value of magnetic field, which is known as the Aharonov-Bohm caging~\cite{aharonov1959significance}.
This phenomenon was proposed to be observable in superconducting wire networks under appropriate conditions.
Subsequently, indirect evidences of flatbands due to Aharonov-Bohm caging was reported~\cite{abilio1999magnetic, naud2001aharonov}.
Moreover, recent advancements in two-dimensional material fabrication enabled the creation of artificial flatband lattices~\cite{tadjine2016from,qiu2016designing}.
In 2017, 2D Lieb lattices were constructed with two methods via scanning tunneling microscopy: creating electron lattices by atom removal~\cite{drost2017topological} and molecule addition to a substrate~\cite{qiu2016designing,slot2017experimental}.
These scanning tunneling microscopy setups allows electron density measurements, offering observations into flat and dispersive band Bloch waves.

\emph{Flatbands in optical lattices}: Progress in cold atom research, facilitated by laser cooling and ion trapping, has enabled precise exploration of flatband lattices, with the Lieb lattice standing out due to its simplicity in atom manipulation.
Conversely, lattices like the dice~\cite{bercioux2009massless, urban2011barrier} and kagomee~\cite{santos2004atomic,ruostekoski2009optical}, though easier to construct, present challenges in generating highly excited flatbands in cold atom systems. 
Early experiments sought to create optical Lieb lattices~\cite{shen2010single}, focusing on the Dirac cone in dispersive bands rather than the flatband.
These optical lattices unavoidably introduced a width to the flatband due to next-nearest neighbor hopping.
However, optimizing laser beam intensities allowed for nearly flat bands with minimal width~\cite{apaja2010flat}.
In 2015, Takahashi et al.~\cite{taie2015coherent} achieved an optical Lieb lattice for bosonic cold atoms by dynamically adjusting the lattice, leading to the observation of interactions that caused condensate decay into lower dispersive bands.
Subsequent experiments in 2017 furthered the understanding of optical Lieb lattices: the introduction of fermionic cold atoms~\cite{taie2015coherent} and the exploration of interaction effects on flatband energy shifts~\cite{ozawa2017interaction}.

\emph{Flatbands in photonic systems}: Flatbands are vital in slow light applications in photonics~\cite{baba2008slow}, enhancing nonlinear effects with suppressed group velocity and possibly useful for pulse buffering.
However, achieving ideal flatbands is challenging due to balancing low group velocity with a useful bandwidth~\cite{letartre2001group,notomi2001extremely}.
Initial efforts with photonic crystal featured by circular rods~\cite{takeda2004flat} faced limitations due to difficulties in fabrication.
Later, in 2017, innovative kagome structures~\cite{schulz2017photonic} offered improved group velocity reduction compared to the previous research~\cite{baba2008slow, li2008systematic, xu2015design}, revitalizing interest in flatbands in slow light engineering in photonic crystal waveguides~\cite{endo2010tight, feigenbaum2010resonant, nakata2012observation, nixon2013observing, kajiwara2016observation}.
Moreover, the femtosecond laser writing technique enabled the fabrication of optical waveguide networks, providing control over coupling and facilitating the study of various lattice types~\cite{szameit2010discrete}.
While early experiments with Lieb lattices indirectly inferred flatbands~\cite{guzman2014experimental} because a superposition of all bands was excited by a single waveguide input, later studies directly excited compact localized states~\cite{mukherjee2015observation, vicencio2015observation}.

\emph{Flatbands in exciton-polariton condensate : }Exciton-polaritons in semiconductor microcavities, created through strong light-matter coupling, offer a unique platform for exploring Bose-Einstein condensation and low-power optical switching, facilitated by their low effective mass and exciton-mediated nonlinearity.
These features extend to the creation of structured potentials, enabling the construction of periodic lattices and flatbands within exciton-polariton systems~\cite{schneider2016exciton}.
Below the condensation threshold, the linear band structure is easily measurable, as reported by Jacqmin et al.~\cite{jacqmin2014direct} in 2014 when they successfully constructed a honeycomb lattice with a flatband.
Overcoming challenges related to precise control over condensation states at higher pump powers, Baboux et al.~\cite{baboux2016bosonic} have introduced solutions involving manipulating the spatial structure of the optical pump to induce condensation into a flatband.
In a 2D Lieb lattice with micropillars, exciton-polariton condensation into a flatband was also realized~\cite{klembt2017polariton,whittaker2018exciton}.
%\clearpage
% =========================================================================== %
% =========================================================================== %
% SECTION : EFFECT OF PERTURBATION AND INTERACTION
% =========================================================================== %
% =========================================================================== %
\section{Effect of perturbation and interaction}\label{ch1:sec3:perturbation}

In recent years, there has been a growing interest in physical systems that exhibit macroscopic degeneracies.
As previously discussed, one key characteristic of these systems is the presence of flatbands.
The intriguing aspects of flatband models lie in their unique behavior by having a charge carrier with zero group velocity, which leads to negligible kinetic energy and extreme suppression of their motion.
Then, the dominant energy scale would be a perturbation and an interaction strength, eventually breaking the macroscopic degeneracy no matter how weak they are.
This intriguing behavior acts as a catalyst for a wide variety of exotic and unconventional correlated phases.

For instance, unconventional Anderson localizations~\cite{derzhko2010low,nita2013spectral,flach2014detangling}, an inverse Anderson transition~\cite{goda2006inverse,nishino2007flat} where the re-entrant of localized phase exists in a sequence of insulator-metal-insulator, and a non-perturbative metal-to-insulator transitions~\cite{cadez2021metal,longhi2021inverse,li2022aharonov} are among the distinct manifestations observed due to various types of perturbations in several flatband systems.
In the realm of many-body flatband systems, the effect of different types of interactions gives rise to several notable phenomena.
These include a flatband ferromagnetism~\cite{tasaki1992ferromagnetism,mielke1999ferromagnetism}, a frustrated magnetism~\cite{ramirez1994strongly, derzhko2015strongly}, an ergodicity breaking~\cite{kuno2020flat,danieli2020many,vakulchyk2021heat,orito2021nonthermalized,danieli2022many}, a quantum caging of particles~\cite{tovmasyan2013geometry,tovmasyan2018preformed,danieli2021nonlinear,danieli2021quantum}, a formation of tightly bounded pairs in attractive interaction~\cite{doucot2002pairing}, a pairing formation in repulsive interaction~\cite{tovmasyan2013geometry,takayoshi2013phase,phillips2015low,mielke2018pair} in bosonic system, and flatband superconductivity~\cite{peotta2015superfluidity, julku2016geometric, tovmasyan2016effective, liang2017band, cao2018unconventional, chan2022pairing, chan2022designer, chan2023superconductivity}.

%\clearpage
% =========================================================================== %
% =========================================================================== %
% SECTION : OUTLINE
% =========================================================================== %
% =========================================================================== %
\section{Contributed work and outline of the thesis}\label{ch1:sec4:published}

In Ref.~\cite{lee2023criticalA, lee2023criticalB}, I studied the effect of quasiperiodic perturbations on the full manifold of one-dimensional two-band all-bands-flat lattice models.
Such networks can be diagonalized by a finite sequence of local unitary transformations parameterized by angles \(\theta_{i}\)s.
Without loss of generality, the case of two bands with bandgap \(\Delta\) is focused.
The primary focus is on a two-band ABF ladder subjected to a quasiperiodic perturbation that is comparatively weak relative to the bandgap.
Weak perturbations lead to an effective Hamiltonian with both on- and off-diagonal quasiperiodic terms that depend on \(\theta_{i}\)s.
For some angle values, the effective model coincides with the extended and the off-diagonal Harper model.
By varying the parameters of the quasiperiodic potentials, localized insulating states and an entire parameter range hosting critical states \updated{over the full spectrum} with subdiffusive transport are observed.
\updated{Furthermore, we identify and refer to the transition between these states as the \emph{critical-to-insulator transition.}}
For finite quasiperiodic potential strength, the critical-to-insulator transition becomes energy-dependent with what we term \emph{fractality edges} separating localized from critical states.
Chapter~\ref{chpt2} provides an overview of a systematic investigation into the effect of quasiperiodic perturbations on one-dimensional ABF networks.

I also studied the effect of on-site interaction of hard-core bosons in one-dimensional and two-dimensional cross-stitch lattices with flatbands.
One of the key findings is that the groundstate energy is only influenced by the CLSs.
\updated{The Wigner crystal arises when CLSs are completely occupied, constrained by the mutual repulsion from hard-core boson constraint.
On the other hand, the band-insulating phase occurs when all lattice sites are fully filled, leading to an insulating state due to the absence of available states for conduction.}
In cases where the lattice is partially filled with the CLSs and with the fully-filled dimers, a mixture of both phases is observed.
Moreover, when a closed barrier of compact localized states is present in both the one- and two-dimensional cross-stitch lattices, hard-core bosons inside the loop become physically trapped in the loop.
\updated{This violation of weak thermalization gives rise to the presence of non-ergodic excited states, causing Hilbert space fragmentation due to the emergence of artificial localization arising from CLS barriers.}
Chapter~\ref{chpt3} provides an overview of the thermalization properties of hard-core bosons in the one- and two-dimensional cross-stitch lattices with on-site repulsive interaction.

\updated{Ref.~\cite{chase2023compact} focuses on creating CLSs in an electrical diamond lattice made solely of capacitors and inductors. An excitation of CLS mode is achieved through local linear driving near the flatband frequency of the lattice.
This research marks a step forward in establishing a versatile circuit platform for creating and manipulating flatband systems.
The comparison between experimental results and numerical simulations reveals a good agreement.
Moreover, the investigation includes the examination of local nonlinear driving by substituting capacitors with varactors.
The introduction of lattice nonlinearity facilitates the generation of a nonlinear CLS continuation.
At the same time, the study considers a one-dimensional stub lattice, which falls in a different flatband class.
An important observation is that local driving cannot isolate a single CLS due to its non-orthogonality, unlike the diamond lattice case.
My main contribution to this project include a numerical modeling of an appropriate nonlinear capacitance of a varactor to achieve good agreement with the experiment and the general manuscript work.
Additionally, I investigate the one-dimensional stub chain, demonstrating the inability to isolate CLS with local driving.}
Chapter~\ref{chpt4} presents an overview of generating one-dimensional flatband electrical lattices through capacitors and inductors.

\chapter[Generating Critical States from Quasiperturbed Flatbands]{Generating Critical States \\ from Quasiperturbed Flatbands}\label{chpt2}

\iffalse
\vfill
In the field of condensed matter physics, an area of significant interest and exploration in recent years has revolved around comprehending the effects of various perturbations on single-particle localized states.
While it is widely recognized that states can become exponentially localized through the influence of random disorder or quasiperiodic potential~\cite{anderson1958absence, kramer1993localization, aubry1980analyticity}, localization can also be attained even in the absence of disorder/perturbation as it is mentioned in the introduction, which is a compact localization in a flatband system.
The fascination with flatbands stems from their extreme sensitivity to perturbations that breaks its macroscopic degeneracy.
Different types of perturbation lead to peculiar phenomena and a diverse range of intriguing and unconventional phases.
In this chaper, we focus on the effect of quasiperiodic perturbation on flatband systems with two bands being flat.
The related works are published in Ref.~\cite{lee2023criticalA,lee2023criticalB}.
\fi

%========================================================================================
%\clearpage
% =========================================================================== %
% =========================================================================== %
% SECTION : MODEL
% =========================================================================== %
% =========================================================================== %
\section{Introduction}

In condensed matter physics, an area of significant interest and exploration in recent years has revolved around comprehending the effects of various perturbations on single-particle localized states.
While it is widely recognized that states can become exponentially localized through the influence of random disorder or quasiperiodic potential~\cite{anderson1958absence, kramer1993localization, aubry1980analyticity}, localization can also be attained even in the absence of disorder/perturbation as it is mentioned in the introduction, which is a compact localization in a flatband system.
The fascination with flatbands stems from their extreme sensitivity to perturbations that break their macroscopic degeneracy.
Different types of perturbation lead to an evolution of CLSs, making a system that generates a diverse range of intriguing and unconventional phases.

Further fine-tuning of flatband systems can result in \emph{all-bands-flat} (ABF) networks~\cite{vidal1998aharonov, danieli2021nonlinear}, where all dispersive bands are made flat, making them even highly sensitive to perturbations.
A finite sequence of non-commuting local unitary transformations can transform an ABF network into a diagonal Hamiltonian, a \emph{parent network} of the corresponding ABF network.
Any perturbation applied to the original ABF network leads to a nontrivial perturbation in the parent network, forming non-zero non-diagonal elements.
Recently, the effect of random disorder on ABF networks has been systematically explored~\cite{cadez2021metal}.
Nonperturbative metal-to-insulator transitions were discovered in three-dimensional ABF systems, while in the one-dimensional case, only Anderson localization occurred across the entire spectrum~\cite{cadez2021metal}.
We can also question about what happens if we consider quasiperiodic perturbation on the one-dimensional ABF networks.

In this chapter, we summarize a systematic study of the impact of quasiperiodic perturbations on the one-dimensional ABF networks, published in Ref.~\cite{lee2023criticalA,lee2023criticalB}.
Without loss of generality, we focus on a two-band ABF ladder and introduce a quasiperiodic perturbation, which is weak compared to the bandgap.
We use the smallness of the perturbation to project the Hamiltonian onto a single sublattice, thereby deriving a new effective projected Hamiltonian.
By varying the parameters of the perturbation, we find that the entire spectrum of the projected Hamiltonian is either localized or critical---the \emph{critical-to-insulator transition} (CIT), but never extended, using a mapping to the extended Harper model~\cite{avila2017spectral, xiao2021observation}.
Increasing the strength of the quasiperiodic potential, the critical eigenstates are gradually replaced by localized ones via the appearance of an energy-dependent CIT~\cite{liu2022anomalous, zhang2022lyapunov, wang2022quantum, shimasaki2022anomalous, lin2023general} which is called as \emph{anomalous mobility edges} \updated{(but, in this chapter, those edges are dubbed as \emph{fractality edges})}.
The emergence of regions of critical states in one-dimensional systems is quite different compared to the conventional Aubry-Andr\'e model, which features a metal-to-insulator transition.

The chapter is organized as follows.
We start by defining and discussing the construction of the all-bands-flat models in Sec.~\ref{ch2:sec:model} and the numerical methods in Sec.~\ref{ch2:sec:nummethod}.
In Sec.~\ref{ch2:sec:weak}, we derive an effective model valid in the limit of weak quasiperiodic perturbation and use it to chart the phase diagram, confirmed numerically.
The transport properties at the weak perturbation is discussed in Sec.~\ref{ch2:sec:spreading}.
The properties of the full model at finite perturbation strength are investigated in Sec.~\ref{ch2:sec:finite}, followed by conclusions in Sec.~\ref{ch2:sec:conclusion}.
%\clearpage
% =========================================================================== %
% =========================================================================== %
% SECTION : MODEL
% =========================================================================== %
% =========================================================================== %
\section{Model}\label{ch2:sec:model}

\begin{figure}
\centering
  \includegraphics[width = \columnwidth]{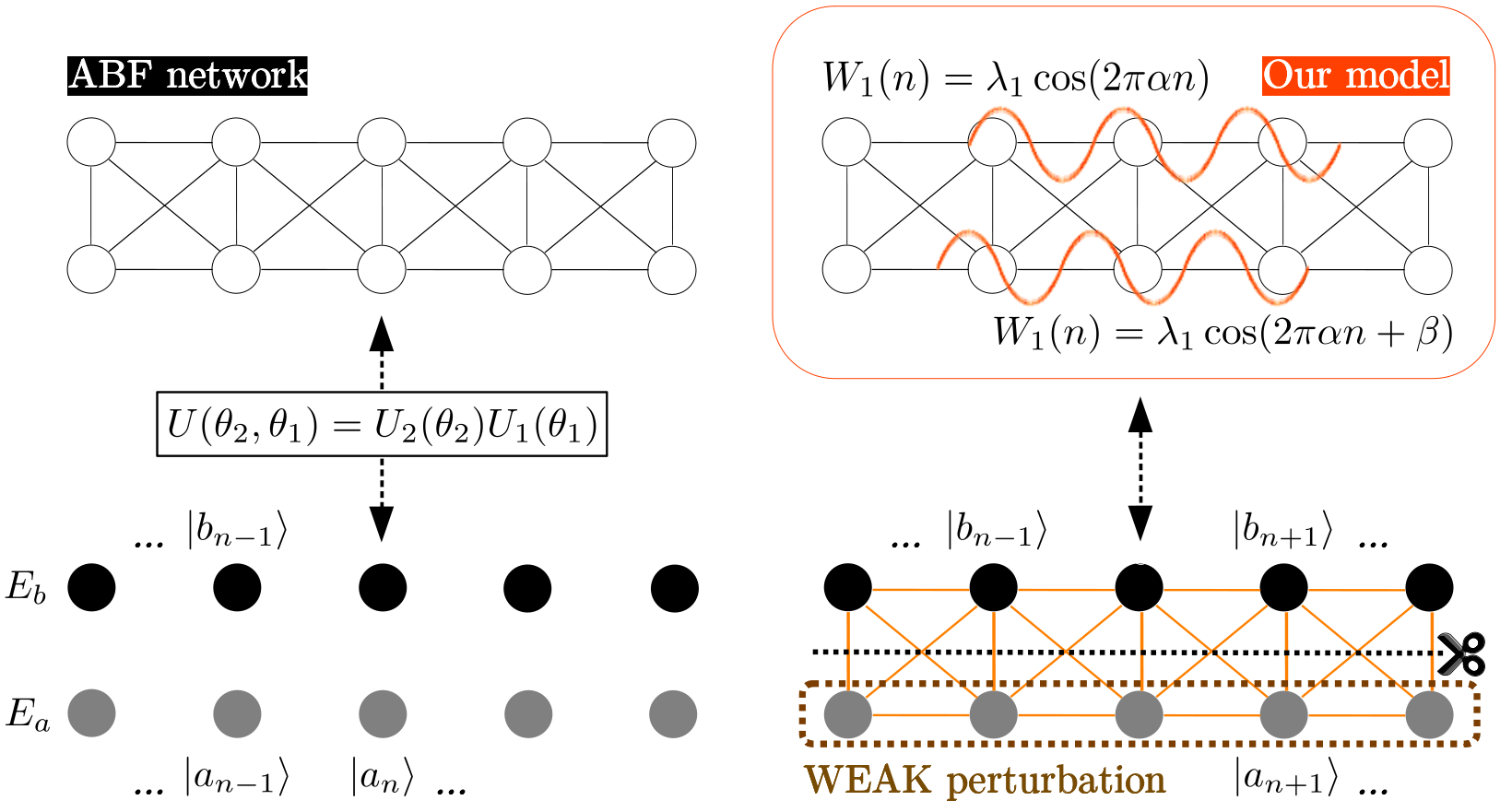}
  \caption[Perturbed ABF network and projected model]{
    The ABF model is fully detangled/diagonalized into a diagonal Hamiltonian via local unitary transformations \(U_{1}\) and \(U_{2}\) with appropriate angles \(\theta_{1}\) and \(\theta_{2}\) (left).
    By adding quasiperiodic perturbation \(W\) made up of two quasiperiodic fields \(W_{1,2}\) in Eq.~\eqref{eq:fields}, nontrivial hoppings are created in the fully detangled basis (right).
    For weak quasiperiodic fields, the first-order degenerate perturbation theory is used to derive an effective projected model.
    In the detangled basis, this corresponds to only keeping the hopping terms coupling the sites on the same sublattices.
  }
  \label{fig:scheme}
\end{figure}

We focus on the one-dimensional ABF ladder, which consists of two flatbands with the nearest-neighbor unit cells hopping~\cite{maimaiti2019universal}.
An illustration of the model and the scheme is provided in Fig~\ref{fig:scheme}.

% =========================================================================== %
% SUBSECTION : Local unitary transformation and ABF lattice
% =========================================================================== %
\subsection{Local unitary transformation and ABF lattice}

ABF Hamiltonian \(\mhabf\) can be constructed from a macroscopically degenerate diagonal matrix \(\mhfd\) with onsite energies \(\varepsilon_a\) and \(\varepsilon_b\) on the two sublattices having the bandgap \(\Delta = \abs{\varepsilon_a - \varepsilon_b}\)~\cite{danieli2021nonlinear},
\begin{gather}
    \mhfd = \sum_{n} \varepsilon_{a}\ketbra{a_{n}}{a_{n}} + \varepsilon_{b}\ketbra{b_{n}}{b_{n}}.
\end{gather}
The state vector \(\ket{\Psi}\) of \(\mhfd\) is written as linear combinations of orthonormal unit basis \(\{\ket{a_n},\ket{b_n} : n\in \mathbb{Z}\}\) where \(a_n, b_n\) are the amplitudes at the two sublattices,
\begin{gather}
    \ket{\Psi} = \sum_{n\in\mathbb{Z}}a_{n}\ket{a_n} + b_{n}\ket{b_n}, \quad\text{where }\sum_{n}|a_{n}|^2 + |b_{n}|^2 = 1.
\end{gather}
We term \(\mhfd\) as the parent Hamiltonian for the manifold of ABF systems and refer to it as \emph{fully detangled}.~\cite{flach2014detangling,danieli2021nonlinear,danieli2020many}.
The construction of ABF systems is based on a sequence of unitary transformations \(U_i\)~\cite{flach2014detangling,danieli2021nonlinear,danieli2020many,cadez2021metal,lee2023criticalA, lee2023criticalB} applied to the fully detangled Hamiltonian.
It is a way of achieving a set of flatband lattices \updated{that is distinct from} the method mentioned in Sec.~\ref{ch1:sec1:flatband} in Chapter~\ref{chpt1}.
Each transformation \(U_{1,2}\) is a direct sum of local transformations, \updated{i.e.}, it takes a block diagonal form.
For a one-dimensional system with the nearest-neighbor unit cell hopping, only two local unitary transformations \(U=U_2 U_1\) are enough to produce a connected hopping network.
For a one-dimensional tight-binding system with two sites per unit cell, the most general \(2\times 2\) block is a \(\mathrm{SU}(2)\) matrix,
\begin{gather}
  U_{i} \! = \!\!\!\!\!\sum_{n, n^{\prime}\in\mathbb{N}}\!\!\!\!
  z_{i}\vert{a^{(i)}_{n}}\rangle\langle{a^{(i-1)}_{n}}\vert \!+\! w_{i}\vert{a^{(i)}_{n}}\rangle\langle{b^{(i-1)}_{{n}^{\prime}}}\vert \!-\! w_{i}^{*}\vert{b^{(i)}_{n}}\rangle\langle{a^{(i-1)}_{{n}^{\prime}}}\vert \!+\! z_{i}^{*}\vert{b^{(i)}_{n}}\rangle\langle{b^{(i-1)}_{n}}\vert. 
  \label{eq:unitary}
\end{gather}
The index \(i\) denotes the \(i\)-th local unitary transformation, and the indices \(n, n^\prime\) label the unit cells.
In the simplest case, the blocks are parameterized by only two angles: \(\theta_{1,2}\) for \(U_{1,2}\), respectively, producing real \(U_{1,2}\in\mathrm{SO(2)}\).
The \(2\times 2\) blocks can act on the \(2\) sites within the same unit cell, \(n^\prime = n\), or different unit cells, \(n^\prime \neq n\).
This generates different, non-commuting transformations.
Then, we obtain additional tunneling between sites, giving a connected ABF network.
Throughout the chapter, ABF Hamiltonian always refers to the connected ABF network.
\updated{The energy spectra is preserved due to the property of a unitary transformation}, and the local rotation allows us to maintain the localized/delocalized property of a given eigenstate.
Specifically, we choose the \(U_1 \in \mathrm{SO(2)}\) blocks to act within the same unit cell, 
while for the \(U_2 \in \mathrm{SO(2)}\) blocks, one of the sublattice sites is taken from the neighboring unit cell, \(n^\prime = n - 1\).
Next, we add a quasiperiodic perturbation to the ABF Hamiltonian with \(\mh = \mhabf + W\).
Here, \(W\) is defined as a direct sum of \(2\times 2\) matrices \(W(n)\) of all \(1\leq n \leq L\), where \(L\) is the number of unit cells,
\begin{gather}
    W = \bigoplus_{n = 1}^{L}W(n), \qquad
    W(n) =
    \begin{bmatrix}
        W_{1}(n) & 0 \\ 0 & W_{2}(n)
    \end{bmatrix}.
\end{gather}
The generic form of \(W(n)\) is given by two quasiperiodic fields \(W_{1}\) and \(W_{2}\),
\begin{gather}
    \label{eq:fields}
    W_{1}(n) = \lambda_{1}\cos(2\pi\alpha n + \phi) \quad\text{and}\quad W_{2}(n) = \lambda_{2}\cos(2\pi\alpha n + \beta + \phi). 
\end{gather}
Here, the spatial frequency \(\alpha\) is an irrational number, \(\alpha \in \mathbb{R}\setminus\mathbb{Q}\), \(\beta\) is the phase difference between \(W_{2}\) and \(W_{1}\),
\(\phi\) is the phase used for averaging and \(\lambda_{1}\) and \(\lambda_{2}\) are the strengths of the quasiperiodic potentials.
Taking two angles \(\theta_{1,2}\) into account, we can tune five total parameters to understand how this perturbation affects different ABF networks.

% =========================================================================== %
% SUBSECTION : Geometric symmetry of ABF lattice
% =========================================================================== %
\subsection{Geometric symmetry of ABF lattice}

The ABF Hamiltonian \(\mhabf = U_{2}U_{1}\mhfd(U_{2}U_{1})^{\dagger}\) has a rotational symmetry and it allows us to obtain the irreducible domain of angles \((\theta_{1},\theta_{2})\).
Suppose \(\theta_{1} \to \theta_{1} + \pi\) and \(\theta_{2}\) remains the same. Then \(U_{1}\) becomes \(-U_{1}\).
The ABF network with shifted unitary transformation \((-U_{2}U_{1})\mhfd (-U_{2}U_{1})^{\dagger}\) is equal to \(\mhabf\).
The same situation occurs for \(\theta_{2}\) shifting to \(\theta_{2} + \pi\).
If both \(\theta_{1}\) and \(\theta_{2}\) are shifted, then we retrieve the original ABF system as well.

The second case is when \(\theta_{1}\) is reversed and shifted by \(\pi\), hence \(\pi \!-\!\theta_{1}\).
\updated{Initially, \(I \!=\! U^{\dagger}_{1}(\theta_{1})U_{1}(\theta_{1})\) is applied on the right and the left end of \(U_{1}(\pi \!-\! \theta_{1})\mhfd U^{\dagger}_{1}(\pi \!-\! \theta_{1})\).}
By using the relation \(U_{1}(\pi \!-\! \theta_{1})U_{1}(\theta_{1}) \!=\! -I\), it is easy to verify that \(U_{1}(\pi \!-\! \theta_{1})\mhfd U^{\dagger}_{1}(\pi \!-\! \theta_{1})\) and \(U_{1}(\theta_{1})\mhfd U^{\dagger}_{1}(\theta_{1})\) are equivalent.
%\begin{align}
%    U_{1}(\pi - \theta_{1})\mhfd U^{\dagger}_{1}(\pi - \theta_{1}) \! &= \! U^{\dagger}_{1}(\theta_{1})U_{1}(\theta_{1})U_{1}(\pi - \theta_{1})\mhfd U^{\dagger}_{1}(\pi - \theta_{1})U^{\dagger}_{1}(\theta_{1})U_{1}(\theta_{1}) \notag \\
%    &= \! U_{1}(\theta_{1})H_\text{FD}U^{\dagger}_{1}(\theta_{1}).
%\end{align}
Hence, the overall ABF system will not change.
The similar result is obtained for \(\theta_{2}\) being shifted to \(\pi \!-\! \theta_{2}\) as well.
We can further think for the case when \(\theta_{1} \to \pi/2 - \theta_{1}\) and \(\theta_{2}\) remains at it is.
However, such a case will not lead to a symmetry.
As a conclusion, it is enough to consider \(\theta_{1},\theta_{2} \in [0, \pi/2]\).

%\clearpage
% =========================================================================== %
% =========================================================================== %
% SECTION : WEAK PERTURBATION
% =========================================================================== %
% =========================================================================== %
\section{Numerical methods}\label{ch2:sec:nummethod}

The relation between localization and statistical properties of the energy spectrum and eigenstates in disordered systems is widely recognized.
Eigenstates that exhibit broad spatial distribution suggest an electronic transport, with the localization length being inversely related to the exponent describing the decay of eigenstate components across system sites.
An exact formula for the localization length in the one-dimensional lattice model remains unknown, although field-theoretic methods have provided expressions for extreme cases of strong and weak random disorder~\cite{evers2008anderson}.
Therefore, numerical approaches are often employed to approximate the localization length.

% =========================================================================== %
% SUBSECTION : inverse participation ratio
% =========================================================================== %
\subsection{Inverse participation ratio}

The generalized inverse participation ratio (GIPR) is a commonly used numerical quantity. It has been utilized to measure the localization/delocalization of the normalized eigenstate \(\psi(E)\) of energy \(E\) in the analysis of a tight-binding system,
\begin{gather}
    \operatorname{GIPR}_{q}(E) = \sum_{n\in\mathbb{z}} |\psi_{n}(E)|^{2q} \sim L^{-\tau_{q}}.
\end{gather}
Especially when \(q=2\), the physical interpretation is well-established, and the quantity denoted as \(\operatorname{IPR}(E)\) is commonly known as the inverse participation ratio (IPR), serving as a measure of spatial extension for a given state.
To illustrate, consider two extreme examples.
In the case of a homogeneous state \(\psi(E)\), we obtain \(|\psi(E)|^{2} \sim 1/L\) where \(L\) is the size of a lattice.
This leads to \(\operatorname{IPR}(E) \sim 1/L\).
On the other hand, if \(\psi(E)\) is localized at a single site, then IPR is equal to 1,
\begin{gather}
    \operatorname{IPR}(E) = \begin{dcases}
        1/L\quad&\tau_{2} = 1,\\
        1 \quad&\tau_{2} = 0.
    \end{dcases}
\end{gather}

% =========================================================================== %
% SUBSECTION : inverse participation ratio
% =========================================================================== %
\subsection{Defining a linear model for localized states}

The quantity \(\tau = \tau_{2}\) is a scaling exponent of the \(\mathrm{IPR}\).
In the thermodynamic limit, \(\tau\) is defined explicitly as follows at energy \(E\)~\cite{vicsek1992fractal}, 
\begin{gather}
  \label{eq:tauexplicit}
  \tau = \lim_{L\to\infty}\frac{1}{\ln(1/L)}\ln\operatorname{IPR}(E).
\end{gather}
However, we cannot use this definition for a single eigenstate as a function of system size \(L\).
%The exponent \(\tau\) is defined for a single eigenstate and eigenenergy.
Changing the system sizes changes the eigenenergies and does not allow their smooth continuation from one system size to another.
Instead, we calculate the average of \(\tau\) over a small energy bin to extract the scaling behavior of \(\tau\).
This is achieved by rescaling the eigenspectrum into the interval \([0, 1]\) and splitting it into equidistant bins \(e\).
This allows us to calculate the average of \(\tau\) in a single bin \(e\), denoted as \(\mbeet\) as a function of lattice size \(L\). 
It follows:
\begin{gather}
  \label{eq:asymptotic}
  \lim_{L\to\infty}\mbeet =
  \begin{dcases}
    0\text{, all states localized in \(e\),} \\
    \tau_{0}\text{, some states not localized in \(e\)}.
  \end{dcases}
\end{gather}
In the thermodynamic limit, all eigenstates in the bin \(e\) are localized if \(\mbeet(L\to\infty) = 0\).
Otherwise, \(\mbeet(L\to\infty)\) takes a finite value \(\tau_{0}\) between \(0\) and \(1\).
If we take the entire spectrum to get the average \(\tau\), the dependence of \(e\) is dropped and it is written as \(\mbet\).
To confirm the localization of an eigenstate, we use the fact \updated{that} \(\tau = 0\) in the thermodynamic limit
and \updated{introduce the following linear model \(\langle\hat{\tau}(e)\rangle (L)\) to be optimized with given data of \(\mbeet\), inspired by Eq.~\eqref{eq:tauexplicit}},
\begin{gather}
  \label{eq:linearmodel}
  \langle\hat{\tau}(e)\rangle (L) = \frac{\hat{\kappa}(e)}{\ln(1/L)} + \mbeet(L\to\infty), \quad \mbeet(L\to\infty) = 0.
\end{gather}
The slope \(\hat{\kappa}(e)\) defines the participation number (an inverse of inverse participation ratio, \(\mathrm{IPR}^{-1}(e)\)) averaged over \(e\), and it is obtained by taking \(\exp(-\hat{\kappa}(e))\). 

An important quantity in such a statistical approach is the \(R^2\) \updated{obtained from the goodness-of-fit test}~\cite{steel1960principles}.
\(R^2\) reflects how well the data is described with a given statistical model; \(R^2 \approx 1\) implies that the linear model is well optimized, and the states in bin \(e\) are localized in the thermodynamic limit. 
\updated{On the other hand}, \(R^2 \leq 0\) indicates that the set of numerical data does not follow the linear model in Eq.~\eqref{eq:linearmodel} at all and cannot be appropriately optimized.
Hence, a different fitting model should be used for extended or critical states.
The definition of \(R^2\) at an energy bin \(e\) is given as
\begin{gather}
    R^{2}(e) = 1 - \sum_{L} \left(\mbeet (L) - \langle\hat{\tau}(e)\rangle (L) \vphantom{\overline{\mbeet}} \right)^{2} \Big/ \sum_{L} \left(\mbeet (L) - \overline{\mbeet} \right)^{2},
\end{gather}
where \(L\) is an index of different lattice sizes.
\(\langle\hat{\tau}(e)\rangle (L)\) represents \updated{an optimal value of \(\mbeet (L)\) obtained from numerically optimized \(\hat{\kappa}(e)\)}, and the numerator of the fraction corresponds to the residual sum of squares.
On the other hand, \(\overline{\mbeet}\) is an average of \(\mbeet (L)\) across all lattice sizes.
The denominator of the fraction is the total sum of difference squares.

The calculation of \(R^{2}\) is essential and straightforward when identifying localized states because we expect the IPR scaling exponent \(\tau\) to be zero in the thermodynamic limit.
Alternatively, the more common approach can be used to directly extract the coefficients \(\tau\) themselves.
This can be achieved, for instance, using the following scaling ansatz,
\begin{gather}
    \langle\mathrm{IPR}\rangle \sim L^{-\tau_m}.
\end{gather}
In this way, the exponents for the mean IPR, \(\tau_{m}\), can be extracted.
Localized states are identical by \(\tau_{m}=0\), delocalized states by \(\tau_{m}=1\), and critical states by \(0<\tau_{m}<1\).
This approach also allows us to distinguish between delocalized and critical states.
However, the \(R^{2}\) based approach is more convenient for identifying the localized states because we know that the IPR scaling exponent (the intercept) \(\tau\) must be zero in the thermodynamic limit.
However, diagnosing states other than the localized states is challenging or inapplicable unless one knows the exact value of \(\tau\) in the thermodynamic limit of some states.
%\clearpage
% =========================================================================== %
% =========================================================================== %
% SECTION : WEAK PERTURBATION
% =========================================================================== %
% =========================================================================== %
\section{Weak perturbation}\label{ch2:sec:weak}

We start by examining the limit of weak quasiperiodic perturbation, i.e., vanishing \(\lambda_{1,2}\), as compared to the bandgap \(\Delta\) by applying the first-order degenerate perturbation theory.
We apply the inverse unitary transformation \(U^\dagger\) to \(\mh\) so that the Hamiltonian in a fully detangled basis is \(\tilde{\mh} = \mhfd + U^{\dagger} W U\).
The second term represents hoppings solely due to the quasiperiodic perturbation.
The strongest enhancement of these hoppings occurs for \(\theta_{1,2} = \pi/4\).
%This is the case we focus on below as we expect the strongest delocalizing effect due to the maximal enhancement of the hoppings~\cite{cadez2021metal}.
Without loss of generality, we set \(\lambda_{2} \leq \lambda_{1}\).
Factorizing out \(\lambda_{1}\) from \(\tilde{\mh}\) gives us the perturbation \(\tilde{W} = W/\lambda_{1}\), which depends on the ratio \(\lambda_{2}/\lambda_{1} \in [0,1]\) only.
Then, we get the following eigenvalue problem via first-order degenerate perturbation theory,
\begin{gather}
    P_{a}U \tilde{W} U^{\dagger}P_{a}\ket{a_{n}} = \lambda_{1}\varepsilon^{(1)}_{a,n}\ket{a_{n}}.
\end{gather}
Here \(\ket{a_n}\) are the states in the fully detangled basis, and \(P_{a}\) is a projection operator onto flatband \(\varepsilon_a\) that is local.
The above equation describes an effective 1D tight binding problem called the \emph{projected model}~\cite{cadez2021metal,lee2023criticalA, lee2023criticalB}.
Fig.~\ref{fig:scheme} presents the schematics of obtaining the projected model:
without perturbation, reverting the unitary transformation \(U\) gives two sublattices of decoupled sites, while one gets onsite energies and hoppings from \(U^\dagger W U\) due to the perturbation.
The projected model, valid for weak interactions, neglects the couplings between the sublattices, producing two decoupled projected models.
The flatband energy we want to focus on dictates the choice of the sublattice/projected model.
On general grounds, we expect the effective model to feature both quasiperiodic onsite energies and finite-range hopping, \updated{given that \(\tilde{W}\) is local and quasiperiodic, and \(U^\dagger P_a\) is a local operator.}
\updated{In the fully detangled basis, the effective problem for a particular sublattice (such as the lower leg in Fig.~\ref{fig:scheme}) is as follows,}
\begin{gather}
    \label{eq:pm}
    E a_{n} = v_{n}a_{n} + t_{n-1}a_{n-1} + t_{n}^{*}a_{n+1},
\end{gather}
where the onsite potential \(v_{n}\) and the hopping \(t_{n}\) are indeed quasiperiodic,
\begin{align}
    \label{eq:coefficients:v}
    v_{n} &= v_{s}\sin(2\pi\alpha n - \pi\alpha) + v_{c}\cos(2\pi\alpha n - \pi\alpha), \\
    \label{eq:coefficients:t}
    t_{n} &= t_{s}\sin(2\pi\alpha n) + t_{c}\cos(2\pi\alpha n).
\end{align}
\updated{The expressions for the amplitudes \(v_{s,c}\) and \(t_{s,c}\) of \(v_{n}\) and \(t_{n}\) are quite intricate, as they contain details from both quasiperiodic fields and the local unitary transformation. 
Those amplitudes are obtained as follows (no derivation shown),}
\begin{align}
    v_{s} &= \sin(\pi\alpha)\left(\sin^2\theta_{1}\sin^2\theta_{2} - \cos^2\theta_{1}\cos^2\theta_{2}\right) \vphantom{\frac{\lambda_{2}}{\lambda_{1}}} \\
    &+ \frac{\lambda_{2}\cos\beta}{\lambda_{1}}\sin(\pi\alpha)\left(\sin^2\theta_{1}\cos^2\theta_{2} - \cos^2\theta_{1}\sin^2\theta_{2}\right) \notag \\
    &- \frac{\lambda_{2}\sin\beta}{\lambda_{1}}\cos(\pi\alpha)\left(\cos^2\theta_{1}\sin^2\theta_{2} + \sin^2\theta_{1}\cos^2\theta_{2}\right), \notag \\
    v_{c} &= \cos(\pi\alpha)\left(\cos^2\theta_{1}\cos^2\theta_{2} + \sin^2\theta_{1}\sin^2\theta_{2}\right) \vphantom{\frac{\lambda_{2}}{\lambda_{1}}} \\
    &+ \frac{\lambda_{2}\sin\beta}{\lambda_{1}}\sin(\pi\alpha)\left(\sin^2\theta_{1}\cos^2\theta_{2} - \cos^2\theta_{1}\sin^2\theta_{2}\right) \notag \\
    &+ \frac{\lambda_{2}\cos\beta}{\lambda_{1}}\cos(\pi\alpha)\left(\cos^2\theta_{1}\sin^2\theta_{2} + \sin^2\theta_{1}\cos^2\theta_{2}\right), \notag \\
    t_{s} &= \frac{1}{4}\sin 2\theta_{1}\sin 2\theta_{2}\frac{\lambda_{2}\sin\beta}{\lambda_{1}} \label{eq:weakhopping_sin}\\
    t_{c} &= \frac{1}{4}\sin 2\theta_{1}\sin 2\theta_{2}\left(1 - \frac{\lambda_{2}\cos\beta}{\lambda_{1}}\right). \label{eq:weakhopping_cos}
\end{align}

% =========================================================================== %
% SUBSECTION : Extended Harper model
% =========================================================================== %
\subsection{Extended Harper model}

The projected model in Eq.~\eqref{eq:pm} features both a quasiperiodic onsite potential and hopping.
The most generic model of this type is defined by a self-adjoint quasiperiodic Jacobi operator \(J\) acting on a vector of \(l^2 (\mathbb{Z})\) space~\cite{teschl2000jacobi},
\begin{gather}\label{eq:jacobi}
	(Ju)_{n} = v(\alpha n + \phi)u_{n} + c_{\rho}(\alpha n + \phi)u_{n+1} + c_{\rho}^{*}(\alpha(n-1) + \phi)u_{n-1},
\end{gather}
where \(\alpha\) is an irrational spatial frequency and \(\phi\) is a fixed phase, and \(\rho\) is some additional parameter controlling the hopping \(t\).
No complete phase diagram or transport properties --- numerical or analytical --- have been established for this most generic case.
However, the projected model can be mapped onto the already studied extended Harper model for specific parameter values, as discussed below.
The definition of the extended Harper model is given as follows, 
\begin{align}
  \label{eq:EHM}
  (Ju)_{n} &= 2\cos(2\pi\alpha n - \pi\alpha)u_{n} \\
  & + 2\rho\left[\cos(2\pi\alpha n)u_{n+1} + \cos(2\pi\alpha (n-1))u_{n-1}\right]. \notag
\end{align}
Its spectral properties have been characterized completely in Ref.~\cite{avila2017spectral}.
For \(2\rho < 1\), the eigenstates are all localized, as guaranteed by the RAGE theorem~\cite{ruelle1969remark,amrein1973characterization,enss1978asymptotic} \updated{(see the introduction of Chapter~\ref{chpt1})}.
Otherwise, for \(2\rho \geq 1\), the energy spectrum has a fractal structure~\cite{avila2017spectral}, similar to the Aubry-Andr\'e-Harper model at its critical point.
\updated{In this chapter, such transition is called as} a critical-to-insulator transition analogous to a metal-to-insulator transition.
Moreover, the corresponding eigenstates show multifractal behavior, as demonstrated numerically in~\cite{chang1997multifractal}.

% =========================================================================== %
% SUBSECTION : Off-diagonal Harper model
% =========================================================================== %
\subsection{Off-diagonal Harper model}

The onsite potential can be neglected for effectively infinite \(2\rho\), and we are left with the hopping terms only.
The model in this limit is called the off-diagonal Harper model (OHM)~\cite{han1994critical,kraus2012topological},
\begin{gather}
    \label{eq:OHM}
    (Ju)_{n} = 2\rho\left[\cos(2\pi\alpha n)u_{n+1} + \cos(2\pi\alpha (n-1))u_{n-1}\right].
\end{gather}
It is self-dual under a modified Fourier transformation similar to the duality of the Aubry-Andr\'e model.~\cite{lee2023criticalA}.
\begin{equation}
\ket{k} = \frac{1}{\sqrt{L}}\sum_{n}e^{-i2\pi\alpha n k}\ket{a_{n}}.
\end{equation}
Substituting the above transformation to the projected model, we obtain the Hamiltonian has a similar structure in real-space and Fourier-space.
Consequently, the entire spectrum of the model is critical, and the eigenstates are multifractal,
\begin{align}\label{Fstquantized}
    H &= 2\rho\sum_{n}\cos(2\pi\alpha n)\left( \vphantom{\sum}{\ket{a_{n}}}\bra{a_{n+1}} + \ket{a_{n+1}}\bra{a_{n}} \right) \notag \\
    % &= \frac{\rho}{L} \sum_{n, k, k^{\prime}}\left(e^{i2\pi\alpha n (k - k^{\prime} +1)} + e^{i2\pi\alpha n (k - k^{\prime} -1)}\right) \ket{k}\bra{k^{\prime}} e^{-i2\pi\alpha k^{\prime}} \notag \\
    % &+ \frac{\rho}{L} \sum_{n, k, k^{\prime}}\left(e^{i2\pi\alpha n (k - k^{\prime} +1)} + e^{i2\pi\alpha n (k - k^{\prime} -1)}\right) \ket{k}\bra{k^{\prime}} e^{i2\pi\alpha k} \notag \\
    %&= \rho \sum_{k}\left( e^{-i2\pi\alpha (k+1)} + e^{i2\pi\alpha k} \right)\ket{k}\bra{k+1} + \left( e^{-i2\pi\alpha(k-1)} + e^{i2\pi\alpha k} \right) \ket{k}\bra{k-1} \notag \\
    %&= \rho \sum_{k}\left( e^{-i2\pi\alpha (k+1)} + e^{i2\pi\alpha k} \right)\ket{k}\bra{k+1} + \left( e^{-i2\pi\alpha k} + e^{i2\pi\alpha (k+1)} \right) \ket{k+1}\bra{k} \notag \\
    &= 2\rho \sum_{k} \cos(\pi\alpha (2k+1)) \left( \vphantom{\sum} e^{-i\pi\alpha} \ket{k}\bra{k+1} + e^{i\pi\alpha} \ket{k+1}\bra{k} \right)
\end{align}

% =========================================================================== %
% SUBSECTION : Lyapunov exponent and localization length
% =========================================================================== %
\subsection{Lyapunov exponent and localization length}

After first realizing that the distribution of eigenvalues and the electron localization in a disordered system are related in Ref.~\cite{herbert1971localized}, Thouless came up with a simple argument that can be applied to any random one-dimensional tight-binding system with next nearest neighbor hopping~\cite{thouless1972relation},
\begin{gather}
    E_{\beta}u_{n}^{\beta} = v_{n}u_{n}^{\beta} - t(u_{n+1}^{\beta} + u_{n-1}^{\beta}).
\end{gather}
His approach was to take the retarded Green function \(G^{+}(E)\) of the system and obtain the matrix element \(G^{+}_{1N}(E)\), which gives the transition amplitude of an electron from the first to the last site using the fact that the corresponding cofactor (upper triangular matrix) is the product of diagonal elements,
\begin{gather}
    G^{+}(E) = \left(E - H + i\varepsilon\right)^{-1} = \frac{\operatorname{adj}(E - H + i\varepsilon)}{\det(E - H + i\varepsilon)}, \\
    G_{1N}^{+}(E) = \langle 1 \vert G^{+}(E) \vert N \rangle = t^{N-1}\left[\prod_{\alpha = 1}^{N}(E - E_{\alpha} + i\varepsilon)\right]^{-1}.
\end{gather}
\updated{Now, let us focus on a specific eigenvalue \(E_{\beta}\).}
Then, the explicit calculation of \(|G_{1N}^{+}(E_{\beta})|\) gives the following equation,
\begin{gather}
     |G_{1N}^{+}(E_{\beta})| = |u_{1}^{\beta}u_{N}^{\beta}| = |t^{N-1}|\left[ \prod_{\alpha \neq \beta}^{N} \big\vert E_{\beta} - E_{\alpha} + i\varepsilon)\big\vert\right]^{-1}.
\end{gather}
If \(u_{1}^{\beta}\) and \(u_{N}^{\beta}\) are exponentially decaying with its maximum value at the site \(m\), the product of \(u_{1}^{\beta}\) and \(u_{N}^{\beta}\) is the following,
\begin{gather}
    |u_{1}^{\beta}u_{N}^{\beta}| = \exp(-\lambda_{\beta}(m-1) - \lambda_{\beta}(N-m)) = \exp(-\lambda_{\beta}(N - 1)).
\end{gather}
The exponent \(\lambda_{\beta}\) is obtained as follows, and it is called \emph{Thouless formula},
\begin{gather}
    \lambda_{\beta} = \!\! \lim_{N\to\infty}\frac{1}{N - 1}\sum_{\alpha \neq\beta}\ln |E_{\beta} - E_{\alpha}| - \ln t
     = \!\! \int_{\mathbb{R}} \!\! \ln|E_{\beta} - E^{\prime}|\rho(E^{\prime})dE^{\prime} - \ln t.
\end{gather}
In one-dimensional case, especially, an inverse of \(\lambda_{\beta}\) defines the localization length \(\xi(E)\) at the energy \(E_{\beta}\). 
For the continuum case, the summation becomes the integral described with a (integrated) density of state \(\rho(E)\).
Later on, several mathematicians rigorously argued the Thouless formula that \(\lambda_{\beta}\) turned out to be the Lyapunov exponent of the associated transfer matrix.

The preceding analysis of a Lyapunov exponent was focused on a random system.
However, the Thouless formula may not be applicable for a quasiperiodic system.
To address this challenge, Avila came up with a theory that extends the formula for a Lyapunov exponent to one-dimensional quasiperiodic systems with nearest neighbor hoppings~\cite{avila2015global}, so-called the \emph{global theory}.
First, we write a transfer matrix for a quasiperiodic Jacobi operator (E.~\eqref{eq:jacobi}) in the following form,
\begin{gather}
    \label{eq:transfer}
    T(E;\varphi) = \frac{1}{c_{\rho}(\varphi)}B(E;\varphi) \quad\text{and}\quad
    B(E;\varphi) = \begin{bmatrix}
        E - v(\varphi) & -c_{\rho}^{*}(\varphi - \alpha) \\
        c_{\rho}(\varphi) & 0
    \end{bmatrix}.
\end{gather}
The angle \(\varphi\) is \(\alpha n + \phi\).
\(B(E;\varphi)\) is a \emph{non-singular}, ensuring that the matrix is always well-defined and avoids issues with division by zero for each element~\cite{jitomirskaya2012analytic}.
Then, the product of the transfer matrices at a given energy \(E\) gives us the update of the corresponding wave function elements,
\begin{gather}
    T^{(n)}(E;\varphi) = T(E;\varphi+(n-1)\alpha)\cdots T(E;\varphi),
\end{gather}
and the averaged asymptotic of the quasiperiodic cocycle can also be quantified with Lyapunov exponent by definition,
\begin{align}
    \lambda(E) &= \lim_{n\to\infty}\frac{1}{n}\int_{\mathbb{T}}\ln \lVert T^{(n)}(E;\varphi) \rVert d\varphi \\
    &= \gamma(E) - \lim_{n\to\infty}\frac{1}{2\pi n}\int_{0}^{2\pi} \!\!\! \ln \vert c(\varphi) \vert d\varphi \\
    \gamma(E) &= \lim_{n\to\infty}\frac{1}{2\pi n}\int_{0}^{2\pi} \!\!\! \ln \lVert B^{(n)}(E;\varphi) \rVert d\varphi
\end{align}
The matrix norm is defined as the largest Lyapunov exponent.
The main strategy of the global theory is to complexify the Lyapunov exponent by making \(\varphi \to \varphi + i\varepsilon\).
By then considering the asymptotic limit \(|\varepsilon| \to 0\), we can effectively compute \(\lambda(E)\) and perform extrapolation to \(\varepsilon = 0\).
This method enables us to evaluate the Lyapunov exponent accurately for the one-dimensional quasiperiodic system,
\begin{gather}
    \lambda_{\varepsilon}(E) = \lim_{n\to\infty} \frac{1}{2\pi n}\int_{0}^{2\pi} \!\!\! \ln \lVert T^{(n)}(\varphi + i\varepsilon) \rVert d\varphi.
\end{gather}
The integral mentioned above is split into two distinct terms.
The first term represents the Lyapunov exponent of the complexified \(B(E;\theta)\), while the second term is the contribution of quasiperiodic hopping, as evident from Eq.~\eqref{eq:transfer},
\begin{gather}
    \lambda_{\varepsilon}(E) = \lambda(\alpha, B(E;\varphi+i\varepsilon)) - \int_{\mathbb{T}} \! \ln|c(\varphi)| d\varphi.
\end{gather}

For the extended Harper model defined in Eq.~\eqref{eq:EHM}, the Lyapunov exponent of the complexified \(B(E;\theta)\) is given in Ref.~\cite{jitomirskaya2012analytic} and the contribution of quasiperiodic hopping is explicitly obtained.
When \(\rho\) is equal to or larger than \(1/2\), the Lyapunov is zero; otherwise, we have a positive and finite value,
\begin{gather}
    \label{eq:lypEHM}
    \lambda(E) = \left( \ln \Big\vert \frac{1 + \sqrt{1 - 4\rho^{2}}}{2} \Big\vert + \cancel{2\pi\varepsilon} \right) - \left(\ln\vert \rho \vert + \cancel{2\pi\varepsilon} \vphantom{\sum}\right).
\end{gather}
Consequently, the Lyapunov exponent remains independent of energy.
This observation suggests that we can consider the entire spectrum as our energy bin, and then calculate the average of \(\tau\), which we will denote as \(\mbet\).

\begin{figure}
\centering
  \includegraphics[width = \columnwidth]{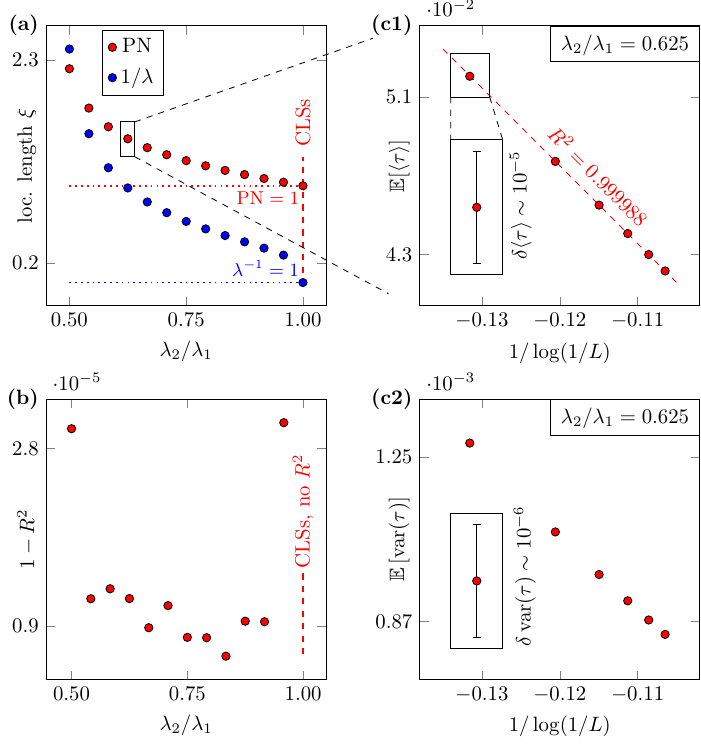}
  \caption[Localization lengths of extended Harper models]{
    (a) Two different localization length are plotted.
    The inverse of the Lyapunov exponent (\(1/\lambda\)) [Eq.~\eqref{eq:lypEHM}] and the participation number (PN) diverges as \(\lambda_{2}/\lambda_{1}\) approaches to the transition point \((\lambda_{2}/\lambda_{1})_{c} \approx 0.468\)
    (b) Plot of \(1-R^{2}\). It shows that data follows the linear model in Eq.~\eqref{eq:linearmodel} very well.
    (c1) and (c2) show the detailed statistics at \(\lambda_{2}/\lambda_{1} = 0.625\).
    \updated{\(\mathbb{E}[\langle\tau\rangle]\) is an average of mean scaling exponent over different realizations, and \(\mathbb{E}[\operatorname{var}(\tau)]\) is an average of its variance over different realizations.}
  }
  \label{fig:lyapunov}
\end{figure}

Through numerical simulations, we calculate the inverse participation ratio with the energy bin encompassing the entire spectrum for lattice sizes \(L = 2000, 4000, \dots 12000\) in steps of \(2000\). 
\(R^{2}\) values are found to be closely approximating \(1\), as depicted in Fig.~\ref{fig:lyapunov}.
Notably, the standard error of \(\mbet\) is approximately \(10^{-5}\) with a magnitude one thousand times smaller than \(\mbet\) itself, as highlighted in Figure \ref{fig:lyapunov} (c1).
In Fig.~\ref{fig:lyapunov} (c2), the variance of \(\tau\)s and its fluctuations are extremely small compare to \(\mbet\).

Some important mathematical properties have been established in the existing literature for the off-diagonal Harper model described in Eq.~\eqref{eq:OHM}.
Notably, it has been proven that there is no absolute continuous spectrum~\cite{jitomirskaya2019critical}, implying the absence of extended states.
Additionally, it is known that there exists no pure point spectrum~\cite{avila2017spectral}, signifying the nonexistence of localized states.
Based on these findings, we expect the Lyapunov exponent to be zero, indicating critical behavior for the corresponding eigenstate.
Explicit calculations of the Lyapunov exponent for the off-diagonal Harper model are possible.
First, the complexified non-singular transfer matrix is defined as follows with \(c_{\rho}(\varphi + i\varepsilon) = 2\rho \cos(2\pi(\varphi + i\varepsilon))\).
At \(\varepsilon \to \infty\), \(B(E;\varphi+i\varepsilon)\) is further simplified,
\begin{gather}
    B(E;\varphi + i\varepsilon) = \begin{bmatrix}
    E   & c_{\rho}(\varphi - \alpha + i\varepsilon) \\
    c_{\rho}(\varphi + i\varepsilon) & 0
    \end{bmatrix} \approx \rho e^{2\pi\varepsilon}M_{\varepsilon}(\varphi,\varepsilon) \\
    M_{\varepsilon}(\varphi,\varepsilon) = e^{-i2\pi(\varphi - \alpha/2)}\begin{bmatrix}
    \mathcal{O}\left(\exp(-\varepsilon)\right) & -e^{i\pi\alpha} \\ e^{-i\pi\alpha} & 0
    \end{bmatrix}.
\end{gather}
Then the corresponding complexified Lyapunov exponent is obtained as follows,
\begin{align}
    \gamma_{\varepsilon}(E) = \lim_{n\to\infty}\frac{1}{2\pi n}\int_{0}^{2\pi} \!\!\!\! d\theta \ln \lVert B^{(n)}(E,\varphi+i\varepsilon)\rVert \approx \ln\vert\rho\vert + 2\pi\varepsilon.
\end{align}
Finally, the Lyapunov exponent of the off-diagonal Harper model at energy \(E\) is zero.
This result indicates a diverging localization length, signifying critical behavior for the corresponding eigenvector,
\begin{gather}
    \label{eq:lypOHM}
    \lambda(E) = \left(\cancel{\ln\vert\rho\vert} + \cancel{2\pi\varepsilon} \vphantom{\sum}\right) - \left(\cancel{\ln\vert\rho\vert} + \cancel{2\pi\varepsilon} \vphantom{\sum}\right) = 0.
\end{gather}

% =========================================================================== %
% SUBSECTION : Case 1
% =========================================================================== %
\subsection{Case 1 -- \texorpdfstring{\(\beta = 0\)}{zero beta} and \texorpdfstring{\(\lambda_{1}\neq\lambda_{2}\)}{lambda1 neq lambda2} at \texorpdfstring{\(\theta_{1,2} = \pi/4\)}{theta12 eq pi/4}}

\begin{figure}
\centering
  \includegraphics[width = \columnwidth]{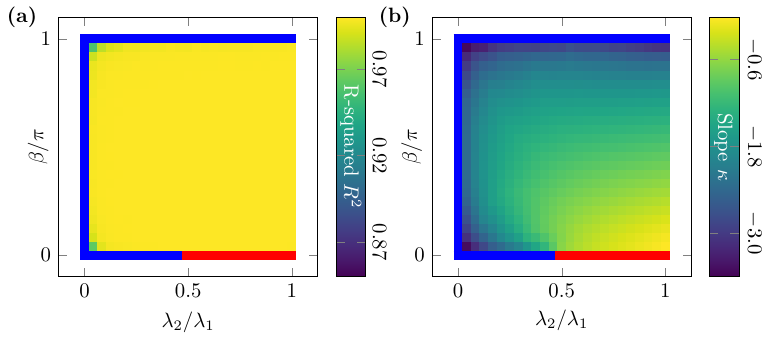}
  \caption[Phase diagram at \(\theta_{1,2} = \pi/4\)]{
    Phase diagram of the projected model based on the values of the exponent \(\tau\) averaged over the spectrum, \(\mbet\), computed for \(25\times 25\) values of parameters \(\lambda_2/\lambda_1, \beta/\pi\).
    The blue and red lines on the borders indicate the critical and localized regimes, respectively.
    (a) \(R^2\) values for parameters \(\lambda_2/\lambda_1, \beta/\pi\).
    All points have values very close to \(1\), implying that all eigenstates are localized in the thermodynamic limit in this entire region.
    (b) Slope \(\hat{\kappa}\)~\eqref{eq:linearmodel} for the parameters away from the border of the phase diagram.
    The absolute value of the slope increases closer to the border, implying larger localization length.
  }
  \label{fig:phasediagram}
\end{figure}

Let us connect the projected model with the extended and off-diagonal Harper models.
We first start our mapping by choosing \(\beta = 0\) and \(\lambda_{1}\neq\lambda_{2}\). 
The onsite potential and hopping in Eq.~\eqref{eq:pm} are reduced to the following expressions, which gives us the extended Harper model,
\begin{align}
    v_{n} &= \left[\frac{1}{4}\left(\frac{\lambda_{2}}{\lambda_{1}} + 1\right)\cos(\pi\alpha)\right]2\cos(2\pi\alpha n - \pi\alpha),\\
    t_{n} &= \left[\frac{1}{4}\left(1 - \frac{\lambda_{2}}{\lambda_{1}}\right)\right]\cos(2\pi\alpha n).
\end{align}
\(2\rho\) in Eq.~\eqref{eq:EHM} can be either larger or smaller than \(1\) depending on the values of \(\alpha\) and \(\lambda_{2}/\lambda_{1}\) where the critical-to-insulator transition is at \(2\rho=1\). \updated{In this case, the formula for \(2\rho\) is easily obtained as follows,}
\begin{gather}
  \label{eq:transition}
  2\rho = \left|\frac{1}{\cos(\pi\alpha)}\right| \left|\frac{1 - \lambda_{2}/\lambda_{1}}{1 + \lambda_{2}/\lambda_{1}}\right|.
\end{gather}
For \(\alpha = (\sqrt{5}-1)/2\), the phase transition point is \((\lambda_{2}/\lambda_{1})_c \approx 0.468\).
As we have seen from Eq.~\eqref{eq:lypEHM}, the localization length remains independent of energy.
Then, we can take the entire spectrum as our energy bin and calculate the average of \(\tau\), denoted as \(\mbet\).
In Fig.~\ref{fig:phasediagram}, we compare the participation number and the Lyapunov exponents of the projected model.

We point out that when the quasiperiodic amplitudes are equal (\(\lambda_1=\lambda_2\)) in, the hopping \(t_n\) vanishes in the projected model, which makes it diagonal and leaves only the non-zero onsite potential.
Hence, the eigenstates of the projected model exhibit compact localization with eigenstates occupying a single site.
Since the \updated{diagonal values are all different}, the eigenenergies have no degeneracy.
In numerical perspective, \(R^{2}\) is not defined in this case.

% =========================================================================== %
% SUBSECTION : Result : \(\lambda_{2}/\lambda_{1} = 0\) for any phase difference \(\beta\)
% =========================================================================== %
\subsection{Case 2 -- \texorpdfstring{\(\lambda_{2}/\lambda_{1} = 0\)}{zero ratio} for any phase difference \texorpdfstring{\(\beta\)}{beta}}\label{sec:mappingEHM}

If we only apply the quasiperiodic potential~\eqref{eq:fields} to one leg of the ladder, e.g. for example \(\lambda_{2}/\lambda_{1} = 0\),
then the projected model~\eqref{eq:pm} simplifies: \(\tau_{s}\) is zero and \(v_{s}\) takes the following form,
\begin{gather}
    v_{s} = \sin(\pi\alpha)\cos(\theta_1 - \theta_2)\cos(\theta_2 + \theta_1).
\end{gather}
Then, the model~\eqref{eq:pm} maps to the extended Harper model for \(v_{s} \equiv 0\) only, which produces the following constraint on the angles \(\theta_{1,2}\),
\begin{gather}
    \label{eq:multifractalrelation}
    \theta_{2} + \theta_{1} = \frac{\pi}{2}.
\end{gather}
Using this to eliminate \(\theta_{2}\), we obtain the following expressions for \(v_{c}\) and \(\tau_{c}\),
\begin{align}
    v_{c} &= 2\cos(\pi\alpha)\sin^2\theta_{1}\cos^2\theta_{1},\\
    t_{c} &= \sin\theta_{1}\cos\theta_{1}\sqrt{1 - \cos 4\theta_{1}}/\sqrt{8}.
\end{align}
Note that both quantities vanish for \(\theta_1=0,\pi/2\), producing a set of decoupled sites.
Then for \(\theta_1\neq 0,\pi/2\) the control parameter, \(\rho\) of the EHM~\eqref{eq:EHM} is expressed as follows in terms of \(\theta_1\),
\begin{gather}
    \label{eq:commonhopping}
    2\rho = \left|\frac{\sqrt{1 - \cos 4\theta_{1}}/\sqrt{8}}{\cos(\pi\alpha)\sin\theta_{1}\cos\theta_{1}}\right| = \left|\frac{1}{\cos(\pi\alpha)}\right| \geq 1.
\end{gather}
The absolute value above ensures that the negative \(t_{c}\) is also covered (the sign can be trivially gauged away) and also to match with the definition of the extended Harper model~\cite{avila2017spectral} where non-negative \(2\rho\) is assumed.
\begin{figure}[ht]
\centering
    \includegraphics[width = \columnwidth]{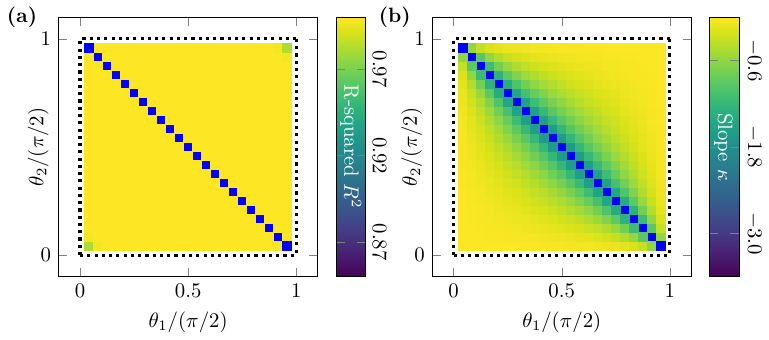}
    \caption[More mapping to extended Harper model]{
        Phase diagram \((\theta_1, \theta_2)\) of the projected model~\eqref{eq:pm} for \(\lambda_2/\lambda_1 = 0\), e.g. one of the quasiperiodic fields is absent.
        The yellow dots indicate the fully localized spectrum.
        The diagonal blue line indicates the region of critical spectrum given by Eq.~\eqref{eq:multifractalrelation}.
        The black dotted lines on the border correspond to the case of disconnected dimers: all eigenstates are compactly localized.
        (a) \(R^2\) values for parameters \(\theta_{1},\theta_{2}\): All eigenstates are localized in the thermodynamic limit except the diagonal.
        (b) Slope \(\hat{\kappa}\)~\eqref{eq:linearmodel} for the parameters away from the diagonal of the phase diagram.
        The absolute value of the slope \(\kappa\) increases upon approaching the diagonal, implying  a larger localization length.
    }
    \label{fig:phasediagramEHM}
\end{figure}
For all the other \(\theta_{1,2}\), e.g. not satisfying the above condition~\eqref{eq:multifractalrelation}, we establish the character of the spectrum --- localized or not --- numerically.
We scanned the full parameter region, \(0<\theta_{1,2}<\pi/2\), discretized into the \(25\times 25\) grid.
The exponent \(\mbet\) averaged over the entire spectrum is computed via Eq.~\ref{eq:linearmodel} for lattice sizes \(L = 2000, 4000, \dots 12000\) in steps of \(2000\).
The results are summarized in Fig~\ref{fig:phasediagramEHM}: Apart from the diagonal line, all points have the \(R^2\)-values very close to \(1\).
That is, all the eigenstates of the projected model~\eqref{eq:pm} are localized in the thermodynamic limit whenever it does not map onto the extended Harper model, e.g., away from the diagonal \(\theta_1 + \theta_2 = \pi/2\).
The increase of the absolute value of the slope \(\hat{\kappa}\)~\eqref{eq:linearmodel} towards the diagonal line is related to the increase of the localization length as we approach the transition to critical states.

% =========================================================================== %
% SUBSECTION : Result : \(\lambda_{2}/\lambda_{1} = 0\) for any phase difference \(\beta\)
% =========================================================================== %
\subsection{Case 3 -- \texorpdfstring{\(\lambda_{2} = \lambda_{1}\)}{one ratio} and \texorpdfstring{\(\beta = \pi\)}{pi beta}}
\label{sec:mappingOHM}

%\noindent\textcolor{blue}{Mapping to OHM}
For \(\theta_2=\pi/4\), equal potential strengths \(\lambda_{2} = \lambda_{1} = 1\) and phase difference \(\beta = \pi\), \(v_{s,c} = t_{s} = 0\) and we are left with only the hopping terms \(t_{c} = \sin 2\theta_{1}/2\) in the model~(\ref{eq:pm}, \ref{eq:coefficients:t}).
The projected model becomes precisely equivalent to the off-diagonal Harper model, and the role of \(t_{c}\) is simply to rescale the energy,
\begin{gather}
    \label{eq:OHM1}
    Ea_{n} = t_{c}\left[\cos(2\pi\alpha n)a_{n+1} + \cos(2\pi\alpha (n-1))a_{n-1}\right].
\end{gather}
%We have established in Ref.~\cite{lee2023criticalA} an almost diffusive spreading of an initially localized wavepacket in the off-diagonal Harper model.
%We discuss soon in the next section in detail.
%Now the factor \(t_c\) only sets the global timescale for the wavepacket spreading.
\begin{figure}
\centering
    \includegraphics[width = \columnwidth]{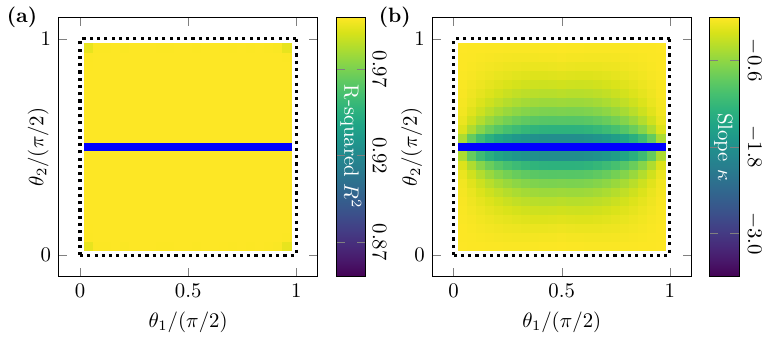}
    \caption[More mapping to off-diagonal Harper model]{
        Phase diagram \((\theta_1,\theta_2)\) of the projected model~\eqref{eq:pm} for equal strength of the quasiperiodic potentials \(\lambda_2=\lambda_1\) and phase difference \(\beta = \pi\).
        The yellow dots indicate the fully localized spectrum.
        The horizontal blue line, \(\theta_{2} = \pi/4\), the models with critical spectrum, which map to the off-diagonal Harper mode.
        The black dotted lines on the border correspond to the system of disconnected dimers and compactly localized eigenstates.
        (a) \(R^2\) values for parameters \(\theta_{1},\theta_{2}\).
        Aside from the horizontal blue line, all points have the values very close to \(1\): all  eigenstates are localized in the thermodynamic limit.
        (b) Slope \(\hat{\kappa}\)~\eqref{eq:linearmodel} for the parameters away from the horizontal of the phase diagram.
        The absolute value of the slope $\kappa$ increases upon approaching the diagonal, implying  a larger localization length.
    }
    \label{fig:phasediagramOHM}
\end{figure}
For \(\theta_{2}\neq \pi/4\), we carried out a numerical scan of the spectrum of the projected model over the region \(0<\theta_{1,2}<\pi/2\), discretized into \(25\times 25\) grid.
The fitting procedure of \(\mbet\) (over the entire spectrum) is performed via Eq.~\ref{eq:linearmodel} for lattice sizes \(L = 2000, 4000, \dots 12000\) in steps of \(2000\).
The observed results follow closely those of the case \(\lambda_2/\lambda_1 = 0\):
all the eigenstates of the projected model~\eqref{eq:pm} are localized in the thermodynamic limit away from the horizontal line of the phase diagram, 
e.g., when the projected model does not map onto the off-diagonal Harper model, as shown in Fig.~\ref{fig:phasediagramOHM}.
The increase of the absolute value of the slope towards the horizontal line is related to the increase of the localization length towards the line of critical states.

% =========================================================================== %
% SUBSECTION : Lyapunov exponent and localization length
% =========================================================================== %
\subsection{Remark -- Near zero angles \texorpdfstring{\(\theta_{1,2}\to 0\)}{theta12 to zero}}

In this section, we focus on the case of \(\theta_{1,2}\to 0\) and demonstrate that all the eigenstates of the projected model~\eqref{eq:pm} are effectively compactly localized.
To see this we approximate \(\sin\theta \approx \theta\) and \(\cos\theta \approx 1\), so that the coefficients~(\ref{eq:coefficients:v}-\ref{eq:coefficients:t}) simplify to
\begin{align}
    v_{s} &= \sin(\pi\alpha)(\theta_{1}^{2}\theta_{2}^{2} - 1) \\
    &+ \frac{\lambda_{2}}{\lambda_{1}}\cos\beta \sin(\pi\alpha)(\theta_{1}^{2} - \theta_{2}^{2}) - \frac{\lambda_{2}}{\lambda_{1}}\sin\beta \cos(\pi\alpha)(\theta_{2}^{2} + \theta_{1}^{2}) \notag \\
    v_{c} &= \cos(\pi\alpha)(1 + \theta_{1}^{2}\theta_{2}^{2}) \\
	&+ \frac{\lambda_{2}}{\lambda_{1}}\sin\beta\sin(\pi\alpha)(\theta_{1}^{2} - \theta_{2}^{2}) + \frac{\lambda_{2}}{\lambda_{1}}\cos\beta\cos(\pi\alpha)(\theta_{2}^{2} + \theta_{1}^{2}) \notag \\
    t_{s} &= \theta_{1}\theta_{2}\frac{\lambda_{2}\sin\beta}{\lambda_{1}} \\
    t_{c} &= \theta_{1}\theta_{2}\left(1 - \frac{\lambda_{2}\cos\beta}{\lambda_{1}}\right)
\end{align}
Keeping only linear terms in \(\theta_{1,2}\), so that \(\theta_{1}^{2} = \theta_{2}^{2} = \theta_{1}\theta_{2} = 0\), 
we see that the hoppings all vanish \(t_{s} = t_{c} = 0\) and only the onsite potential terms are left.
Then, all the eigenstates are effectively compactly localized, but their energies are non-degenerate because of the onsite potential.
A similar argument implies that all states are compactly localized for \(\theta_{1,2} \approx \pi/2\).

%\clearpage
% =========================================================================== %
% =========================================================================== %
% SECTION : WAVEPACKET SPREADING
% =========================================================================== %
% =========================================================================== %
\section{Wavepacket spreading in projected model}\label{ch2:sec:spreading}

The critical states are exhibited in both the extended and off-diagonal Harper models.
Nonetheless, during our investigation of their transport properties, there are notable qualitative distinctions in the eigenstates of each model.
Firstly, we provide an overview of the established theoretical findings concerning the lower and upper bounds of diffusion exponents in quasiperiodic systems.
Subsequently, we see the numerical results of the projected models at \(\theta_{1,2} = \pi/4\), which are mapped to the extended and the off-diagonal Harper model.

% =========================================================================== %
% SUBSECTION : Lower bound of diffusion exponent
% =========================================================================== %
\subsection{Lower bound of diffusion exponent}

The diffusion exponent \(\gamma\) is obtained from the deviation of the particle's position at time \(t\), \(\hat{X}_{\phi}(t)\).
The explicit definition is given below~\cite{hiramoto1988dynamics,geisel1991new,passaro1992anomalous}, and we keep the dimension of a system to be one since we only deal with one-dimensional systems,
\begin{gather}
    \lim_{t\to\infty}\langle 0 \vert |\hat{X}_{\phi}(t) - \hat{X}(0)|^2 \vert 0 \rangle \sim t^{2\gamma}, \\
    \hat{X}_{\phi}(t) = \exp(iH_{\phi}t)\hat{X}\exp(-iH_{\phi}t).
\end{gather}
The cases of \(\gamma = 1\) signify ballistic motion, \(\gamma = 1/2\) corresponds to regular diffusion, and \(\gamma = 0\) represents a localization of an initial wavepacket.
When \(\gamma\) is between \(0\) and \(1/2\), it represents a subdiffusion.
If \(\gamma\) is between \(1/2\) and \(1\), it corresponds to a superdiffusion.
In the context of a one-dimensional lattice with a fractal-like energy spectrum~\cite{guarneri1989spectral,combes1993connections, last1996quantum}, there exists a rigorous lower limit for the diffusion exponent, denoted as \(\gamma^{-}\), which falls within \(0 < \gamma^{-} \leq 1/2\).

% =========================================================================== %
% SUBSECTION : Upper bound of diffusion exponent
% =========================================================================== %
\subsection{Upper bound of diffusion exponent}

Concerning the Aubry-Andr\'e-Harper model at its critical point, the energy spectrum has a fractal-like structure.
In this scenario, some numerical observations claimed numerically that the corresponding diffusion exponent is almost near to \(1/2\),
without any theoretical proof~\cite{hiramoto1988dynamics, wilkinson1994spectral}.
Using the translational covariance relation that quasiperiodic systems have~\cite{bellissard1995noncommutative}, \(U_{n}H_{\phi}U_{n}^{-1} = H_{\tau^{n}\phi}\), it is possible to obtain the upper bound of the corresponding diffusion exponent.
The operator \(\tau\) is a group action that constructs an orbit, and \(U_{n}\) is a shift operator which shifts to the \(+n\) site.
\(H(\theta)\) is the Hamiltonian.
In the case of the projected model, \(\tau^{n}\phi\) equals \(\phi + \alpha n\).
Furthermore, the two-point correlation measure strongly connects with the current-current correlation measure \(m\)~\cite{bellissard1995noncommutative},
\begin{gather}
    m(dE,dE^{\prime}) = \frac{1}{L}\sum_{n,m}\delta(E - E_{n})\delta(E^{\prime} - E_{m})  |\langle E_{n} \vert J \vert E_{m} \rangle|^2 dE dE^{\prime},
\end{gather}
where \(J = i[H,x] = \nabla H\).
In Ref.~\cite{schulz1998anomalous}, the Stieltjes transform of the current-current correlation \(m\) allows us to calculate the diffusion exponent.
\begin{gather}
    S_{m}(z_{1},z_{2}) \approx \int_{\mathbb{R}^{2}}\frac{m(dE,dE^{\prime})}{(E - z_{1})(E^{\prime}-z_{2})}.
\end{gather}
The theorem 10 in Ref.~\cite{schulz1998anomalous} states that the real value of the Stieltjes form \(S\) and the diffusion exponent are interrelated as follows,
\begin{gather}
    \lim_{\varepsilon\to 0^{+}}\operatorname{Re}\left( \int_{\mathbb{R}^{2}}\!\!\!\! da S_{m}(a + i\varepsilon, a- i\varepsilon) \right) \sim \varepsilon^{1 - 2\gamma}.
\end{gather}
By means of the above theorem, we can find the upper bound by assuming \(|\langle E \vert [H, x] \vert E^{\prime} \rangle|\) is bounded by some positive constant \(M\),
\begin{gather}
    \int_{\mathbb{R}}S_{m}(a + i\varepsilon, a - i\varepsilon) da < \int_{\mathbb{R}^{3}}\frac{M dE dE^{\prime}da}{(E - (a + i\varepsilon))(E^{\prime} - (a - i\varepsilon))}.
\end{gather}
Applying Taylor expansion on the integrand with respect to small \(\varepsilon\) and taking the real part gives us a constant leading order term,
\begin{gather}
    \Re\left(\int_{\mathbb{R}^{3}}\frac{M dE dE^{\prime}da}{(E - (a + i\varepsilon))(E^{\prime} - (a - i\varepsilon))}\right) \sim \epsilon^{0}.
\end{gather}
Then, we obtain an upper bound \(\gamma^{+}\) of the diffusion exponent of the projected model under the assumption.
Hence, the actual diffusion exponent \(\gamma\) is smaller than or equal to \(\gamma^{+} = 1/2\).

% =========================================================================== %
% SUBSECTION : Upper bound of diffusion exponent
% =========================================================================== %
\subsection{Numerical results}

To further quantify the transport properties, we analyze the spreading of an initially localized wavepacket for weak quasiperiodic perturbation.
We use an initial state localized on a single site in the center of the lattice.
For convenience, we assign the lattice center to be the zero coordinate.
To quantify the spreading of the initial state, we compute the root-mean-square of the displacement \(\sigma(t)\)
where \(\langle n\rangle\) is the average position,
\begin{gather}
  \label{eq:sigma-t}
  \sigma(t) = \left[\sum_{n\in\mathbb{Z}} (n - \langle n\rangle(t))^2 \abs{\psi_{n}(t)}^2 \right]^{1/2} \!\!\!\!\!\!\!\! \propto t^{\gamma}, \hspace{0.5em}\text{where } \langle n \rangle(t) = \sum_{n\in\mathbb{Z}} n \abs{\psi_n(t)}^2.
\end{gather}
We also average the results over \(40\) values of the phase \(\phi\)~\eqref{eq:fields} sampled from the range \([0, \pi]\).
The displacement \(\sigma(t)\) provides the deviation of the particle's position from its average at time \(t\).
The absence of spreading indicates localization.
Also, the spreading stops in finite systems once the boundaries are reached.
At an intermediate time, i.e., a period before the boundaries are reached, \(\sigma(t)\) is fitted by a power law with the exponent \(\gamma\) whose value indicates the type of transport: diffusive, subdiffusive, or ballistic~\cite{dominguez2019aubry}.

\begin{figure}
\centering
    \includegraphics[width = \columnwidth]{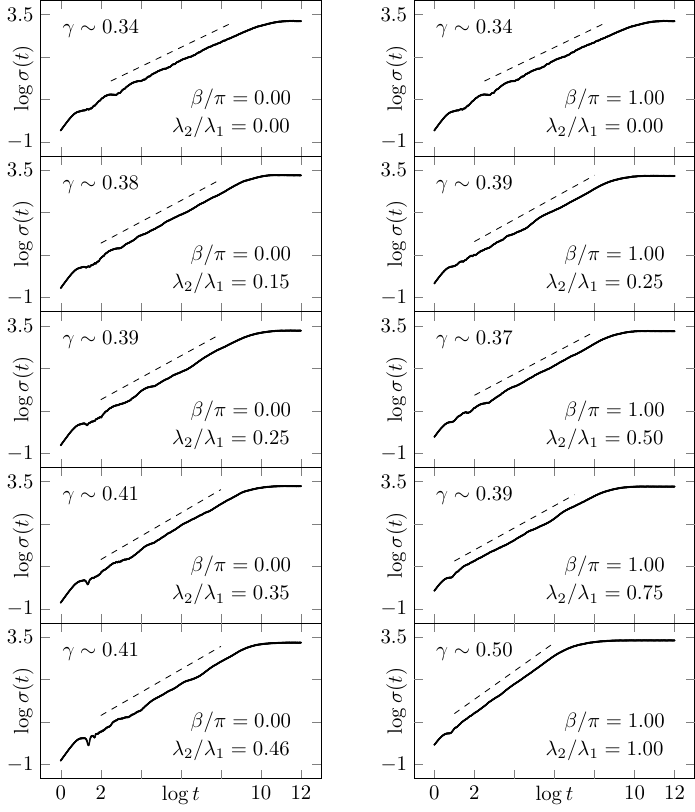}
    \caption[Log-log plots of the spreading of a wavepacket]{
    Log-log plots of the spreading of a wavepacket initially localized on a single site using \(\log_{10}\).
    The order of the plots follows the data in Table~\ref{table:subdiffusion}.
    } \label{fig:diffusion}
\end{figure}

%  Table and Figure
{
\begin{table}
    \centering
    \setlength{\tabcolsep}{17pt}
    \renewcommand{\arraystretch}{1.1}
    \begin{tabular}{c|c||c|c}
    \noalign{\smallskip}\noalign{\smallskip}\hline\hline
    \((\lambda_{2}/\lambda_{1}, \beta/\pi)\) & \(\gamma \pm \Delta\gamma\) & \((\lambda_{2}/\lambda_{1}, \beta/\pi)\) & \(\gamma \pm \Delta\gamma\) \\
    \hline
    \((0.00, 0.00)\) & \(0.34\pm 0.01\) & \((0.00, 1.00)\) & \(0.34\pm 0.01\) \\
    \hline
    \((0.15, 0.00)\) & \(0.38\pm 0.01\) & \((0.25, 1.00)\) & \(0.39\pm 0.04\) \\
    \hline
    \((0.25, 0.00)\) & \(0.39\pm 0.04\) & \((0.50, 1.00)\) & \(0.37\pm 0.01\) \\
    \hline
    \((0.35, 0.00)\) & \(0.41\pm 0.03\) & \((0.75, 1.00)\) & \(0.39\pm 0.02\) \\
    \hline
    \((0.46, 0.00)\) & \(0.41\pm 0.03\) & \((1.00, 1.00)\) & \(0.50\pm 0.01\) \\
    \hline\hline
    \end{tabular}
    \caption{
    Diffusion exponent \(\gamma\) in Eq.~\eqref{eq:sigma-t} for various points on the border of the phase diagram in Fig.~\ref{fig:phasediagram}.
    Except for \(\lambda_{2}/\lambda_{1} = 1\) and \(\beta = \pi\), all results show that the critical states support clear subdiffusive transport.
    For the off-diagonal Harper model, \(\beta=\pi,\lambda_1=\lambda_2\), the diffusion exponent is either diffusive or subdiffusive but very close to diffusive, 
    similar to the case of the Aubry-Andr\'e-Harper model at the critical point~\cite{hiramoto1988dynamics,wilkinson1994spectral}.
    } \label{table:subdiffusion}
\end{table}
\begin{figure}
\centering
    \includegraphics[width = \columnwidth]{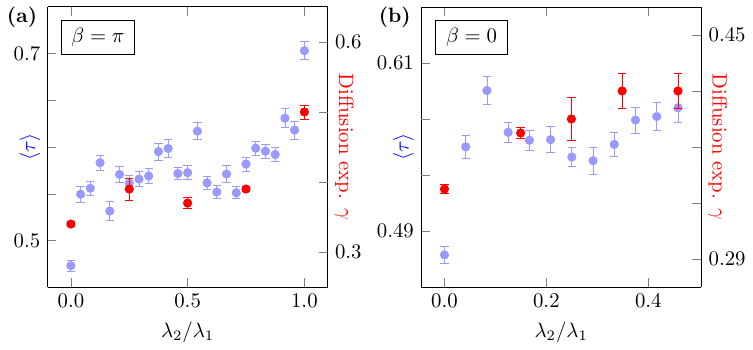}
    \caption[Correlation of diffusion exponents and averaged IPR exponents]{
    Positive correlation between the exponent \(\gamma\) in Eq.~\eqref{eq:sigma-t} (red points) and the exponent \(\tau\) averaged over the spectrum, \(\mbet\), in Eq.~\eqref{eq:linearmodel} (blue points).
    The values of \(\gamma\) are taken from Table~\ref{table:subdiffusion}.
    (a) \(\beta=\pi\).
    (b) \(\beta=0\).} \label{fig:taugamma}
\end{figure}
}
We sampled the boundary of the phase diagram given in Fig.~\ref{fig:phasediagram}, which corresponds to the extended Harper model and features critical eigenstates (for localized cases, the spreading stops as discussed above).
The results of the fitting are provided in Table~\ref{table:subdiffusion}, and the details of the wavepacket spreading are shown in Fig.~\ref{fig:diffusion} for system size \(L = 12801\).
In all cases, we see a clear subdiffusion~\cite{amini2017spread,detomasi2019survival,khaymovich2021dynamical}, whose exponent \(\gamma\) depends on the position on the border of the phase diagram, i.e., the values of \(\lambda_2/\lambda_1, \beta/\pi\).
For \(\lambda_2/\lambda_1=0\), \(\beta\) drops out of the Hamiltonian in Eq.~\eqref{eq:pm}, and consequently, the diffusion exponent \(\gamma\) does not depend on \(\beta\).
Furthermore, we observe a positive correlation between \(\gamma\) and \(\mbet\) in Fig.~\ref{fig:taugamma}, which suggests that the transport properties are strongly affected by the details of the profiles of the critical eigenstates.

In Sec.~\ref{sec:mappingEHM}, we observe that when the condition \(\theta_{1} + \theta_{2} = \pi/2\) holds, the projected model becomes equivalent, up to an overall scaling factor, to the specific scenario where \(\theta_{1,2} = \pi/4\).
Consequently, we can infer that for values of \(\theta_{1,2}\) satisfying Eq.~\eqref{eq:multifractalrelation}, the wavepacket spreading exhibits subdiffusive behavior.
Similarly, in Sec.~\ref{sec:mappingOHM}, when we keep \(\theta_{2} = \pi/4\), we obtain the off-diagonal Harper model.
Then, the factor \(t_c\) in Eq.~\eqref{eq:OHM1} serves as the global timescale for the almost diffusive spreading.

%\clearpage
% =========================================================================== %
% =========================================================================== %
% SECTION : WEAK PERTURBATION
% =========================================================================== %
% =========================================================================== %
\section{Finite perturbation}\label{ch2:sec:finite}

In the limit of weak interaction, we saw the emergence of entirely critical spectra for specific values of the local unitary transformation parameters \(\theta_{1,2}\), the relative potential strength \(\lambda_2/\lambda_1\), and the phase difference \(\beta\).
As we increase the perturbation strength and make it finite, we expect the system to localize for large enough values of \(\lambda_{1,2}\).
The open question is what happens to the critical states for moderate values of \(\lambda_{1,2}\) and whether any extended states might emerge.
Also, we do not expect any change in the localization properties of the states that were already localized for weak quasiperiodic perturbation.

For finite potential strengths, the projected model description is no longer valid; thus, we have to focus on the original Hamiltonian, \(\mh = \mhabf + W\).
Then, the original Hamiltonian can be expressed as a non-Abelian Aubry-Andr\'e-Harper model~\cite{wang2016phase} with a quasiperiodic block \(A_{n}\) and a fixed hopping block \(B\), \updated{where they are not commutative in general: \([A_{n}, B] \neq 0\),}
\begin{gather}
  E\psi_{n} = A_{n}\psi_{n} + B\psi_{n-1} + B^{\dagger}\psi_{n+1},
\end{gather}
where \(\psi_n = (p_n, f_n)^T\), and \(A_{n}\) is a block matrix with the quasiperiodic fields and \(\theta_{1,2}\) and \(B\) is a matrix only depends on \(\theta_{1,2}\).
As an example, when \(\theta_{1,2} = \pi/4\), the original model looks like as follows.
\begin{gather}
  \label{eq:Exact}
  A_{n} = \frac{4}{\Delta}
  \begin{bmatrix}
    W_{1}(n) & 0 \\
    0 & W_{2}(n)
  \end{bmatrix},
  \quad B =
  \begin{bmatrix}
    1 & -1 \\
    1 & -1
  \end{bmatrix}.
\end{gather}
This section focuses on the parameter regions where the projected model [Eq.~\eqref{eq:pm}] for weak disorder hosts critical states.
% =========================================================================== %
% SUBSECTION : Representation of perturbed ABF on different basis
% =========================================================================== %
\subsection{Representation in semi-engtangled basis}

Sometimes it is convenient to work with the \emph{semi-detangled} Hamiltonian,~\cite{danieli2021nonlinear,cadez2021metal,lee2023criticalA,lee2023criticalB} which is defined by inverting only the second unitary transformation \(U_{2}\) (see Sec.~\ref{ch2:sec:model}).
This defines the semi-detangled basis \(\{u_{n}, d_{n}\}\) and gives the new, semi-detangled Hamiltonian,
\begin{gather}
    \label{eq:ham-sd}
    \mhsd = U_{1} \mhfd U_{1}^{\dagger} + U_{2}^{\dagger}WU_{2}.
\end{gather}
The semi-detangled wavefunction amplitudes \(u_{n}\) and \(d_{n}\) are related to the basis states of the ABF Hamiltonian \(\{p_{n}, f_{n}\}\) as follows,
\begin{gather}
    u_{n} = p_{n}\cos\theta_{2} - f_{n}\sin\theta_{2} \quad\text{and}\quad d_{n} = p_{n}\sin\theta_{2} + f_{n}\cos\theta_{2}.
\end{gather}
The model~\eqref{eq:ham-sd} takes the following form in the semi-detangled basis:
\begin{align}
    \label{eq:semi}
    Eu_{n} &= \left[\varepsilon_{b}\sin^{2}\theta_{1} + \varepsilon_{a}\cos^{2}\theta_{1}\right] u_{n} + \left[W_{2}(n)\sin^{2}\theta_{2} + W_{1}(n)\cos^{2}\theta_{2}\right] u_{n} \notag \\
    &+ \left[(W_{1}(n) - W_{2}(n))\cos\theta_{2}\sin\theta_{2}\right] d_{n} +\left[\Delta\cos\theta_{1}\sin\theta_{1}\right] d_{n+1}, \\
    Ed_{n} &= \left[\varepsilon_{a}\sin^{2}\theta_{1} + \varepsilon_{b}\cos^{2}\theta_{1}\right] d_{n} + \left[W_{1}(n)\sin^{2}\theta_{2} + W_{2}(n)\cos^{2}\theta_{2}\right] d_{n} \notag \\
    &+ \left[(W_{1}(n) - W_{2}(n))\cos\theta_{2}\sin\theta_{2}\right] u_{n} + \left[\Delta\cos\theta_{1}\sin\theta_{1}\right] u_{n-1}.
\end{align}
Throughout the section, we consider the system in semi-detangled basis.

% =========================================================================== %
% SUBSECTION : Critical-to-insulator transition in finite perturbation
% =========================================================================== %
\subsection{Critical-to-insulator transition in finite perturbation\label{claim}}

Let us set \(\theta_{2} = \pi/4\) and \(\theta_{1} = \theta\).
In the case when both quasiperiodic fields have same amplitude \(\lambda_{2}=\lambda_{1}=\lambda\) and the phase difference \(\beta = \pi\), we have simply \(W_{2} = -W_{1} = -W\).
With these conditions, the semi-detangled Hamiltonian in Eq.~\eqref{eq:semi} takes the following form,
\begin{align}
    E_{u}(E,\theta)u_{n} &= W(n)d_{n} + \Delta\cos\theta\sin\theta d_{n+1}, \\
    E_{d}(E,\theta)d_{n} &= W(n)u_{n} + \Delta\cos\theta\sin\theta u_{n-1},
\end{align}
where \(E_{u}(E,\theta)\) and \(E_{d}(E,\theta)\) are defined as
\begin{align}
    E_{u}(E,\theta) &= E - \varepsilon_{b}\sin^{2}\theta - \varepsilon_{a}\cos^{2}\theta, \\
    E_{d}(E,\theta) &= E - \varepsilon_{a}\sin^{2}\theta - \varepsilon_{b}\cos^{2}\theta.
\end{align}
We can eliminate one of the amplitudes \(u_n\) or \(d_n\) by substituting one of the equations into the other. 
If we choose to eliminate \(u_n\), we get the following effective equation, resulting in equations with \(d_n\) only:
\begin{align}
    E_{d}(\theta)E_{u}(\theta)d_{n} &= W^{2}(n)d_{n} + \Delta^{2}\cos^{2}\theta\sin^{2}\theta d_{n} \notag \\
    &+ \Delta\cos\theta\sin\theta\left[W(n)d_{n+1} + W(n-1) d_{n-1}\right].
\end{align}
By rearranging the above equation, we get the following eigenequation,
\begin{align}
    \label{eq:EHMlike}
    \tilde{E}(\theta, \lambda) d_{n} &= \cos(4\pi\alpha n)d_{n} \notag \\
    &+ K(\theta, \lambda)) \left[\cos(2\pi\alpha (n-1))d_{n-1} + \cos(2\pi\alpha n)d_{n+1}\right],
\end{align}
where \(\tilde{E}(\theta,\lambda)\) and \(K(\theta,\lambda)\) are defined as,
\begin{gather}
    \label{eq:CIT}
    \tilde{E}(\theta, \lambda) = \frac{2}{\lambda^{2}}E_{d}(\theta)E_{u}(\theta) - \frac{1}{2}\quad\text{and}\quad K(\theta,\lambda) = \frac{2\Delta}{\lambda}\cos\theta\sin\theta.
\end{gather}
This eigenequation looks very similar to the extended Harper model, Eq.~\eqref{eq:EHM}, except for the spatial frequency of the onsite potential that is double that of the hopping.
Also, the potential is twice as large as in the extended Harper model.

For instance, when both \(\theta_{1}\) and \(\theta_{2}\) are equal to \(\pi/4\), Eq.~\eqref{eq:EHMlike} is reduced into the following equation:
\begin{gather}
  \tilde{E} d_{n} = \cos(4\pi\alpha n)d_{n} \vphantom{\frac{2t}{\lambda}} +\frac{2t}{\lambda}\left[ \cos(2\pi\alpha (n-1))d_{n-1} + \cos(2\pi\alpha n)d_{n+1} \right],
\end{gather}
where the eigenvalue \(\tilde{E}\) is given as
\begin{gather}
    \tilde{E} = \frac{2}{\lambda^2}\left[(E - \varepsilon)^2 - t^2 - \frac{\lambda^{2}}{2}\right].
\end{gather}
The non-Abelian Aubry-Andr\'e-Harper model for this case has been studied numerically~\cite{degottardi2013majorana, wang2016phase}.
The critical eigenstates over the entire spectrum show up until \(\lambda\) reaches the value of the flatband bandgap \(\Delta=|\varepsilon_{b} - \varepsilon_{a}|\).
Then, the CIT occurs once \(\lambda\) is equal to \(\Delta\) for the entire spectrum.
We remark that this result coincides with the extended Harper model estimate of the transition, \(2t/\lambda=1\) if we neglect the differences between our model, Eq.~\eqref{eq:EHMlike}, and the true extended Harper model, Eq.~\eqref{eq:EHM}.
Then, we get the CIT for the generalized case in Eq.~\eqref{eq:EHMlike} when \(K(\theta,\lambda) = 1\) as well.

An example for \(\theta = 0.1\pi/2\) is shown and summarized in Fig.~\ref{fig:OHM0P1}.
The flatband energies are taken to be \(\varepsilon_{a} = -1\) and \(\varepsilon_{b} = 2\).
The entire spectrum is rescaled for each \(\lambda_{1}\). 
Then we apply the linear model introduced in Eq.~\eqref{eq:linearmodel} introducing \(50\) energy bins \(\tilde{e}\) for the spectrum rescaled to fit in \([0, 1]\).
Lattice sizes (in unit cells) are \(L = 3283, 4181, 4832, 5473, 6765, 10946\).
The exact model has two sites for each unit cell, and the size of the Hamiltonian matrix is \(2L\times 2L\).
The transition point \(\lambda_{c} \approx 0.927\) based on Eq.~\eqref{eq:CIT} is indicated with the black dashed line.
It matches perfectly with the numerical results.
All states are critical to the left of \(\lambda_{c}\) and localized to the right of it.
The string of eigenstates between the two broadened flatband energies is due to the open boundary conditions.

% =========================================================================== %
% SUBSECTION : Non-degenerate compact localized states
% =========================================================================== %
\subsection{Non-degenerate compact localized states\label{clspreserve}}

Suppose both \(\theta_{1}\) and \(\theta_{2}\) are \(\pi/4\).
When \(\beta\) is zero and \(\lambda = \lambda_{1} = \lambda_{2}\), we obtain the following Hamiltonian in the semi-detangled basis \(u_n, d_n\),
\begin{align}
  \tilde{E}u_{n} &= \lambda\cos(2\pi\alpha n )u_{n} + t d_{n+1},\\
  \tilde{E}d_{n} &= \lambda\cos(2\pi\alpha n )d_{n} + t u_{n-1},
\end{align}
where \(\tilde{E} = E - \varepsilon\).
Again, we can eliminate one of the amplitudes through substitution.
Choosing to eliminate \(u_n\) leads to a diagonal problem for \(d_n\):
\begin{equation}
  \tilde{E}d_{n} = \left[\lambda\cos(2\pi\alpha n) + \frac{t}{\tilde{E} - \lambda\cos(2\pi\alpha(n-1))}\right]d_{n},
\end{equation}
and a compact localization of the eigenstates for any \(\lambda\).
We recall that we have already found compact localization for weak perturbation.
This compact localization survives for any potential strength.
When \(\theta_{1}\) and \(\theta_{2}\) are arbitrary values, we again obtain non-degenerate compact localization with more complicated \(\tilde{E}\).

% =========================================================================== %
% SUBSECTION : Non-simple fractality edges
% =========================================================================== %
\subsection{Non-simple fractality edges}
%
%  Figure
\begin{figure}
\vfill
\centering
    \includegraphics[width = \columnwidth]{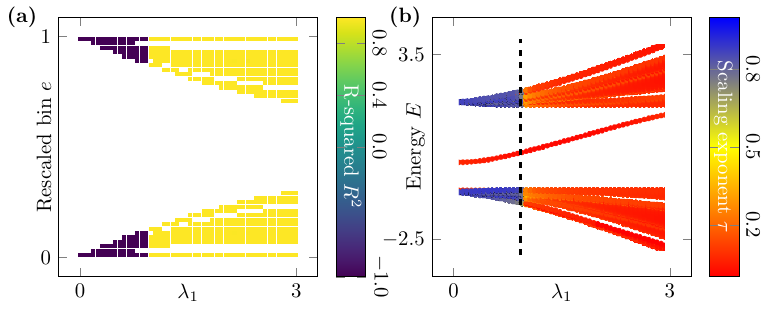}
    \caption[Tunable critical-to-insulator transition]{
        Critical-to-insulator transition at finite potential strength \(\lambda_{2} = \lambda_{1}\), \(\beta = \pi\) at \(\theta = 0.1\pi/2\).
        (a) Rescaled spectrum.
        The negative \(R^{2}\) values are replaced with \(-1\) for better visual separation of localized and critical regions.
        (b) Original, non-rescaled energy spectrum for lattice size \(L = 10946\).
        The vertical dashed black line marks the CIT transition predicted by Eq.~\eqref{eq:CIT}.
    } \label{fig:OHM0P1}
    \vspace{2em}
    \includegraphics[width = \linewidth]{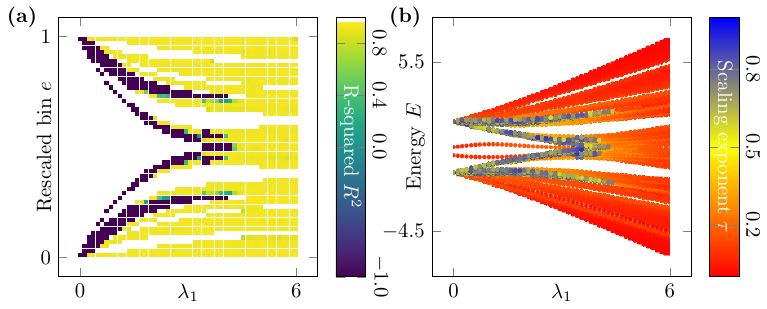}
    \caption[Non-simple fractality edges]{
        (a) Fractality edges in the exact model at \(\beta/\pi = 1, \lambda_{2}/\lambda_{1} =0.5\).
        Every \(R^2\) lower than or equal to zero is revalued to \(-1\) for the purpose of clear distinction between the localized and critical regions in the figure. 
        The \(-1\) \(R^2\) value implies that states are critical.
        For \(R^2 \approx 1\), the states are localized.
        (b) Original, non-rescaled energy spectrum for lattice size \(L = 10946\).
    } \label{fig:BetaPI0P50}
\vfill
\end{figure}
We now consider all the other values of \(\beta, \lambda_{1,2}\) where we observed critical states for weak perturbation.
In the semi-detangled basis \(u_n, d_n\), we obtain the following Hamiltonian,
\begin{align*}
  \tilde{E}u_{n} &= \frac{W_{1}(n) + W_{2}(n)}{2}u_{n} + \frac{W_{1}(n) - W_{2}(n)}{2}d_{n} + td_{n+1}, \\
  \tilde{E}d_{n} &= \frac{W_{1}(n) + W_{2}(n)}{2}d_{n} + \frac{W_{1}(n) - W_{2}(n)}{2}u_{n} + tu_{n-1}.
\end{align*}
\(\tilde{E}\) is \(E - \varepsilon\) and \(W_{1,2}\) are the quasiperiodic fields.
The above effective equations are a mixture of the one given in Sec.~\ref{claim} and Sec.~\ref{clspreserve}, implying that an eigenstate's behavior is at most critical.

To look into the properties of the states, we have to resort to numerical analysis on the full model in Eq.~\eqref{eq:Exact} to analyze their localization properties.
Flatband energies are fixed to \(\varepsilon_{a} \!=\! -1\) and \(\varepsilon_{b} \!=\! 2\) with bandgap \(\Delta=3\).
We consider several values of \(\beta, \lambda_{1,2}\) to probe the different regions of the phase diagram with critical states and repeat the computation of the IPR of the eigenstates.
Then we apply the linear model introduced in Eq.~\eqref{eq:linearmodel} and introduce \(50\) energy bins for the eigenenergy range rescaled to fit between \(0\) and \(1\), which is referred to as the rescaled energy.
Then, it allows us to distinguish localized and critical states as a function of eigenenergy.
The lattice sizes (in unit cells) are \(L \!=\! 2584, 3283, 4181, 4832, 5473\), and \(6765\).
Since the exact model has two sites for each unit cell, the size of the Hamiltonian matrix is \(2L\times 2L\).

The generic observation is that the critical spectrum of the projected model is replaced with a mixed one, comprising partially critical and partially localized states depending on the eigenenergy.
We dub the border between critical and localized states as \emph{fractality edges} by analogy with mobility edges separating localized and extended states~\cite{liu2022anomalous, zhang2022lyapunov, wang2022quantum, shimasaki2022anomalous}.
For instance, let \(\beta \!=\! \pi\) and \(\lambda_{2}/\lambda_{1} \!=\! 0.5\).
Figure~\ref{fig:BetaPI0P50} shows the emergence of fractality edges for a finite strength of perturbation \(\lambda_{1,2}\).
We note that the critical states extend to values \(\lambda_1 \!\geq\! \Delta\), at variance with the case in Sec.~\ref{claim} where all states localize for \(\lambda_1 \!>\! \Delta\).

% =========================================================================== %
% SUBSECTION : Non-degenerate compact localized states
% =========================================================================== %
\subsection{Perturbation independent fractality edges for \texorpdfstring{\(\lambda_{2} = 0\)}{zero lambda2}}
%
%  Figure
\begin{figure}
\vfill
\centering
    \includegraphics[width = \columnwidth]{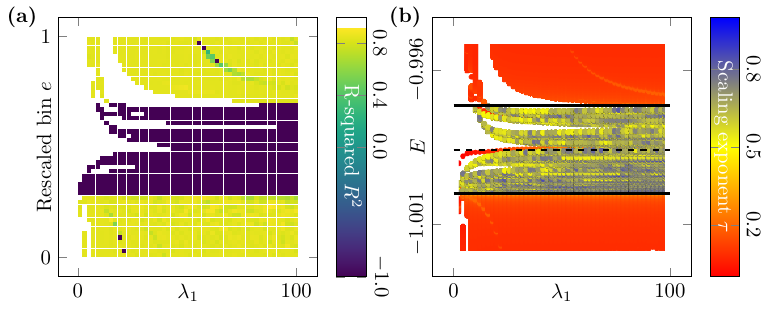}
    \caption[First example of perturbation-independent fractality edges]{
        Fractality edges at finite potential strength \(\lambda_1\) and \(\lambda_{2} = 0\), for \(\theta = 0.01\pi/2\).
        (a) Fractality edges in the rescaled spectrum.
        (b) Fractality edges in the original, non-rescaled energy spectrum with lattice size \(L = 10946\).
        The bottom black line is the flatband \(\varepsilon_{a} = -1\), while the top black black is \(E \approx -0.997\).
        The dashed black line is the upper bound \(E \approx -0.998\) given by Eq.~\eqref{eq:inequalities}.
    } \label{fig:EHM0P01}
    \vspace{2em}
    \includegraphics[width = \columnwidth]{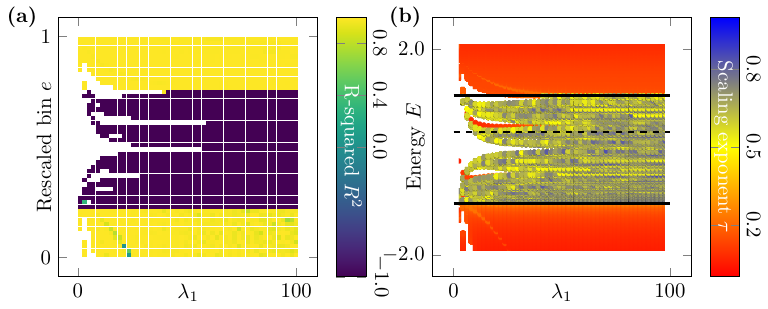}
    \caption[Second example of perturbation-independent fractality edges]{
        Fractality edges at finite potential strength \(\lambda_1\) and \(\lambda_{2} = 0\), for \(\theta = 0.68\pi/2\).
        (a) Fractality edges in the rescaled spectrum.
        (b) Fractality edges in the original, non-rescaled energy spectrum with lattice size \(L = 10946\).
        The bottom black line is the flatband \(\varepsilon_{a} = -1\), while the top black line is \(E \approx 1.05\).
        The dashed black line is the upper bound \(E \approx 0.393\) given by Eq.~\eqref{eq:inequalities}.
    } \label{fig:EHM0P68}
\vfill
\end{figure}
Now, let us consider a case where we remove one of the quasiperiodic potentials, \(\lambda_2 = 0\) and set \(\theta_2 = \pi/2 - \theta, \theta_1 = \theta\).
The model in the semi-detangled basis takes the following form,
\begin{align}
    \frac{E u_{n}}{\cos\theta\sin\theta} &= \left[\frac{\varepsilon_{a}}{\tan\theta} + \frac{\left(\varepsilon_{b} + W_{1}(n)\right)}{1/\tan\theta}\right]u_{n} + W_{1}(n)d_{n} + \Delta d_{n+1}, \label{eq:semiFE1} \\
    \frac{E d_{n}}{\cos\theta\sin\theta}  &= \left[\frac{\left(\varepsilon_{b} + W_{1}(n)\right)}{\tan\theta} + \frac{\varepsilon_{a}}{1/\tan\theta}\right]d_{n} + W_{1}(n)u_{n} + \Delta u_{n-1}. \label{eq:semiFE2}
\end{align}
After multiplying Eq.~\eqref{eq:semiFE2} by \(\tan\theta\) and taking the difference of the above equations,
we get a single equation without quasiperiodic potentials:
\begin{gather}
    E_{u}(E,\theta) u_{n} + \tan\theta E_{g}u_{n-1} = \tan\theta E_{d}(E,\theta)d_{n} + E_{g}d_{n+1}, \notag
\end{gather}
where \(E_{u}(E,\theta)\) and \(E_{d}(E,\theta)\) are
\begin{align}
    E_{u}(E,\theta) &= \left[\frac{E}{\cos\theta\sin\theta} - \left(\frac{E_{a}}{\tan\theta} + E_{b}\tan\theta\right)\right], \\
    E_{d}(E,\theta) &= \left[\frac{E}{\cos\theta\sin\theta} - \left(\frac{E_{b}}{\tan\theta} + E_{a}\tan\theta\right)\right].
\end{align}
For \(\theta_{1,2} = \pi/4\), we obtain a simpler model.
There are fractality edges at the flatband energies \(\varepsilon_{a,b}\) for any perturbation strength \(\lambda_1\) and all the eigenstates in between are critical:
\begin{gather}
    (2E - \sigma)u_{n} + E_{g}u_{n-1} = (2E - \sigma)d_{n} + E_{g}d_{n+1},
\end{gather}
where \(\sigma\) is defined as \(\varepsilon_{a} + \varepsilon_{b}\).
This result can be rationalized with a simple assumption that the critical states appear when the hopping is larger than the onsite potential,
\(|2E - \sigma| \leq E_{g}\),
implying \(\varepsilon_{a} \leq E \leq \varepsilon_{b}\) for the critical states.
We can apply the same logic for \(\theta \neq \pi/4\).
Then, the above inequality for energy \(E\) is modified as follows,
\begin{gather}
    \label{eq:inequalities}
    |E_{u}| < E_{g}\tan\theta \quad\text{and}\quad |E_{d}|\tan\theta < E_{g}.
\end{gather}
However, the simple constraint does not explain the entire range of fractality edges but only covers the narrower range of the critical states.
Examples are shown and summarized in Fig.~\ref{fig:EHM0P01} and~\ref{fig:EHM0P68}, where \(\theta = 0.01\pi/2\) and \(\theta = 0.68\pi/2\), respectively.
We fixed the flatband energies to \(\varepsilon_{a} = -1\) and \(\varepsilon_{b} = 2\).
The narrowness of the critical region in Fig.~\ref{fig:EHM0P01} is due to the smallness of the angle \(\theta_1\); 
remember that for \(\theta_1 = 0\) we have trivial flatbands due to the model decoupling into disconnected dimers.
Since the fractality edges are potential strength independent, we only considered part of the spectrum around the edges and used it for the rescaled spectrum.
We apply the linear model introduced in Eq.~\eqref{eq:linearmodel}, and introduce \(50\) energy bins \(e\) for the spectrum rescaled to fit in \([0, 1]\).
The lattice sizes (in unit cells) are \(L = 3283, 4181, 4832, 5473, 6765, 10946\).
The exact model has two sites for each unit cell, and the size of the Hamiltonian matrix is \(2L\times 2L\).
In both examples, two fractality edges are independent of the potential strength \(\lambda_1\), with critical states in between and localized states outside.
The lower fractality edge coincides with the flatband energy \(\varepsilon_{a}\).
By symmetry, setting \(\lambda_1=0\) and varying \(\lambda_2\), one would get the similar spectral behavior with the upper fractality edge would be at \(\varepsilon_{b}\).

%\clearpage
% =========================================================================== %
% =========================================================================== %
% SECTION : WEAK PERTURBATION
% =========================================================================== %
% =========================================================================== %
\section{Conclusion}\label{ch2:sec:conclusion}

In this chapter, we summarized our study on the effect of quasiperiodic perturbation, composed of two potentials, on a general model of the ABF manifold with two bands. 
First, we considered the case of weak quasiperiodic perturbation.
We identified the ABF submanifolds of models with spectra containing critical states and exhibiting subdiffusive and almost diffusive transport for weak quasiperiodic perturbation.
The critical states with subdiffusive transport were found for the ABF models satisfying the condition \(\theta_{1} + \theta_{2} = \pi/2\) for the angles of the unitary transformation generating the model.
The other constraint is the absence of one of the quasiperiodic fields.
The other submanifold of models featuring critical states with almost diffusive transport is given by \(\theta_{2} = \pi/4\) and arbitrary \(\theta_1\).
The quasiperiodic potentials must have equal strengths and the relative phase \(\beta = \pi\), e.g., be of opposite signs.
All such models can be mapped onto the off-diagonal Harper model.
Outside of these manifolds, all the states are localized in the thermodynamic limit, as suggested by numerical analysis.
Specifically, for small angles \(\theta_{1,2}\), we demonstrated compact localization of the spectrum.

For a finite quasiperiodic potential, the perturbed ABF models, which map to the extended Harper model, display fractality edges in the spectrum separating critical from localized states.
These edges are independent of the potential strength.
On the other hand, the models that map onto the off-diagonal Harper model show no fractality edges.
Instead, they exhibit a critical-to-insulator transition.
The transition point is derived analytically and confirmed numerically and depends on the angle \(\theta_1=\theta\) with fixed \(\theta_{2} = \pi/4\).

The presented ABF manifold, perturbed by quasiperiodic potentials, could be potentially implemented using ultracold atomic gases loaded onto optical lattice potentials where ABF networks have been realized, 
e.g. in the Creutz ladder~\cite{kang2020creutz,he2021flat} or the diamond chain~\cite{li2022aharonov,martinez2023interaction}.
A second promising platform is light propagation in photonic lattices where the diamond chain with its Aharonov-Bohm cages was experimentally obtained~\cite{caceres2022controlled}.
Another promising future platform is electric circuits.
The main issue for the experimental realization of ABF systems is achieving both positive and negative hopping.
A possible candidate is a stack of LC electric circuits~\cite{zhao2018topological} in which the topology of wiring can generate positive and negative hopping, even complex hopping.
We also expect by adding extra LC resonators to each site might create the onsite quasiperiodic modulation of the ABF system. 

\chapter[Trapped hard-core bosons in flatband systems]{Trapped Hard-core Bosons \\ in Flatband Systems}~\label{chpt3}

\iffalse
\vfill
Recently, there has been considerable interest in physical systems with macroscopic degeneracies.
One particular example of such system is a flatband system~\cite{maksymenko2012flatband,leykam2018artificial,rhim2021singular} with compact localized states, where the eigenstates are confined to a finite number of sites~\cite{sutherland1986localization,aoki1996hofstadter}.
This is a result of destructive interferences caused by the network geometry.
The extreme sensitivity of macroscopically degenerate flatbands to perturbations and interactions gives rise to a diverse range of interesting and exotic phases.
One example is an ergodicity breaking by considering a fine-tuned interaction~\cite{kuno2020flat,danieli2020many,danieli2021nonlinear,danieli2021quantum,vakulchyk2021heat}.
This suggests that flatband systems offers a platform for investigating intriguing phenomena related to quantum ergodicity breaking.
In this chapter, we focus on how many-body hard-core bosons behave in the one-dimensional cross-stitch lattices and relate with non-ergodic excited states.
The related work is published in Ref.~\ref{xx}.
\fi
%========================================================================================
%\clearpage
% =========================================================================== %
% =========================================================================== %
% SECTION : Introduction
% =========================================================================== %
% =========================================================================== %
\section{Introduction}

Recently, there has been considerable interest in physical systems with macroscopic degeneracies.
One particular example of such a system is a flatband system~\cite{maksymenko2012flatband,leykam2018artificial,rhim2021singular} with compact localized states, where the eigenstates are confined to a finite number of sites~\cite{sutherland1986localization,aoki1996hofstadter}, which is a result of destructive interferences caused by the network geometry as we have mentioned in Chapter~\ref{chpt1}.
The extreme sensitivity of macroscopically degenerate flatbands to perturbations and interactions gives rise to a diverse range of interesting and exotic phases.
One example is ergodicity breaking by considering a fine-tuned interaction~\cite{kuno2020flat,danieli2020many,danieli2021nonlinear,danieli2021quantum,vakulchyk2021heat}.
In this contenxt, flatband systems offer a platform for investigating intriguing phenomena related to quantum ergodicity breaking.

%\subsection{Previous research on weak thermalization}
Thermalization has fascinated physicists as it describes the evolution of quantum many-body systems from reversible microscopic dynamics toward equilibrium.
One intriguing aspect is the tendency of all pure states within a specific energy shell to exhibit thermal-like behavior~\cite{srednicki1994chaos}.
In search of an explanation, an eigenstate thermalization hypothesis (ETH) was proposed~\cite{deutsch2018eigenstate}, suggesting that the thermalization in isolated quantum systems can be attributed to the assumption that every energy eigenstate possesses thermal properties.
ETH has been widely discussed, but weak ETH emerged where \emph{nearly} all energy eigenstates exhibit thermal properties to some extent~\cite{biroli2010effect}.
It should be noted that weak ETH alone cannot determine the presence or absence of thermalization for physically realistic initial states~\cite{biroli2010effect}.
Nevertheless, weak ETH does assure that an initial state lacking significant overlap with rare states will undergo thermalization.
It is known that weak ETH holds even for a wide range of translation-invariant systems, regardless of their integrability~\cite{alba2015eigenstate, ikeda2013finite}.
However, many-body localization induced by the disorder can violate weak ETH~\cite{imbrie2016many}.

%\subsection{General summary of the chapter}
This chapter summarizes the work on the thermalization behavior of hard-core bosons in the one- and two-dimensional cross-stitch lattice with on-site repulsive interaction having the strongest collision effect.
It is known that the presence of macroscopic degeneracy in the flatband energy allows for the amplification of the effects of interactions and perturbations.
We specifically focus on the interplay between compact localized states (CLSs) and the extreme limit of strong repulsion imposed by hard-core constraints.
Our findings reveal the band insulating phase, the Wigner crystal phases, and the violation of weak ETH.
The non-ergodic behaviors observed in the absence of disorder are of particular significance, highlighting the presence of strictly non-ergodic excited states. This non-ergodic behavior is closely linked to the concept of Hilbert space fragmentation.
The related work is in preparation.

%\subsection{Outline of the chapter}
The chapter is organized as follows.
We start by defining one- and two-dimensional cross-stitch lattices and briefly mention the properties of hard-core bosons in Sec.~\ref{ch3:sec:models}.
In Sec.~\ref{ch3:sec:groundstate}, we explore the process of site filling to obtain the band insulating phase and Wigner crystal phases.
Moving on to Sec.~\ref{ch3:sec:nonergodic}, we investigate the non-ergodic excited states in a closed CLS barrier.
We briefly point out that, in the diamond lattice, similar phenomena occur as the cross-stitch lattice in Sec.~\ref{ch3:sec:diamond}.
Then, we make a conclusion of the chapter in the last section.

%\clearpage
% =========================================================================== %
% =========================================================================== %
% SECTION : Models
% =========================================================================== %
% =========================================================================== %
\section{Models}\label{ch3:sec:models}

This section defines the one- and two-dimensional cross-stitch lattices, which this chapter focuses on.
Additionally, the properties of hard-core bosons are going to be discussed.

\subsection{1D and 2D cross-stitch lattices}
The Hamiltonian for the one-dimensional cross-stitch chain, which includes vertical hopping denoted as \(t\), is expressed as follows, with the intra-unit cell hopping term \(\hat{h}_{n}\) and the inter-unit cell hopping term \(\hat{v}_{n}\) defined accordingly,
\begin{gather}\label{eq:def1dcs}
    \mh = -\sum_{n}\hat{h}_{n} + \hat{h}_{n}^{\dagger} + t(\hat{v}_{n} + \hat{v}_{n}^{\dagger}), \\
    \hat{h}_{n} = (\hat{a}_{n}^{\dagger} + \hat{b}_{n}^{\dagger})(\hat{a}_{n+1} + \hat{b}_{n+1}) \quad\text{and}\quad \hat{v}_{n} = \hat{a}_{n}^{\dagger}\hat{b}_{n}.
\end{gather}
\updated{Here, \(\hat{a}_{n}^{\dagger}\) and \(\hat{b}_{n}^{\dagger}\) represent the creation operators at site \(A_{n}\) and \(B_{n}\), respectively.}
When \(t\) is sufficiently small, it results in a ground state characterized by a flatband. The corresponding energy bands are depicted in Fig.~\ref{fig:1d_cs_bands}.
Likewise, the two-dimensional cross-stitch lattice can be defined as
\begin{gather}
    \mh = -\sum_{n,m}\hat{h}_{n,m} + \hat{h}_{n,m}^{\dagger} + t(\hat{v}_{n,m} + \hat{v}_{n,m}^{\dagger}), \\
    \hat{h}_{n,m} = (\hat{a}_{n,m}^{\dagger} \! + \hat{b}_{n,m}^{\dagger})(\hat{a}_{n+1,m} \!+ \hat{b}_{n+1,m} \!+ \hat{a}_{n,m+1} \!+ \hat{b}_{n,m+1}), \\
    \hat{v}_{n,m} = \hat{a}_{n,m}^{\dagger}\hat{b}_{n,m}.
\end{gather}

\subsection{Properties of hard-core bosons}
A hard-core boson is a particle that obeys a set of mixed commutation relations, such that
\(
    [\hat{c}_{i},\hat{c}_{j}] = [\hat{c}_{i}^{\dagger},\hat{c}_{j}^{\dagger}] = [\hat{c}_{i}, \hat{c}_{j}^{\dagger}] = 0,
\)
for all \(i\neq j\), and exhibits fermion-like behavior at the same site, as described by the anti-commutation relation \(\{\hat{c}_{i},\hat{c}_{i}\} = \{\hat{c}_{i}^{\dagger},\hat{c}_{i}^{\dagger}\} = 0\) and \(\{\hat{c}_{i},\hat{c}_{i}^{\dagger}\} = 1\).
\updated{\(\hat{c}_{n}^{\dagger}\) is a hard-core boson creation operator at site \(n\).}
If we merge all the commutation properties, we get the following single expression,
\begin{gather}
    [\hat{a}_{i},\hat{a}_{j}^{\dagger}] = \delta_{ij}(1 - 2\hat{a}_{i}^{\dagger}\hat{a}_{i}).
\end{gather}
These commutation relations define a bosonic system characterized by strong repulsion at short distances, making it theoretically unfavorable for states with more than two particles to occupy a single site.
Such particle can be achieved at low densities and low temperature in certain regimes of large scattering length in quasi-one-dimensional systems~\cite{olshanii1998atomic, petrov2000regimes, dunjko2001bosons}.
It is known that, in the one-dimensional hard-core bosonic gas system, the true condensation at zero temperature cannot be achieved~\cite{lenard1964momentum, lenard1966one, vaidya1979one}.
Instead, due to the presence of a quasi-long-range order in the system, quasi-condensates at
finite momentum can emerge~\cite{rigol2004emergence, rigol2005free}.
In three-dimensional systems, Mott insulating phase and Bose-Einstein condensation can exist~\cite{aizenman2004bose}.
Our study focuses specifically on the on-site interaction, disregarding the interaction term.
%\clearpage
% =========================================================================== %
% =========================================================================== %
% SECTION : CLS groundstates
% =========================================================================== %
% =========================================================================== %
\section{CLS groundstates}\label{ch3:sec:groundstate}

This section demonstrates two findings on the one-dimensional cross-stitch lattice.
Firstly, we establish that occupying CLSs results in the groundstate configuration.
Additionally, we illustrate that fully-filled dimers do not contribute to the overall groundstate eigenenergy.
The subsequent discussion will provide a comprehensive understanding of these observations.

% :: Subsection :: -- Filling the flatband CLS
\subsection{\texorpdfstring{\(\nu \leq 1/2\)}{nu leq 1/2} -- filling of the flatband CLS}
First, consider a CLS in the one-dimensional cross-stitch lattice at the \(m\)-th unit cell.
To describe the ground state of the flatband, we propose an ansatz, where \(P_{\mathrm{FB}}\) represents the set of all CLSs occupying certain unit cells,
\begin{gather}\label{eq:groundstate}
    \kgsfb = \!\!\!\prod_{n\in P_\mathrm{FB}}\!\!\!\cls_{n} \quad\text{and}\quad \cls_{m} = \frac{\hat{a}_{m}^{\dagger} - \hat{b}_{m}^{\dagger}}{\sqrt{2}}\vac.
\end{gather}
\updated{Applying the ansatz to the Hamiltonian} gives the following filling procedure,
\begin{gather}
    \hat{h}_{n}\cls_{m} = \hat{h}_{n}^{\dagger}\cls_{m} = 0, \\
    (\hat{v}_{n} + \hat{v}_{n}^{\dagger})\cls_{m} = -\delta_{nm}\cls_{m}.
\end{gather}
Then, the proposed ansatz in Eq.~\eqref{eq:groundstate} represents the ground state.
The energy of the flatband groundstate is derived as follows,
\begin{gather}\label{eq:gsenergy}
    E_\mathrm{GS} = t|P_{\mathrm{FB}}| = 2t\nu N,
\end{gather}
where \(\nu\) is a filling fraction which is \(\nu \leq 1/2\), and \(N\) is the total number of unit cells.
Moreover, \(E_\mathrm{GS}\) is macroscopically degenerated.
\updated{At \(\nu = 1/2\), CLSs are completely occupied, and the mutual repulsion from hard-core boson constraint prevents hopping to other sites.
Hence, a Wigner crystal configuration is attained.}

% :: Subsection :: -- Filling the dimers CLS
\subsection{\texorpdfstring{\(\nu \leq 1\)}{nu leq 1} -- filling the \texorpdfstring{\(a_{n}, b_{n}\)}{an and bn} dimers}

\updated{
For hard-core bosons, when the number of particles in some flatband systems exceeds the critical filling fraction of the flatband-induced Wigner crystal, an extra particle forms a pair with an existing CLS, as discussed in Ref.~\cite{drescher2017hard, mielke2018pair}.
In the context of the cross-stitch lattices, introducing an extra particle above the critical filling fraction has the same effect as replacing a CLS with a fully-filled dimer, as expressed mathematically below,}
\begin{gather}
    \hat{a}^{\dagger}(\hat{a}^{\dagger} - \hat{b}^{\dagger}) \propto \hat{a}^{\dagger}\hat{b}^{\dagger} \propto \hat{b}^{\dagger}(\hat{a}^{\dagger} - \hat{b}^{\dagger}).
\end{gather}
Now, let us replace some of these CLSs with fully-filled dimers.
Let \(P_{d}\) denote the set of unit cells occupied by these fully-filled dimers.
\updated{Then the flatband groundstate \(\kgsfb\) is composed of states that are not elements of \(P_{d}\).}
From that, our new groundstate ansatz is proposed as follows,
\begin{gather}\label{eq:dimerground}
    \kgs = \prod_{k\in P_{d}}\hat{a}_{k}^{\dagger}\hat{b}_{k}^{\dagger}\kgsfb.
\end{gather}
The size of \(P_{d}\) is expressed with the filling fraction \(\nu\) where \(1/2 < \nu \leq 1\) and \(N\) is total number of unit cells, \(|P_{d}| = (2\nu - 1)N\).

\begin{figure}
    \includegraphics[width = \columnwidth]{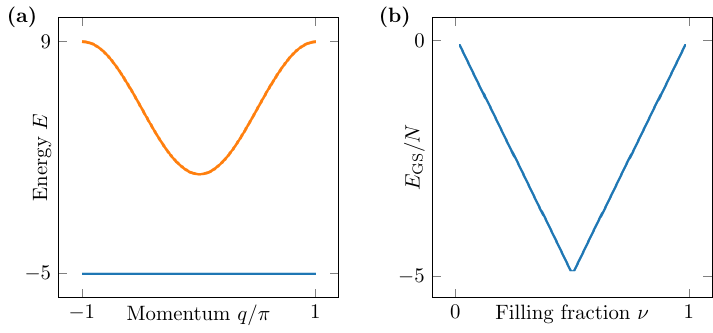}
    \caption[Energy bands and groundstate energy of 1D cross-stitch lattice]{
        (Left) Energy bands of 1D cross-stitch lattice. Here we set \(t = -5\).
        (Right) Changing of the groundstate energy as a function of a filling fraction \(\nu\). Up to \(\nu = 1/2\), we see the decreasing \(E_{\mathrm{GS}}\). Then we have increasing \(E_{\mathrm{GS}}\).
    } \label{fig:1d_cs_bands}
    \vspace{1em}
    \includegraphics[width = \columnwidth]{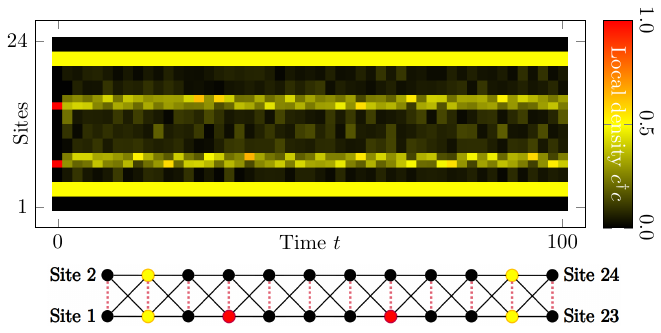}
    \caption[Trapped hard-core bosons in 1D cross-stitch lattice]{
        The time evolution of the initial wavefunction with four hard-core bosons on the one-dimensional cross-stitch lattice is plotted.
        Here, we set \(t = -5\).
        At time zero, two CLSs are located at the second and the 11th unit cells (yellow) with two hard-core bosons positioned at the seventh and the 15th sites (red).
        The CLSs are fixed as time evolves and two hard-core bosons are strictly confined.
    } \label{fig:1d_cs_2hc}
\end{figure}

Next, let us consider the filling procedure for the \(n\)-th unit cell belongs to \(P_{d}\) which include cases either \(n+1\in P_{d}\) or \(n+1\notin P_{d}\).
In either scenario, the ansatz in Eq.~\eqref{eq:dimerground} represents an eigenstate since the following relations hold,
\begin{gather}
    \hat{h}_{n}\kgs = \hat{h}_{n}^{\dagger}\kgs = 0 \quad\text{and} \quad \hat{v}_{n}\kgs = \hat{v}_{n}^{\dagger}\kgs = 0.
\end{gather}
The energy \(E_{\mathrm{GS}}\) is determined solely by the contribution of CLSs and exhibits macroscopic degeneracy,
\begin{gather}
    E_{\mathrm{GS}} = t(N - |P_{d}|) = 2tN(1 - \nu).
\end{gather}
\updated{At \(\nu = 1\), every site is entirely occupied by hard-core bosons, resulting in an insulating phase due to the lack of available states for conduction.
Hence, we see a band insulator phase.}
Otherwise, we observe a combination of the band insulator and the Wigner crystal phases.
The diagram of \(E_\mathrm{GS}\) depending on \(\nu\) is drawn in Fig.~\ref{fig:1d_cs_bands}.

% Figure 1 and 2

%\clearpage
% =========================================================================== %
% =========================================================================== %
% SECTION : Non-ergodic excitation in 1D cross-stitch
% =========================================================================== %
% =========================================================================== %
\section{Non-ergodic excitation in cross-stitch lattices}\label{ch3:sec:nonergodic}

This section presents our finding of non-ergodic excitations on the cross-stitch lattices.
Initially, we examine the scenario where one hard-core boson is situated between two CLSs.
Next, we map our model onto a spin-1 chain.
From that, we aim to explain the non-ergodicity of bosonic excitations.

% :: Subsection :: -- Three particles
\subsection{One hard-core boson between two CLSs}

Let us examine the scenario where two CLSs are positioned at the \(\pm N\)-th unit cells.
Then, we introduce a single hard-core boson between these two CLSs.
For instance, we consider the case where the boson is created at the \(a_{q}\)-th site,
\begin{gather}
    \kpsi_{1} = \hat{a}_{q}^{\dagger}\kfb,\quad \kfb = \cls_{N}\otimes \cls_{-N}, \quad\text{where }  |q| < N.
\end{gather}
\updated{Applying the Hamiltonian to \(\kpsi_{1}\)}, we encounter four distinct cases that require evaluation.
In the first case, when \(n \neq \pm N\) and \(n \neq q\), all hoppings become zero as no particles are present at those sites,
\begin{gather}
    \hat{h}_{n}\kpsi_{1} = \hat{h}_{n-1}\kpsi_{1} = \hat{h}_{n}^{\dagger}\kpsi_{1} = \hat{h}_{n-1}^{\dagger}\kpsi_{1} = 0,\\
    \hat{v}_{n}\kpsi_{1} = \hat{v}_{n}^{\dagger}\kpsi_{1} = 0.
\end{gather} 
In the second case, when \(n = q\) and \(q\pm 1 \neq \pm N\),
the term \(\hat{v}_{q}\kpsi_{1}\) is zero as no particle is created at the \(b_{q}\)-th site and no multiple counting is allowed at the \(a_{q}\)-th site.
Similarly, \updated{acting with \(\hat{h}_{q}\) and \(\hat{h}_{q-1}^{\dagger}\) on \(\kpsi_{1}\)} yields zero values because no particles are created at the \(a_{q\pm 1}\)-th and \(b_{q\pm 1}\)-th sites.
For \(\hat{h}_{q-1}\), \(\hat{h}_{q}^{\dagger}\) and \(\hat{v}_{q}^{\dagger}\), we do observe particle hops to nearby vacant sites,
\begin{gather}
    \hat{h}_{q}\kpsi_{1} = \hat{h}_{q-1}^{\dagger}\kpsi_{1} = 0,\\
    \hat{h}_{q-1}\kpsi_{1} = (\hat{a}_{q-1}^{\dagger} + \hat{b}_{q-1}^{\dagger})\kfb, \\
    \hat{h}_{q}^{\dagger}\kpsi_{1} = (\hat{a}_{q+1}^{\dagger} + \hat{b}_{q+1}^{\dagger})\kfb,\\
    \hat{v}_{q}\kpsi_{1} = 0 \quad\text{and}\quad \hat{v}_{q}^{\dagger}\kpsi_{1} = \hat{b}_{q}^{\dagger}\kfb.
\end{gather} 
Moving to the third case, \(n = q\) and \(q+1 = N\),
\updated{applying \(\hat{h}_{q}\) and \(\hat{h}_{q}^{\dagger}\) on \(\kpsi_{1}\)} results in zero due to the properties of a CLS.
Similarly, \updated{acting with \(\hat{h}_{q-1}^{\dagger}\) on \(\kpsi_{1}\)} results in zero because no particles are present at the \(a_{q-1}\)-th and \(b_{q-1}\)-th sites.
Again, \updated{acting with \(\hat{v}_{q}\) on \(\kpsi_{1}\)} also gives zero, since no particle exists at the \(b_{q}\)-th site and multiple counting is prohibited.
This same logic applies when \(q - 1 = -N\),
\begin{gather}
    \hat{h}_{q}\kpsi_{1} = \hat{h}_{q}^{\dagger}\kpsi_{1} = \hat{h}_{q-1}^{\dagger}\kpsi_{1} = 0, \\
    \hat{h}_{q-1}\kpsi_{1} = (\hat{a}_{q-1}^{\dagger} + \hat{b}_{q-1}^{\dagger}) \kfb, \\
    \hat{v}_{q}\kpsi_{1} = 0 \quad\text{and}\quad \hat{v}_{q}^{\dagger}\kpsi_{1} = \hat{b}_{q}^{\dagger}\kfb.
\end{gather}
In the fourth case, \(n = N\),
\updated{acting with \(\hat{h}_{N}\),\(\hat{h}_{N-1}\),\(\hat{h}_{N}^{\dagger}\), and \(\hat{h}_{N-1}^{\dagger}\) on \(\kpsi_{1}\)} yield zero values, which are attributed to the nature of a CLS.
Additionally, \(\kpsi_{1}\) is the eigenstate of \(\hat{v}_{N} + \hat{v}_{N}^{\dagger}\).
Similar results are observed in the case of \(n = -N\),
\begin{gather}
    \hat{h}_{N}\kpsi_{1} = \hat{h}_{N-1}\kpsi_{1} = \hat{h}_{N}^{\dagger}\kpsi_{1} = \hat{h}_{N-1}^{\dagger}\kpsi_{1} = 0, \\
    (\hat{v}_{N} + \hat{v}_{N}^{\dagger})\kpsi_{1} = -\kpsi_{1}.
\end{gather}
In summary, the presence of CLSs at both ends leads to eliminating the lattice outside of the boundary, resulting \(\kfb\) as an effective vacuum state.
Hence, we are left with an effective finite Hamiltonian \(\tilde{\mh}\), which describes the system confined within two CLSs.
\begin{align}
    \tilde{\mh}\kpsi_{1} &= -(t\hat{b}_{q}^{\dagger} \!+ \hat{a}_{q+1}^{\dagger} \!+ \hat{b}_{q+1}^{\dagger} \!+ \hat{a}_{q-1}^{\dagger} \!+ \hat{b}_{q-1}^{\dagger})\kfb, \\
    \tilde{\mh}\kpsi_{1} &= (t\hat{b}_{q}^{\dagger} + \hat{a}_{q-1}^{\dagger} + \hat{b}_{q-1}^{\dagger})\kfb,\quad q+1 = N \\
    \tilde{\mh}\kpsi_{1} &= (t\hat{b}_{q}^{\dagger} + \hat{a}_{q+1}^{\dagger} + \hat{b}_{q+1}^{\dagger})\kfb,\quad q-1 = -N
\end{align}
To illustrate this, let us consider a scenario where two CLSs are created with three vacant unit cells between them.
Then, we add one particle in one of the vacant unit cells.
In this configuration, the resulting effective Hamiltonian becomes the truncated version of the original Hamiltonian, representing the confined system within this boundary.
The effective vacuum state \( \vert \tilde{\varnothing} \rangle\) is \(\kfb\),
\begin{gather}
    \tilde{H} \cong \begin{bmatrix}
    0 & t & 1 & 1 & 0 & 0 \\
    t & 0 & 1 & 1 & 0 & 0 \\
    1 & 1 & 0 & t & 1 & 1 \\
    1 & 1 & t & 0 & 1 & 1 \\
    0 & 0 & 1 & 1 & 0 & t \\
    0 & 0 & 1 & 1 & t & 0 \end{bmatrix}
    \quad
    \text{and}
    \quad \vert \tilde{\varnothing} \rangle = \kfb.
\end{gather}
\updated{Since a single hard-core boson is unable to escape the confinement imposed by the CLSs, we can expect that when the number of bosons is less than the number of sites within the unoccupied unit cells, these bosons located in the vacant unit cells will also be trapped, and unable to leak out.}
Fig.~\ref{fig:1d_cs_2hc} presents an example of this confinement, depicting two hard-core bosons residing within eight vacant unit cells of the one-dimensional cross-stitch lattice.
\updated{The numerical simulation in Fig.~\ref{fig:1d_cs_2hc}} clearly shows that the hard-core bosons become trapped in the specified region, unable to move beyond the boundaries set by the CLSs as time evolves.

% :: Subsection :: -- Mapping to spin-1 chain
\subsection{Mapping to spin-1 chain}

To provide a more rigorous explanation for the non-ergodic excited states involving multiple particles in the one-dimensional many-body cross-stitch system, we establish a mapping of the hard-core bosons to a spin-1 chain. To accomplish this mapping, we introduce the following ladder operators:
\(S_{n}^{-} = \hat{a}_{n}^{\dagger} + \hat{b}_{n}^{\dagger}\) and \(S_{n}^{+} = \hat{a}_{n} + \hat{b}_{n}\).
After performing the necessary calculations, we derive the expression for \(S^{z}\) in the following form,
\begin{gather}
    S_{z} = 1 - (a_{n}^{\dagger}a_{n} + b_{n}^{\dagger}b_{n}).
\end{gather}
Then, we get the following triplet states at the \(n\)-th unit cell,
\begin{align}
    \ket{+}_{n} := \vert S_{n} = 1, m_{n} &= +1 \rangle = \vac, \\
    \ket{0_{t}}_{n} := \vert S_{n} = 1, m_{n} &= \hphantom{+}0 \rangle = \frac{\hat{a}_{n}^{\dagger} + \hat{b}_{n}^{\dagger}}{\sqrt{2}}\vac, \\
    \ket{-}_{n} := \vert S_{n} = 1, m_{n} &= -1 \rangle = \sqrt{2}\hat{a}^{\dagger}_{n}\hat{b}^{\dagger}_{n}\vac.
\end{align}
To clarify, the state \(\ket{+}\) corresponds to an empty particle state \(\vac\) in the language of a hard-core boson at the \(n\)-th unit cell in the one-dimensional cross-stitch lattice.
Furthermore, CLSs are represented as singlet states \(\ket{0_{s}}\),
\begin{gather}
    S^{-}_{n}\cls_{n} = 0 = S^{+}_{n}\cls_{n}.
\end{gather}
\updated{Then we can rewrite the one-dimensional cross-stitch Hamiltonian (Eq.~\eqref{eq:def1dcs}, but in the many-body context) in spin notation as follows,}
\begin{align}\label{eq:spinchain}
    \mh &= -\!\!\sum_{n}S_{n}^{-}S_{n+1}^{+} \!+ S_{n+1}^{-}S_{n}^{+} \!+ t(S_{n}^{-}S_{n}^{+} \!\!+ S_{n}^{z} -1) \\
    &= -\!\!\sum_{\langle i,j\rangle}\!\left(S_{i}^{x}S_{j}^{x} + S_{i}^{y}S_{j}^{y}\right) \!-\! t\!\sum_{n}\!\left((S_{n}^{z})^{2} - S_{n}^{z} - 1\right).\label{eq:spin1xy}
\end{align}
First, let us examine the (ground) state \(\ket{0_{s} \cdots 0_{s}}\), which occurs when CLSs are present in every unit cell.
This results in the outcome described in Eq.~\eqref{eq:gsenergy},
\begin{gather}
    \mh\ket{0_{s} \cdots 0_{s}} = tL\ket{0_{s} \cdots 0_{s}}.
\end{gather}
Let us focus on the following state \(\ket{+ \cdots +}\) full of triplet states.
This state corresponds to every site being empty, leading to \( \mh\ket{+ \cdots +} = 0\ket{+ \cdots +} \).
Next, suppose we have a sequence \(\Omega\) of triplet states and a linear combination of triplet and singlet states.
Then, for the given initial state,
\(
    \ket{ + \cdots +  0_{s} ,  \Omega ,  0_{s} + \cdots + }
\),
\(\ket{+}\) and \(\ket{0_{s}}\) remain unchanged and \(\Omega\) hops to the another possible sequence \(\Omega^{\prime}\).
The system described by the Hamiltonian in Eq.~\eqref{eq:spin1xy} corresponds to a spin-1 XY model~\cite{moudgalya2022quantum} (and a spin-1/2 XYZ creutz ladder~\cite{buvca2022out}).
What is particularly intriguing about this model is the existence of quantum many-body scars~\cite{schecter2019weak, moudgalya2022quantum,buvca2023unified} and Hilbert space fragmentation~\cite{buvca2022out,moudgalya2022quantum,buvca2023unified}.
The pivotal role played by CLSs involves the decomposition of the Hilbert space into disjoint Krylov subspaces, leading to the emergence of a truncated system, as shown in Fig.~\ref{fig:1d_cs_2hc}.
However, it is important to note that the trapped bosons within the CLS barriers in Fig.~\ref{fig:1d_cs_2hc} do not constitute a quantum many-body scar because the state is a linear combination of triplet and singlet states.
Instead, their presence gives rise to localized-like states induced by the CLS singlet state.

% :: Subsection :: -- Localized excitations in 2D lattices
\subsection{Localized excitations in 2D lattices}

Same phenomena as in the one-dimensional cross-stitch lattice are also observed in the two-dimensional cross-stitch model.
The idea of non-ergodic excited states in the one-dimensional cross-stitch lattice expands to a 2D scenario, resulting in the confinement of particles in a closed loop formed by CLSs.
The calculations for this case are straightforward and equivalent to those performed for the one-dimensional cross-stitch lattice.

Consider a configuration where multiple CLSs are arranged in a closed loop denoted as \(\mathcal{S}\).
Now, let us create a single boson at the \(a_{q,k}\)-th site and examine whether this boson can escape the confines of the loop,
\begin{gather}
    \kpsi_{1} = \hat{a}_{q,k}^{\dagger}\kfb, \quad \kfb = \!\!\!\!\prod_{(i,j)\in\mathcal{S}}\!\!\!\frac{\hat{a}_{i,j}^{\dagger} - \hat{b}_{i,j}^{\dagger}}{\sqrt{2}}\vac.
\end{gather}
\updated{Applying the Hamiltonian on \(\kpsi_{1}\)}, we get four distinct cases that need to be considered.
In the first case, when \((n,m)\notin\mathcal{S}\) and \((n,m) \neq (q,k)\), all hoppings become zero since there are no particles present at those sites,
\begin{gather}
    \hat{h}_{n,m}\kpsi_{1} = \hat{h}_{n-1,m}\kpsi_{1} = \hat{h}_{n,m-1}\kpsi_{1} = 0, \\
    \hat{h}_{n,m}^{\dagger}\kpsi_{1} = \hat{h}_{n-1,m}^{\dagger}\kpsi_{1} = \hat{h}_{n,m-1}^{\dagger}\kpsi_{1} = 0, \\
    \hat{v}_{n,m}\kpsi_{1} = \hat{v}_{n,m}^{\dagger}\kpsi_{1} = 0
\end{gather}
In the second case, when \((n,m) = (q,k)\) and  none of the neighboring sites \((q+1,k),(q,k+1),(q-1,k),(q,k-1)\) are elements of \(\mathcal{S}\),
the term \(\hat{v}_{q,k}\kpsi_{1}\) is zero.
This results from the absence of a particle at site \(b_{q,k}\)-th site and the principle that no multiple counting is allowed.
Similarly, \updated{acting with \(\hat{h}_{q,k}\) and \(\hat{h}_{q-1,k}^{\dagger}\) on \(\kpsi_{1}\)} results zero because there are no particles created at sites \(a_{q\pm 1,k}\)-th and \(b_{q\pm 1,k}\)-th sites.
For \(\hat{h}_{q,k}^{\dagger}\) and \(\hat{v}_{q,k}^{\dagger}\), we obtain the hoppings to the nearby empty sites,
\begin{gather}
    \hat{h}_{q,k}\kpsi_{1} = \hat{h}_{q-1,k}^{\dagger}\kpsi_{1} = \hat{h}_{q,k-1}^{\dagger}\kpsi_{1} = 0, \\
    \hat{h}_{q-1,k}\kpsi_{1} = (\hat{a}_{q-1,k}^{\dagger} + \hat{b}_{q-1,k}^{\dagger})\kfb, \\
    \hat{h}_{q,k-1}\kpsi_{1} = (\hat{a}_{q,k-1}^{\dagger} + \hat{b}_{q,k-1}^{\dagger})\kfb, \\
    \hat{h}_{q,k}^{\dagger}\kpsi_{1} \!= (\hat{a}_{q+1,k}^{\dagger} \!\!+ \hat{b}_{q+1,k}^{\dagger} \!\!+ \hat{a}_{q,k+1}^{\dagger} \!\!+ \hat{b}_{q,k+1}^{\dagger})\kfb, \\
    \hat{v}_{q,k}\kpsi_{1} = 0\quad\text{and}\quad \hat{v}_{q,k}^{\dagger}\kpsi_{1} = \hat{b}_{q,k}^{\dagger}\kfb
\end{gather}
Now, let us move to the third case.
It corresponds to the case when \((n,m) = (q,k)\), and we designate some or all of the unit cells at \((q+1,k)\), \((q,k+1)\), \((q-1,k)\), and \((q,k-1)\) as elements of \(\mathcal{S}\).
Then, the movement of particles to other sites depends on the configuration of \(\mathcal{S}\). 
For particles initially placed within \(\mathcal{S}\), they remain confined within it.
Likewise, \(\hat{v}_{q,k}\kpsi_{1}\) equals zero because there are no particles at site \(b_{q,k}\)-th site and no multiple counting is allowed,
\begin{gather}
    \left\{\begin{aligned}
    &\hat{h}_{q,k}\kpsi_{1},\quad \hat{h}_{q-1,k}\kpsi_{1},\quad \hat{h}_{q,k-1}\kpsi_{1}\\
    &\hat{h}_{q,k}^{\dagger}\kpsi_{1},\quad \hat{h}_{q-1,k}^{\dagger}\kpsi_{1},\quad \hat{h}_{q,k-1}^{\dagger}\kpsi_{1}
    \end{aligned}\right\} = \text{Depends on the shape }\mathcal{S}, \\
    \hat{v}_{q,k}\kpsi_{1} = 0 \quad\text{and}\quad \hat{v}_{q,k}^{\dagger}\kpsi_{1} = \hat{b}_{q,k}^{\dagger}\kfb.
\end{gather}
In the fourth case, where \((n,m)\in\mathcal{S}\),
\updated{acting with \(\hat{h}_{n,m}\),\(\hat{h}_{n-1,m}\),\(\hat{h}_{n,m-1}\), and their conjugates on \(\kpsi_{1}\)} yields zero values.
This is attributed to the nature of a CLS.
Additionally, \(\kpsi_{1}\) serves as an eigenstate of \(\hat{v}_{n,m} + \hat{v}_{n,m}^{\dagger}\),
\begin{gather}
    \hat{h}_{n,m}\kpsi_{1} = \hat{h}_{n-1,m}\kpsi_{1} = \hat{h}_{n,m-1}\kpsi_{1} = 0, \\
    \hat{h}_{n,m}^{\dagger}\kpsi_{1} = \hat{h}_{n-1,m}^{\dagger}\kpsi_{1} = \hat{h}_{n,m-1}^{\dagger}\kpsi_{1} = 0, \\
    (\hat{v}_{n,m} + \hat{v}_{n,m}^{\dagger})\kpsi_{1} = -\kpsi_{1}
\end{gather}
In brief, the presence of the CLS loop \((n,m)\in\mathcal{S}\) effectively erases again the region outside of \(\mathcal{S}\).
Hence, \(\kfb\) behaves as vacuum state, yielding an effective Hamiltonian \(\tilde{\mh}\) that characterizes the system confined within \(\mathcal{S}\).

The non-ergodic excitation can again be explained using the spin-1 representation, where \(\mathbf{n}\) and \(\mathbf{m}\) are unit-cell indices, and the angular bracket means that the indices correspond to the nearest neighbor,
\begin{gather}
    \mh = \!\!\sum_{\langle \mathbf{n},\mathbf{m} \rangle}S_{\mathbf{n}}^{-}S_{\mathbf{m}}^{+} + S_{\mathbf{n}}^{-}S_{\mathbf{m}}^{+} - \sum_{\mathbf{n}} t(S_{\mathbf{n}}^{-}S_{\mathbf{n}}^{+} + S_{\mathbf{n}}^{z} -1),
\end{gather}
Then, \(\ket{+}\) and \(\ket{0_{s}}\) remain unchanged when they are acted on by the cross-stitch Hamiltonian.
Hence, we achieve no-leaking bosons across the loop \(\mathcal{S}\).
%\clearpage
% =========================================================================== %
% =========================================================================== %
% SECTION : One-dimensional diamond lattie
% =========================================================================== %
% =========================================================================== %
\section{1D diamond lattice and non-ergodic excitation}\label{ch3:sec:diamond}

\begin{figure}
    \includegraphics[width = \textwidth]{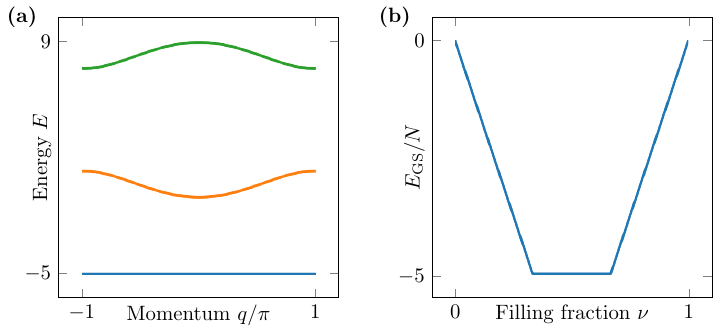}
    \caption[Energy bands and groundstate energy of 1D diamond chain]{
        (Left) Energy bands of 1D diamond lattice. Here we set \(t = -5\).
        (Right) Changing of the groundstate energy~\eqref{eq:gs_1d_dia} as a function of a filling fraction \(\nu\). Up to \(\nu = 1/3\), we see the decreasing \(E_{\mathrm{GS}}\). Then we have constant groundstate energy until \(\nu\) reaches \(2/3\). Then we see an increasing \(E_{\mathrm{GS}}\).
    } \label{fig:1d_dia_bands}
    \vspace{1em}
    \includegraphics[width = \textwidth]{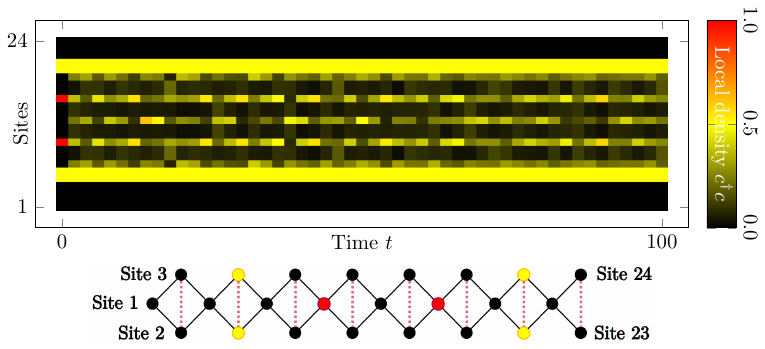}
    \caption[Trapped hard-core bosons in 1D diamond chain]{
        The time evolution of the initial wavefunction with four hard-core bosons on the 1D diamond lattice is plotted.
        Here, we set \(t = -5\).
        At time zero, two CLSs are located at the second and the seventh unit cells (yellow) with two hard-core bosons positioned at the 10th and the 16th sites (red).
        The CLSs are fixed as time evolves and two hard-core bosons are strictly confined.
    } \label{fig:1d_dia_2hc}
\end{figure}

A similar result can be achieved for the one-dimensional diamond lattice.
The value of \(t\) is taken sufficiently low, \(t < -2\), to ensure the flatband is the groundstate energy.
Then, the flatband energy, \(E_{\mathrm{FB}} = t\), is gapped away from the other bands at the single particle level,
\begin{gather}
    \label{eq:dc-ham-hcb}
    \mh = -\sum_{n} \hat{h}_{n} + \hat{h}_{n}^{\dagger} + \hat{t}(\hat{v}_{n} + \hat{v}_{n}^{\dagger}), \\
    \hat{h}_{n} = (\hat{a}_{n}^{\dagger} + \hat{c}_{n}^{\dagger})(\hat{b}_{n} + \hat{b}_{n+1}) \quad\text{and}\quad \hat{v}_{n} = \hat{a}_{n}^{\dagger}\hat{c}_{n}.
\end{gather}
%\alexei{AA: the notation \(a,b,v_n\) for sites is easily confused with \(a,b,c^\dagger_n\)! Please update}
\updated{Here, \(\hat{a}_{n}^{\dagger}\) and \(\hat{c}_{n}^{\dagger}\) represent the creation operators at the dimer sites, \(A_{n}\) and \(C_{n}\), respectively.}
\updated{The creation operator at the bottleneck site, \(B_{n}\), is denoted as \(\hat{b}_{n}^{\dagger}\).}
\updated{To summarize our findings on groundstate energy, we apply methods similar to those used for the one-dimensional cross-stitch lattice.}
%We briefly summarize our observation on the groundstate energy.
%We follow the same procedures of the one-dimensional cross-stitch lattice for the one-dimensional diamond lattice.
Initially, CLSs are filled where the filling fraction should be \(\nu \leq 1/3\).
When \(\nu\) is exactly \(1/3\), \updated{a Wigner crystal emerges.}
\updated{This results from the mutual repulsion from the hard-core boson constraint and also the destructive interference caused by the CLS property.}
Next, we fill the bottleneck sites \(1/3 \leq \nu \leq 2/3\).
Then, some CLSs are replaced with fully-filled dimers.
If all CLSs are replaced with a fully-filled dimer, all sites are filled with each hard-core boson, resulting in the band-insulating phase.
Moreover, the groundstate energy \(E_{\mathrm{GS}}\) is only determined by the contribution of the CLSs, leading to macroscopic degeneracies.
\begin{gather}\label{eq:gs_1d_dia}
    E_\mathrm{GS} = \begin{dcases}
        3t\nu N, &\quad \nu \leq 1/3\\
        tN, &\quad 1/3 \leq \nu \leq 2/3\\
        3tN(1-\nu), &\quad 2/3 \leq \nu \leq 1
    \end{dcases}
\end{gather}
The diagram of \(E_\mathrm{GS}\) depending on \(\nu\) is drawn in Fig.~\ref{fig:1d_dia_bands}.
An example with two hard-core bosons in four vacant unit cells between two CLSs is shown in Fig.~\ref{fig:1d_dia_2hc}.
\updated{From the numerical simulation in Fig.~\ref{fig:1d_dia_2hc},} we see that the hard-core bosons are trapped between two CLSs, and no leakage is observed as time evolves.

In the case of the diamond lattice, there is no well-defined mapping in spin language, as in the cross-stitch lattice, where a well-defined spin-integer chain representation exists.
Instead, it is more accurately represented through a model with (spin-1)-(hard-core boson) coupling,
\begin{gather}
    \mh = -\sum_{n} \hat{T}^{-}_{n}(\hat{b}_{n} + \hat{b}_{n+1}) + (\hat{b}_{n}^{\dagger} + \hat{b}_{n+1}^{\dagger})\hat{T}^{+}_{n} +  t\left(\hat{T}^{-}_{n}\hat{T}^{+}_{n} + \hat{T}^{z}_{n} - 1\right),
\end{gather}
where \(\hat{T}_{n}^{-} = \hat{a}_{n}^{\dagger} + \hat{c}_{n}^{\dagger}\) and \(\hat{T}_{n}^{+} = \hat{a}_{n} + \hat{c}_{n}\).
Then, CLSs are precisely characterized as singlet states.
Moreover, if \updated{\(\hat{b}_{n} + \hat{b}_{n+1}\) acts on the CLS}, we get zero.
The sole contribution arises from the remaining term, \(-t(\hat{T}^{z}_{n} - 1)\), which takes the CLS as an eigenstate.
Hence, the CLSs remain unchanged, and any particles in the loop become confined, similar to what is observed in the cross-stitch lattices.
% Figure 2

%\clearpage
% =========================================================================== %
% =========================================================================== %
% SECTION : Fluxed 1D diamond lattice
% =========================================================================== %
% =========================================================================== %
\section{Fluxed 1D diamond lattice and ergodicity}\label{ch3:sec:fluxeddiamond}

The orthogonality of CLSs is a well-known characteristic in both one- and two-dimensional cross-stitch lattices and the one-dimensional diamond lattice.
However, when introducing flux to the one-dimensional diamond system, the CLSs are no longer orthogonal to each other.
In this case, we question whether we can fill up all CLSs and bottleneck sites to achieve the Wigner crystal and the band-insulating phase.
To preserve the flatness, let us consider the Hamiltonian with zero dimer hopping, \(t = 0\).
This system also has CLSs, \(\cls_{n}^{2}\) \updated{(superscript \emph{2} emphasizes that the CLS occupies two unit cells)}, with the flatband energy \(E_{\mathrm{FB}} = 0\), but they occupy two unit cells,
\begin{gather}
    \mh \!=\! -\sum_{n}(\hat{b}_{n}^{\dagger} \!+ \hat{b}_{n+1}^{\dagger})\hat{a}_{n} \!+ (\hat{b}_{n}^{\dagger} \!+ e^{-i\phi}\hat{b}_{n+1}^{\dagger})\hat{c}_{n} \!+ \mathrm{h.c}., \\
    \cls_{n}^{2} = \frac{1}{\sqrt{4}}\left(\hat{a}_{n}^{\dagger} - \hat{c}_{n}^{\dagger} + e^{-i\phi}\hat{a}_{n+1}^{\dagger} - \hat{c}_{n+1}^{\dagger}\right)\vac.
\end{gather}
The flatband energy is exactly zero and no longer corresponds to the groundstate energy.
The filling procedure is not limited to the groundstate energy; hence, we can fill up CLSs, but they should not overlap.
Then, the filling fraction is \(\nu \leq 1/6\).
The Wigner crystal is obtained when the eigenstate is the following,
\begin{gather}
    \eig = \prod_{n = 1}^{N/2}\cls_{2n}^{2} \hspace{0.5em}\text{or}\hspace{0.5em}\prod_{n = 1}^{N/2}\cls_{2n-1}^{2}.
\end{gather}
The next question is whether we can fill up the empty bottleneck sites in CLSs.
The answer is no; doing so does not yield the eigenstate.
It is easily verified through straightforward calculations using the following state,
\begin{gather}
    \kpsi = \prod_{j=1}^{N/2}\kpsi_{2j},\hspace{0.5em}\text{where }\kpsi_{2j} \!= \hat{b}^{\dagger}_{2j}\cls_{2j-1}^{2}.
\end{gather}
\updated{Acting with \(\mh\) on \(\kpsi\)}, we obtain non-trivial hopping.
This suggests that hard-core bosons are not trapped, as they can escape through the bottleneck sites.
In this case, there are no non-ergodic excited states.
%\clearpage
% =========================================================================== %
% =========================================================================== %
% SECTION : Conclusion
% =========================================================================== %
% =========================================================================== %
\section{Conclusion}\label{ch3:sec:conclusion}

This chapter analyzes the behavior of on-site interacting hard-core bosons in the one- and two-dimensional cross-stitch lattices.
The groundstate of the system is composed of CLSs, which significantly influence the overall ground state energy.
One of our findings are band-insulating and Wigner crystal phases within these systems.
Moreover, a fusion of these phases is also identified.
When a closed CLS loop is present, the bosons that are initially positioned in the loop become trapped, leading to the violation of weak thermalization.
This highlights the unique behavior of the systems in terms of the non-ergodicity, influenced by the properties of the groundstates.
Moreover, the implications of our observations extend to the Hilbert space fragmentation.

Furthermore, it is noteworthy that the non-ergodic phenomenon appears in the one-dimensional and two-dimensional diamond chains.
Our observation suggests that non-ergodic excitations are contingent upon specific hoppings and network geometry conditions, particularly involving bottleneck sites and orthogonal CLSs.
This assumption arises since the same procedure does not work for Kagome, Lieb lattices (not included in the chapter), and the one-dimensional diamond lattice with magnetic flux (Sec.~\ref{ch3:sec:fluxeddiamond}), which have non-orthogonal CLSs.
They do not have (or they do have only partially) the groundstates that are observed in the cross-stitch and the diamond lattices without any flux, which do not lead to the non-ergodic phenomenon.

\chapter[Flatband Electric Circuits]{Flatband Electric Circuits}~\label{chpt4}

\iffalse
\vfill
The first systematic and detailed experimental demonstration of compact localized states, to the best of our knowledge, in one-dimensional flatband electric circuits is provided in this chapter.
The realization of a compact localized state in the experiment is of key importance, as it serves as strong evidence that the experimental setup truly represents a flatband system.
The key feature of our work is to investigate of how flatbands respond to the (local in space) sinusoidal driving.
Furthermore, the robustness of a compact localized state to the addition of nonlinear electronic varactor elements is demonstrated.
To analyze the experimental results thoroughly, both analytical studies and numerical simulations are employed, which show excellent agreement with the experimental data.
This study represents a significant step towards the establishment of a versatile circuit platform for the generation and manipulation of flatbands, and it is expected to be of interest to a wide audience.
The relevant paper is published in Ref.~\cite{chase2023compact}.
\fi
%========================================================================================
%\clearpage
% =========================================================================== %
% =========================================================================== %
% SECTION : Introduction
% =========================================================================== %
% =========================================================================== %
\section{Introduction}

The first systematic and detailed experimental demonstration of compact localized states, to the best of our knowledge, in one-dimensional flatband electric circuits is provided in this chapter.
The generation of a compact localized state (CLS) mode in the experiment is crucial, as it is strong evidence that the experimental setup \emph{really} represents a flatband system.
The key feature in this chapter is to investigate how flatbands respond to the local (in space) sinusoidal driving.
Furthermore, the robustness of a CLS mode to the addition of nonlinear electronic varactor elements is demonstrated.
Analytical studies and numerical simulations are employed to analyze the experimental results thoroughly, which show excellent agreement with the experimental data.
This chapter represents a significant step towards establishing a versatile circuit platform for generating and manipulating flatbands, and it is expected to be of interest to a broad audience.
The relevant paper is uploaded as an \emph{arXiv} preprint in Ref.~\cite{chase2023compact}.

%\subsection{Previous research on realizing flatband lattices}
In the current research on flatband physics, the experimental realization of artificial flatband lattices is a top priority.
However, constructing these lattices is challenging because they require careful fine-tuning to maintain compact localized states (CLSs) over a long time.
Previous attempts to realize flatband lattices in various experimental contexts have faced limitations, such as short observation times, lack of essential relative phase information for a CLS, and insufficient spatial resolution.
These experiments have been conducted using photonic lattices~\cite{nakata2012observation, kajiwara2016observation, mukherjee2015observation, vicencio2015observation, nguyen2018symmetry, ma2020direct}, cold atoms~\cite{taie2015coherent, ozawa2017interaction}, polariton condensates~\cite{baboux2016bosonic, masumoto2012exciton}, electrical circuits~\cite{zhang2023non, wang2022observation, wang2019highly}, topological materials~\cite{kang2020}, and magnonic crystal lattices~\cite{magnonic}.
From options, electrical circuits show particular promise for studying CLSs in detail.
They offer advantages such as the ease of constructing diverse lattices, finely adjusting lattice parameters, and precise experimental control and measurement capabilities.

%\subsection{General summary of the chapter}
In this chapter, we summarize the work on how we build and study one-dimensional flatband electrical lattices using discrete capacitive and inductive circuit elements~\cite{chase2023compact}.
By exciting flatband eigenstates through local sinusoidal driving at the flatband frequency, we explore two distinct lattice structures: the diamond lattice and the stub lattice, each belonging to different flatband categories.
The diamond lattice contains orthogonal CLSs and exhibits resonant modes when driven locally at the flatband frequency.
In contrast, the stub lattice features non-orthogonal CLSs, leading to exponentially localized resonant modes due to overlapping CLSs.
Capacitors are replaced with varactord having voltage-dependent capacitance to introduce nonlinearity to the lattices.
Remarkably, we find that CLSs persist in the diamond lattice even in the highly nonlinear regime, demonstrating the robustness of the underlying linear mechanism across a wide range of excitation amplitudes.

%\subsection{Outline of the chapter}
The chapter is organized as follows.
We start by defining and discussing the diamond and the stub lattice in Sec.~\ref{ch4:sec:models}.
In Sec.~\ref{ch4:sec:results}, the details of the experimental setup is listed, and we look at the experimental and numerical results on generating CLSs in both lattices, followed by conclusions in Sec.~\ref{ch4:sec:conclusion}.

%\clearpage
% =========================================================================== %
% =========================================================================== %
% SECTION : Introduction
% =========================================================================== %
% =========================================================================== %
\section{Models}\label{ch4:sec:models}

In the context of electrical lattices, capacitors and inductors are represented by the vertices and edges of a tight-binding lattice.
The lattices can be seen as discrete examples of electrical transmission lines but with nontrivial geometry.

% =========================================================================== %
% SUBECTION : One-dimensional diamond electric lattice
% =========================================================================== %
\subsection{One-dimensional diamond electric lattice}
\begin{figure}
\centering
    \includegraphics[width = 0.75\columnwidth]{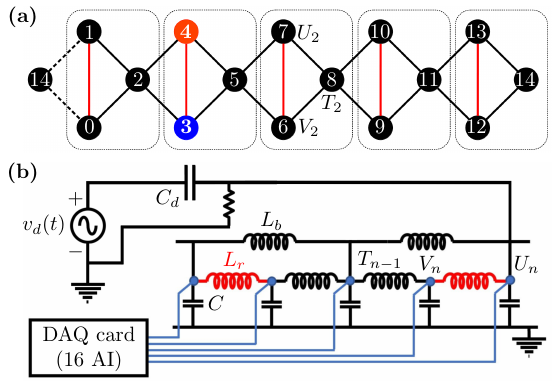}
    \caption[One-dimensional diamond electric circuit]{\updated{
        (a) Tight binding representation of the diamond lattice.
        Grey boxes indicate unit cells, the red and blue circles represent a CLS with opposite amplitudes.
        (b) Schematic representation of the electrical circuit diamond lattice. Each site is grounded with a capacitor, $C$.
        The sites are coupled with inductors of inductance $L_b$ (black) and \(L_r\) (red).
        The lattice is driven at one site via a driving capacitor, \(C_d\), by a sinusoidal voltage signal, \(v_d(t)\), from a signal generator.
        All sites (\(T_0, U_0, V_0, \ldots, T_4, U_4, V_4\)) are monitored simultaneously via a data acquisition card with 16 analog inputs.}
    }
    \label{fig:diamond_scheme}
\end{figure}
In Fig.~\ref{fig:diamond_scheme}, a diamond lattice with two different hopping values, shown as black and red lines, and the corresponding electrical circuit is displayed.
Each node in the lattice has a capacitance of \(C\), and we incorporate inductors with two different inductance values, \(L_{b}\) and \(L_{r}\), as black and red colors, respectively.
The main source of dissipation in the lattice arises from the ferrite core inductors, so we consider these inductors to have an effective serial resistance of \(R\).
At the same time, the capacitors are assumed to be ideal.

The voltages at the three nodes in the \(n\)-th unit cell are represented as \(T_n\), \(U_n\), and \(V_n\).
To drive the lattice, a small driving capacitor with capacitance \(C_d \ll C\) is connected to one site (\(U_m\))  that participates in the two-site CLS situated at the \(m\)-th unit cell.
By applying Kirchhoff's current law at each node, we can derive the equations of motion for the voltages at each node within the \(n\)-th unit cell in the linear level, given as follows,
\begin{align}
    \ddot{T}_n + \beta \dot{T}_n &= -\omega^2_b\left(4T_n - U_n -U_{n+1} - V_n -V_{n+1}\right), \notag \\
    \ddot{U}_n \!+ \beta \dot{U}_n &= -\omega^2_b\left(\left(2 + \alpha\right) U_n -\alpha V_n - T_n-T_{n-1}\right), \label{eq:diamond_undriven} \\
    \ddot{V}_n + \beta \dot{V}_n &= -\omega^2_b\left(\left(2 + \alpha\right) V_n -\alpha U_n - T_n-T_{n-1}\right). \notag 
\end{align}
At the driven site \(U_{m}\), the right-hand side term in Eq~\eqref{eq:diamond_undriven} \updated{needs to be modified} by a factor \(1/(1+\gamma)\) and additional driving force \(v_{d}(t) = v_{d}\sin(\omega_d t)\),
\begin{gather}\label{eq:diamond_driven}
    \ddot{U}_m \!+ \frac{\beta \dot{U}_m}{1 + \gamma} = -\frac{\omega^2_b}{1+\gamma}\left(\left(2 + \alpha\right) U_m \!-\alpha V_m \!- T_m \!-T_{m-1} \vphantom{\sum}\right) + A\sin(\omega_d t), \\
    \omega_{b}^{2} = \frac{1}{L_{b}C}, \hspace{0.5em} \alpha = \frac{L_{b}}{L_{r}}, \hspace{0.5em} \beta = \frac{R_{b}}{L_{b}} = \frac{R_{r}}{L_{r}},\hspace{0.5em} \gamma = \frac{C_{d}}{C},\hspace{0.5em} \text{ and }\hspace{0.5em} A = \frac{\gamma \omega_{d}^{2}}{1+ \gamma}v_{d} \notag .
\end{gather}
Among these, \(\alpha\) is a tunable parameter that can shift the flatband, and \(\gamma\) arises as an artifact due to the driving, which can be minimized to within the experimental tolerance of \(\omega_b^2\).
\(\beta\) accounts for the dissipation effect, which is assumed to be a constant.
It is a reasonable assumption because the larger inductance tends to have larger resistance due to more windings of copper wire.
For the theoretical treatment, we assume the factor \(1/(1+\gamma)\) is approximately equal to one.
The additional local driving is a tiny impurity in this context.
Then, the equation of motion is written using a vector form \(\vert \boldsymbol{\psi} \rangle \!=\! (T_0, U_0, V_0, \ldots)^{t}\), 
\begin{align}\label{eq:diamond_schrodinger_eq}
    \left(\frac{d^2}{dt^2} + \beta \frac{d}{dt} \right) \vert \boldsymbol{\psi} \rangle = \hat{H} \vert \boldsymbol{\psi} \rangle + \vert \mathbf{F}(t) \rangle.
\end{align}
\(\hat{H}\) represents the dynamical Hamiltonian of the diamond lattice, which gives the right-hand side of Eq.~\eqref{eq:diamond_undriven} when acting on \(\vert \boldsymbol{\psi} \rangle\).
The term \(\vert \mathbf{F}(t) \rangle\) corresponds to the driving force, contributing to the last term in Eq.\eqref{eq:diamond_driven}.
Such vectorized form is not limited to the electric diamond system.
In contrast to the hoppings in tight-binding diamond lattices, the inductors also contribute to the ``onsite potential" (diagonal elements of \(\hat{H}\)), ensuring that \(\omega^2\) remains positive.

To determine the eigenfrequencies, we disregard the driving and assume the Bloch waveform \(U_{n} = U(k)\exp(i(\omega t - kn))\) (and similarly for \(V_{n}, T_{n}\)).
Then we obtain the following eigenvalue problem, where \(Q\) is defined as \(-(1 + \exp(ik))\),
\begin{gather}
    \frac{\left(\omega^2 - i\beta \omega\right)}{\omega^2_b}\begin{bmatrix}
        T(k) \\ U(k) \\ V(k)
    \end{bmatrix}
    =
    \begin{bmatrix}
        4 & Q & Q \\
        Q^* & (2+\alpha) & -\alpha \\
        Q^* & -\alpha & (2+\alpha)
    \end{bmatrix}
    \begin{bmatrix}
        T(k) \\ U(k) \\ V(k)
    \end{bmatrix}.
\end{gather}
In the zero \(\beta\) limit, we obtain the following frequency mode:
\begin{gather}\label{eq:diamond_eigenvalues}
    \omega_{\mathrm{FB},0}^{2} = 2\omega^{2}_b(\alpha + 1),
    \hspace{0.5em} \omega_{\mathrm{DB},0}^2 = \omega^2_b(3\pm \sqrt{4\cos(k) + 5}).
\end{gather}
For the non-zero \(\beta\) case, one has to solve \(\omega^2 - i\beta\omega = \omega^2_{\mathrm{FB}/\mathrm{DB},0}\).
Then, it leads to the following solution with a dissipation time \(\tau = 2/\beta\) and shifts the real part:
\begin{align}\label{eq:diamond_eigenvalues_complex}
    \omega_{\mathrm{FB}/\mathrm{DB}} = i\frac{\beta}{2} \pm \sqrt{-\frac{\beta^2}{4} + \omega^2_{\mathrm{FB}/\mathrm{DB},0}}.
\end{align}
We assume only underdamped frequencies, making the square root part always real, and the shifting is less than 1\% in the experiment.
Hence, when \updated{we refer to} \(\omega_{\mathrm{FB}/\mathrm{DB}}\), it is assumed to be \(\omega_{\mathrm{FB}/\mathrm{DB},0}\).

Although the flatband eigenvectors represent Bloch waves, which are spatially spread across the entire lattice, the flatband degeneracy allows CLSs to emerge.
These states can be formed by combining the Bloch waves linearly, as described in Ref.~\cite{rhim2019classification}.
In the case of the diamond lattice, \updated{the translated copies of normalized CLS are expressed as  \(U_{n} = \delta_{nm}/\sqrt{2} = -V_{n}\) and \(T_{n} = 0\), and these states are orthogonal in the flatband subspace.}

% =========================================================================== %
% SSUBECTION : One-dimensional stub electric lattice
% =========================================================================== %
\subsection{One-dimensional stub electric lattice}
\begin{figure}[ht]
\centering
    \includegraphics[width = 0.8\columnwidth]{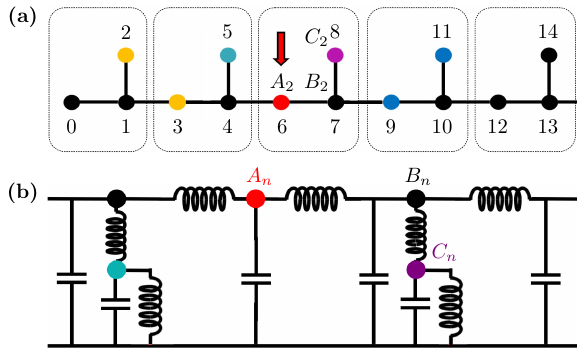}
    \caption[One-dimensional stub electric circuit]{\updated{
        (a) Tight binding representation of the stub lattice.
        In-stub tight-binding diagram, (\(+\)) at sites 5, 8 and (\(-\)) at site 6 represent a CLS.
        The lattice is driven at site 6 via a driving capacitor, \(C_d\), by a sinusoidal voltage signal, \(v_d(t)\), from a signal generator.
        (b) Schematic representation of the electrical circuit stub lattice.
        The general structure is basically the same as for the diamond case.
        The sites are coupled with inductors of inductance \(L_b\).}
    }
    \label{fig:stub_scheme}
\end{figure}
In Fig.~\ref{fig:stub_scheme}, we illustrate the circuit representation and schematic of the stub lattice.
In the stub case, only one type of inductor with an inductance of \(L_{b}\) is required.
As mentioned earlier, the additional inductor for the ``stub" sublattice connected to the ground is necessary to achieve a flatband.
By incorporating these inductors, we can visually confirm that the three capacitors involved in a CLS have two inductor connections each, indicating the same ``onsite potentials" for sites \(A_{n}\) and \(C_{n}\).
In contrast, the connecting nodes, \(B_{n}\), have three inductors each.
A similar analysis can be conducted for the stub lattice flat band.
The equation of motion for the stub lattice is given as follows,
\begin{align}
    \ddot{A}_n +\beta \dot{A}_n &= -\omega^2_b\left[2A_n -B_{n-1} - B_n \right], \notag \\
    \ddot{B}_n +\beta \dot{B}_n &= -\omega^2_b\left[3B_n - C_n - A_{n} - A_{n+1}\right] \label{eq:stub_undriven}, \\
    \ddot{C}_n +\beta \dot{C}_n &= -\omega^2_b\left[2C_n - B_{n}\right]. \notag 
\end{align}
The driven site is \(A_m\), and the equation of motion at the driven site is modified as described in Eq.~\eqref{eq:diamond_driven}.
By applying a similar analysis, we obtain the following eigenfrequencies for the stub lattice, and there exists a flatband:
\begin{gather}\label{eq:stubdisp}
    \omega^2_{\mathrm{FB},0} = 2\omega^2_b \quad\text{and}\quad \omega^2_{\mathrm{DB},0} = \omega^2_b\left(\frac{5}{2} \pm \frac{\sqrt{8\cos(k) + 13}}{2}\right).
\end{gather}
At \(\omega = \omega_{\mathrm{FB}}\), the CLS amplitudes are obtained as follows, and there exists an overlap between two consecutive CLSs, which makes them non-orthogonal.
\begin{gather}\label{eq:stub_cls}
    C_n = \frac{\delta_{nm}}{\sqrt{3}} = C_{n+1} \quad\text{and}\quad A_{n+1} = -C_n.
\end{gather}

%\clearpage
% =========================================================================== %
% =========================================================================== %
% SECTION : Results
% =========================================================================== %
% =========================================================================== %
\section{Experimental and numerical results}\label{ch4:sec:results}

\updated{
This section presents the results of local driving (both linear and nonlinear) applied to the diamond lattice.
The diamond lattice, characterized by orthogonal CLSs, exhibits resonant modes under local driving at the flatband frequency with the persistence of CLSs  even in the highly nonlinear regime.
Subsequently, the impacts of local linear driving on the stub lattice is also presented.
Exponentially localized resonant modes are observed in the stub lattice, featuring non-orthogonal CLSs, due to the overlap among CLSs.
}

% =========================================================================== %
% SUBECTION : Results on the one-dimensional diamond lattice
% =========================================================================== %
\subsection{Results on the one-dimensional diamond lattice}
\begin{figure}
\centering
    \includegraphics[width = \columnwidth]{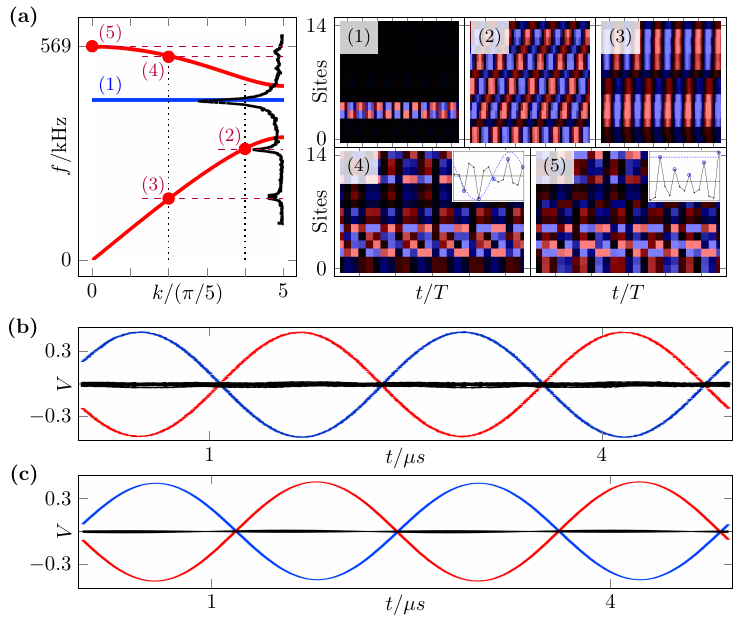}
    \caption[Linear driving on the one-dimensional diamond lattice]{
        (a) The response to local driving (at node 4) as a function of driving frequency. 
        The system features two dispersive bands - one acoustic and one optical branch, as well as one flatband. 
        %Driving near the bottom of the optic branch (or inside it) results in a spatially extended pattern (see top panel). 
        In panels (1) -- (5), a driver frequency is chosen according the labels in (a). Driving near the flatband yields a CLS (1),  
        driving in the acoustic branch yields a spatially extended response (2),(3), and two examples in the optical branch are shown in (4), (5). There the inset shows the spatial mode configuration (consistent with the corresponding k-value).
        The time axis is in units of $T = 1/f_d$.
        (b) response at flatband corresponding to the highest peak in (a), $f_d = 401$ kHz, see panel (1). 
        Blue and red colored lines correspond to the experimental CLS sites 3, 4.
        (c) Simulation result of Eq.~\eqref{eq:diamond_undriven} %,~\eqref{eq:diamond_driven} 
        with experimental parameters at $423$ kHz.
    }
    \label{fig:diamond_result}
\end{figure}
An electrical circuit diamond lattice comprising 5 unit cells is considered with periodic boundary conditions.
The lattice incorporates a total of 15 capacitors, with a capacitance value of \(C=\text{1} \pm \text{0.01 }\mathrm{nF}\).
For the driving capacitor, we used \(C_d=\text{15 }\mathrm{pF}\) resulting in  \(\gamma = 0.015\).
The two distinct inductance values utilized were \(L_b=\text{466 }\mu\mathrm{H}\) and \(L_r=\text{674 }\mu\mathrm{H}\)  both falling within a 1\% tolerance.

The primary sources of dissipation in the inductors are attributed to the ferrite cores and the coil-wire resistance, which collectively affect the quality factor \(Q\), defined as \(Q=\omega L/R_\mathrm{eff}\).
While an inductor's \(Q\) factor remains approximately constant, the effective series resistance fluctuates with the resonant frequency.
The \(Q\) factor is approximately 55 at \(\text{232 }\mathrm{kHz}\) corresponding to an effective series resistance of about \(\text{23 }\Omega\) for the \(L_{b}\) inductor.
To further enhance the circuit's performance, a \(\text{10 }\mathrm{k}\Omega\) resistor is introduced in parallel with the lattice to suppress any DC voltage component and prevent charge buildup on the capacitors.
With these given values, we compute the band structure using Eq.~\eqref{eq:diamond_eigenvalues}, represented by red (dispersive) and blue (flatband) lines in Fig.~\ref{fig:diamond_result} (a).
Notably, the flat-band frequency is determined to be \(f_\mathrm{FB} = \text{429 }\mathrm{kHz}\), falling in the spectral gap situated between the two dispersive bands.

To experimentally investigate the presence of a flatband and excite its CLS, an energy is locally applied in the form of a sinusoidal voltage, as depicted in Fig.~\ref{fig:diamond_scheme}.
The lattice excitation and measurement procedures are shown schematically in Fig.~\ref{fig:diamond_result}.
The response voltage at each lattice site is monitored simultaneously, ranging from 0 to 14, corresponding to \(U_0, V_0, T_0, \ldots, U_{4}, V_{4}, T_{4} \).
The driving voltage is introduced at the fourth site (indexed with \(U_{1}\)), allowing for a prominent display of the flatband response while also potentially exciting other extended wave modes.
It is important to note that, while \updated{the band structure is continuous in an unbounded chain}, there are only \(N\!=\!5\) resonance mode per band (with peaks at \(|k| \!=\! 2\pi/5, 4\pi/5\) for the dispersive bands) due to the finite number of sites, \updated{hence \(\lceil N/2 \rceil=3\) peaks per band due to two-fold degeneracy.}

Let us now focus on the effect of the local driving frequency.
\updated{The function generator was considered} in sweep mode, spanning \(\text{200 -- 600 }\mathrm{kHz}\) in \(\text{25 }\mathrm{ms}\), to acquire the steady-state amplitude responses of site 4, which were obtained using an oscilloscope (no DAQ card).
These responses are represented by the black trace along the right vertical axis in Fig.~\ref{fig:diamond_result} (a).
The prediction for the flat-band frequency is accurately matched.
The strength of this resonance peak depends on various parameters, including dissipation, driving voltage, and the amplitude of the resonant eigenvector at \(U_{m}\).
Our observations show that the largest peak reaches \(\text{0.44 }\mathrm{V}\) at \(\text{423 }\mathrm{kHz}\).
Additionally, two other prominent peaks in the acoustic branch are visible at \(|k|\!=\! 2\pi/5, 4\pi/5\), whereas the resonance modes in the upper dispersive band are not so clearly visible.

Next, \updated{the function generator was adjusted to} align with the frequencies of the observed resonance peaks.
The color-image panels (1)--(5) of Fig.~\ref{fig:diamond_result} (a) display the spatial patterns observed at different drive frequencies once a steady state is reached.
The response voltage is depicted by colors, with red indicating positive, blue representing negative, and black denoting zero voltage.
The color map is re-scaled to the peak amplitude for clarity.
\updated{For panels (4) and (5), inset shows the spatial voltage profile (consistent with the corresponding \(k\)-value of the Bloch eigenfunction) at a specific moment in time.}

At the frequency \(f_d = \text{423 }\mathrm{kHz}\), corresponding to the largest resonance peak located at the flatband, the associated CLS is expected to exist at the two dimer sites \(U_{n}\) and \(V_{n}\) of a single unit cell, with their respective excitations being out of phase.
Fig.~\ref{fig:diamond_result} (b) and Fig.~\ref{fig:diamond_result} (c) are the voltage-time profiles of all 15 sites, both the experimental and numerical results, respectively.
The red trace represents the response of the driven site, while the blue trace represents the other CLS site.
In Fig.~\ref{fig:diamond_result} (b), the two traces are exactly out-of-phase, leading to destructive interference at  neighboring bottleneck sites (\(T_{n}\) and \(T_{n-1}\)).
However, a slight leakage is observed in the rest of the sites (black traces) due to the experimental imperfections introducing disorder and dissipation, broadening the dispersive resonance peaks.
Nonetheless, a very high degree of energy localization is observed.
It is important to note that to generate the CLS, it is essential to drive at a site that participates in the CLS.
For example, if we apply the driving at \(T_{n}\), the CLS cannot be obtained at any driver frequency.
In Fig.~\ref{fig:diamond_result} (c), we see excellent agreement between experiment and simulation.
In the simulation, as the parameter \(\beta\) approaches zero, all sites except the CLS sites tend to have identically zero voltages, resulting in a true CLS that is perfectly compact localized.
The CLS sites oscillate at the flatband frequency of \(\text{429 }\mathrm{kHz}\).
It is worth noting that in the experiment, the frequency is slightly shifted down to \(\text{401 }\mathrm{kHz}\) due to the presence of small parasitic capacitances associated with the measurement apparatus, such as ribbon cables and DAQ board.

% =========================================================================== %
% SUBECTION : Nonlinear driving and robustness of diamond CLS
% =========================================================================== %
\subsection{Nonlinear driving and robustness of diamond CLS}
\begin{figure}
\centering
    \includegraphics[width = \columnwidth]{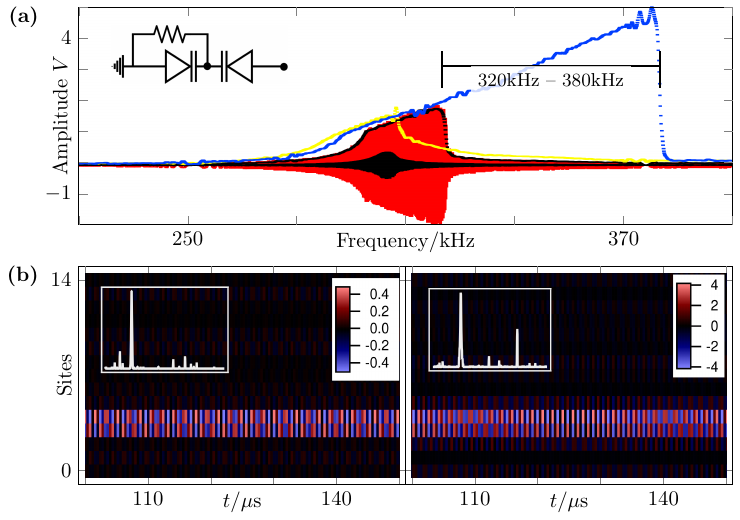}
    \caption[Nonlinear driving on the one-dimensional diamond lattice]{
        Introducing nonlinearity to the system by (a) replacing the capacitors with the element (shown in the inset) comprised of two oppositely facing varactor and a resistor.
        This configuration exhibits symmetrically decreasing capacitance with respect to zero voltage.
        This leads to symmetric hard-type nonlinearity, as revealed by resonance curves (amplitude response vs. driving frequency) when combining the element with an inductor, taken at different driving amplitudes. 
        At the highest amplitude (blue, yellow), a significant hysteresis window emerges.
        Driving the diamond at site \(3\) excites a CLS mode, both at low amplitude, \(v_d = \text{1 }\mathrm{V}\) and \(f_d= \text{560 }\mathrm{kHz}\) left panel of (b), and at high amplitude, \(v_d = \text{7 }\mathrm{V}\) and \(f_d= \text{617 }\mathrm{kHz}\) right panel of (b). 
        The insets show the FFT of response at site \(3\).
        In the right panel of (b), higher harmonics at \(3 f_d\) becomes prominent.
    }
    \label{fig:nonlinear}
\end{figure}
Let us consider the impact of nonlinearity on the diamond CLS.
Prior research has shown that linear CLSs can extend into the nonlinear regime for flatbands that are separated from the dispersive spectrum~\cite{danieli2018compact}.
The capacitors \updated{is replaced} with varactor to introduce nonlinearity in the lattice shown in Fig.~\ref{fig:diamond_scheme}~(b).
For our specific purposes, it is advantageous to use two diodes facing each other for each capacitor, as depicted in the inset in Fig.~\ref{fig:nonlinear} (a).
Additionally, a resistor is connected from the junction between the diodes to the ground to prevent a DC charge buildup.
This arrangement leads to a hard-type nonlinearity because the effective capacitance decreases symmetrically as the voltage becomes either more positive or negative.
Assuming the nonlinearity strength is not too strong, the modification in Eq.~\eqref{eq:diamond_undriven} for the lattice variables is described by replacing \(\omega^2_b\) with \(\omega^2_{b0}(1 \!+\! gU^2_n \!+\! \ldots)\), where \(g > 0\) corresponds to hard-type nonlinearity.

In the experimental demonstration, to construct an RF-resonator, we use varactor diodes along with a \(\text{680 }\mu\mathrm{H}\) inductor.
The resonator is driven by a sweep generator and a linear capacitor, and the resulting resonance curves \updated{are increased} illustrated in Fig.~\ref{fig:nonlinear}~(a).
At low driving amplitudes (black curve), the sweep generates a \updated{symmetric response, relatively.}
However, due to the effective capacitance reduction of the diodes in series to approximately \(\text{400 }\mathrm{pF}\), the \updated{peak} frequency is slightly shifted up.
As we increase the driving amplitude (red curve), the resonance curve shifts towards higher frequencies.
\updated{When we reach the largest driving amplitude \(\text{10 }\mathrm{V}\), a significant bistability window appears, (340 -- 500~{kHz}), displaying hysteresis in the up and down sweeps (yellow and blue) which is a signature of a symmetric hard-type nonlinearity.}

We now present the demonstration of the CLS mode in the diamond lattice in the nonlinear regime.
Fig.~\ref{fig:nonlinear}~(b) shows the CLS at both small (\(v_d=\text{1 }\mathrm{V}\), left panel) and large (\(v_d=\text{7 }\mathrm{V}\), right panel) driving amplitudes.
In the latter case (large driving amplitude), the driving frequency needs to be increased by approximately 10\% (\(\text{560 -- 617 }\mathrm{kHz}\)) due to the nonlinear frequency shift.
However, the spatial structure of the CLS remains largely unaffected.
Notably, the spectral composition of the CLS shows the emergence of harmonics with larger amplitude, indicated in the insets.
This is a clear indication of the impact of nonlinearity on the structure.
At \(v_d=\text{7 }\mathrm{V}\), the third harmonic \(3f_{d}\) becomes prominent, and additional harmonics appear in the spectrum, confirming the nonlinear structure of the response.
It is worth noting that decreasing the amplitude while remaining at \(\text{617 }\mathrm{kHz}\) destroys the CLS.
In the undriven case, it can be theoretically shown that the CLS remains an exact solution in this type of symmetric nonlinearity in the form of Jacobi elliptic function (solution of the duffing oscillator),
\begin{align} \label{eq:duffing}
    \ddot{U}_n &= -\omega^2_{b,0}( 1 + gU_n^2 )(2+2\omega_{r}^2/\omega_{b,0}^2)U_n \propto -(U_{n} + gU_{n}^{3}),\\
    \ddot{V}_n &= -\omega^2_{b,0}(1 + gV_n^2)(2+2\omega_{r}^2/\omega_{b,0}^2)V_n \propto -(V_{n} + gV_{n}^{3}).
\end{align}
This is due to the symmetric nonlinearity, which still allows for the out-of-phase solution \(U_{n} = -V_{n}\), leading to perfect destructive interference at the neighboring bottleneck sites \(T_{n}\) and \(T_{n+1}\), giving zeros.

\subsection{Results on the one-dimensional stub lattice}
\begin{figure}
\centering
    \includegraphics[width = \columnwidth]{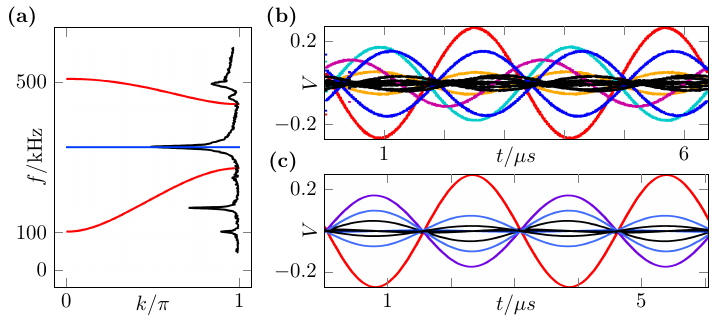}
    \caption[Linear driving on the stub chain]{
        (a) The spectrum of solutions for the stub lattice, according to Eq.~\eqref{eq:stubdisp} --- the red curves indicate dispersive bands and the blue is for the flat band.
        The experimental spectrum is displayed by the black trace along the right vertical axis, obtained by frequency-sweeping the local driver at a CLS site.
        Panel (b) shows the experimental result of local driving at site \(6\), where the trace color assignment and driving location (red arrow) is given in Fig.~\ref{fig:stub_scheme}. 
        Panel (c) displays the corresponding numerical result.
        The colors of the curves in panel (c) pertain to the respective ones used in Fig.~\ref{fig:stub_scheme}.
        The plot of site \(5\) (cyan) is hidden behind the plot of site \(8\) (purple).
        This is also the case for sites \(2\) and \(3\) which are hidden behind the plot of sites \(9\) and \(11\) due to the symmetric geometry. 
    }
    \label{fig:stub_result}
\end{figure}
Theoretically, the arrangement of the stub CLS involves three lattice sites as outlined in Fig.~\ref{fig:stub_scheme}~(a).
However, a local driver cannot exclusively target a single CLS within the lattice.
In this stub chain, two adjacent CLSs share a common site, which makes them non-orthogonal.
For instance, if we introduce a driver at site 6 (indicated in red arrow) as depicted in Fig.~\ref{fig:stub_scheme}~(a), it is expected to induce partial excitation in \updated{both CLSs which shares the site 6}.
\updated{This situation experimentally shown in Fig.~\ref{fig:stub_result}~(b), using a driver frequency of \(\text{312 }\mathrm{kHz}\) with a driving amplitude of \(\text{11 }\mathrm{V}\).}
The similar result is shown in the numerical simulation as shown in Fig.~\ref{fig:stub_result}~(c)
The voltage-time profiles of all 15 sites are presented for two periods.
When site 6 (red) is driven, neighboring CLSs are also excited (depicted in yellow and blue).
This phenomenon arises because sites 5 and 8 are shared with adjacent CLSs, even though site 6 was originally part of a distinct CLS.
It is important to note that slight inhomogeneities result in uneven excitation amplitudes between the two neighboring CLSs.

Analytically, it is possible to show that local driving excites all CLS states with exponential decay of their amplitudes by computing the Green function projected onto the flatband.
In the frequency domain, we have the following form, and \(G(\omega)\) is the Green function,
\begin{gather}
    \left(-\omega^2 + i\beta\omega - \hat{H}\right) G(\omega) = I, \quad\text{where }G(\omega) = \sum_n \frac{\ket{\psi_n}\bra{\psi_n}}{-\omega^2 + i\beta\omega + \omega^2_n}.
\end{gather}
Here, we employ the bra-ket notation for convenience, and \(n\) is the eigenvalue index having eigenvalue \(-\omega^2_n\) and eigenvector \(\ket{\psi_n}\).
The Green function can be divided into flatband and dispersive band cases,
\begin{gather}
    G(\omega) \!=\! \frac{\hat{P}_{\mathrm{FB}}}{-\omega^2 \! + \! i\beta\omega \! + \! \omega^2_{\mathrm{FB}}} + \!\!\!\! \sum_{j \in \mathrm{DB}, k} \!\! \frac{\ket{\psi_j(k)}\bra{\psi_j(k)}}{-\omega^{2} \! + \! i\beta\omega \! + \! \omega^{2}_j(k)} \!=\! G_{\mathrm{FB}}(\omega) \!+\! G_{\mathrm{DB}}(\omega).
\end{gather}
The index \(j \!\in\! \mathrm{DB} \!=\! \{1, 2\}\) is the band index for the dispersive bands.
We separate the frequency responses of flatband and dispersive bands.
\(P_{nm}\) is the flatband projector and \(S_{nm}\) is the overlap matrix,
\begin{gather}
    \hat{P}_{\mathrm{FB}} = \sum_{n,m} S^{-1}_{nm} \ket{\mathrm{CLS}_n}\bra{\mathrm{CLS}_m} \quad\text{and}\quad S_{nm} = \braket{\mathrm{CLS}_n}{\mathrm{CLS}_m}.
\end{gather}
We assume ideal local sinusoidal driving at \(U_n\), then we get
\begin{gather}\label{eq:flatband_response}
    \vert \mathbf{F}(t) \rangle = A\cos(\omega_d t)\ket{U_n} \hspace{0.5em}\text{and}\hspace{0.5em} \vert \mathbf{F}(\omega) \rangle = \frac{A}{2}(\delta(\omega \!-\! \omega_d) + \delta(\omega \!+\! \omega_d))\ket{U_n}.
\end{gather}
Suppose the dispersive bands are sufficiently far from the flatbands compared to the width of resonance peaks.
\(X_{m}\) is a site at \(m\)-th unit cell of \(X\) sublattice, which could be the same as or different from \(U_{n}\).
If we ignore the dispersive term \(G_{\mathrm{DB}}\), which is reasonable when \(\omega_d \approx \omega_{\mathrm{FB}}\), then the spatial profile is given as follows,
\begin{gather}\label{eq:flatband_response_to_local_driving}
    \langle X_m \vert \Psi(\omega_{d}) \rangle \approx \langle {X_m} \vert G_{\mathrm{FB}} \vert {U_n} \rangle \propto \langle {X_m} \vert \hat{P}_{\mathrm{FB}} \vert {U_n} \rangle,
\end{gather}
where \(\vert \Psi(\omega) \rangle = {G}(\omega)\ket{\mathbf{F}(\omega)}\) is the solution of the inhomogeneous coupled ODEs written in Eq.~\eqref{eq:diamond_schrodinger_eq} for the stub case (Eq.~\eqref{eq:stub_undriven}).
When we consider the overlap between nearest neighboring CLSs, \(\sigma\) is nonzero (for instance, see Eq.~\eqref{eq:stub_cls}).
Then, we can write for \(S_{nm}\) in a tridiagonal matrix form,
\begin{gather}\label{eq:overlap_matrix_nearest_neighbor}
    S_{nm} = \langle {\mathrm{CLS}_n} \vert {\mathrm{CLS}_m} \rangle = \delta_{nm} + \sigma \delta_{n \pm 1, m}.
\end{gather}
We can conclude that the overlap of CLSs determines the spatial profile.
If the CLS are orthogonal (\(\sigma = 0\)), \(S\) is a diagonal matrix.
Hence, the projector is compact in real space, as was the case for the diamond lattice. 
On the other hand, when there is an overlap, one can work out the details to obtain the inverse of \(S\).
Using the fact that \(S\) is a tridiagonal matrix with translation-invariance, its inverse can be obtained in the Bloch basis,
\begin{gather}
    S^{-1}_{nm} = -\frac{1}{2\pi}\int^{\pi}_{-\pi} \!\!\!\! dk\frac{\exp(ik|n-m|)}{1 + 2\sigma \cos(k)}.
\end{gather}
Here, we assume that the system size is infinite.
The integration can be solved in a complex plane using Cauchy's integral formula with a substitution, \(\omega = \exp(ik)\).
The integration range becomes a unit circle \(\mathcal{C}\) in a complex plane,
\begin{gather}
    S^{-1}_{nm} = -\frac{1}{2\pi i}\oint_{\mathcal{C}} d\omega \frac{\omega^{|n-m|}}{\sigma\omega^2+ \omega + \sigma} \propto \exp({|n-m|/\xi})
\end{gather}
This leads to the formula for localization length \(\xi\) of the flatband projector \(\hat{P}_{\mathrm{FB}}\) hosting non-orthogonal CLSs and the projector is exponentially localized,
\begin{align}
    \frac{1}{\xi} = \ln \left|\frac{2\sigma}{-1 + \sqrt{1 - 4\sigma^2}}\right|.
\end{align}
For stub, from Eq.~\eqref{eq:stub_cls}, we have \(\sigma = 1/3\) and \(\xi \approx 1.03\). 
This localized resonance mode is qualitatively different from what is expected in resonance modes of dispersive bands with high dissipation since the flatband localization exists even when the dissipation is absent.

%\clearpage
% =========================================================================== %
% =========================================================================== %
% SECTION : Conclusion
% =========================================================================== %
% =========================================================================== %
\section{Conclusion}\label{ch4:sec:conclusion}

In this chapter, we constructed one-dimensional flatband lattices employing complicated electrical circuits and conducted experiments involving local sinusoidal driving to observe resonant modes.
The results align remarkably well with theoretical predictions and numerical simulations.
The investigations confirmed that driving at the flatband frequency effectively triggers a CLS for the diamond lattice.
Conversely, the absence of orthogonality among neighboring CLSs in the stub lattice precluded the direct observation of individual CLSs.
Instead, this leads to resonance modes characterized by exponentially localized spatial profiles.
Lastly, we found the persistence of CLSs in the diamond lattice even when nonlinearity is introduced.
Naturally, this chapter gives rise to several open questions.
For instance, it prompts questions about the feasibility of generating well-defined CLSs in stub lattices, as well as the potential to engineer more intricate quasi-one-dimensional architectures~\cite{morales2016simple} and even two-dimensional structures like the Lieb lattice~\cite{vicencio2015observation}.
In these contexts, generating the linear and nonlinear flatband states remains appealing and challenging for future exploration.

\chapter{Final remarks}\label{chpt5}

%\noindent\textcolor{blue}{Emphasizing the scientific achievement}\\
\section{Scientific contribution and summary}
My main scientific contributions during the Ph.D. program have revolved around several key findings.
I identified the critical-to-insulator transition and discovered fractal edges within flatband systems subjected to quasiperiodic perturbations.
I unveiled the phenomenon of non-ergodic excitation in cross-stitch lattices inhabited by many-body hard-core bosons,
which is closely related to Hilbert space fragmentation.
An orthogonal compact localized state in the one-dimensional diamond lattice using electric circuits has been generated successfully, which opens doors to potential applications in future quantum information.
These achievements constitute significant advancements in our understanding of phase transition and quantum dynamics in perturbed flatband systems and hold promise for various scientific applications.

In chapter~\ref{chpt2}, we summarized the investigation of the impact of quasiperiodic perturbations on the two-bands ABF manifold.
Initially, we focused on the scenario of a weak quasiperiodic perturbation.
In this framework, we identified specific ABF submanifolds that display spectra with critical states and a phenomenon called the critical-to-insulator transition.
For a finite quasiperiodic potential, we observed fractality edges (energy-dependent critical-to-insulator) in the spectrum separating critical from localized states.

In chapter~\ref{chpt3}, we summarized the investigation of the behavior of on-site interacting hard-core bosons in the one- and two-dimensional cross-stitch lattices.
By setting the lowest energy band as a flatband, we found that the groundstate of the system is composed of compact localized states, which significantly influence the overall ground state energy.
Moreover, the interplay between compact localized states and the strong repulsion imposed by hard-core boson constraints highlights the emergence of non-ergodic excitations related to Hilbert space fragmentation in the spin-1 XY model.

In chapter~\ref{chpt4}, we summarized the one-dimensional electric flatband lattices investigation. Both experimental and numerical simulations are discussed when a local sinusoidal driving at the flatband resonance mode was applied to the circuits.
In the diamond chain,the local (linar and nonlinear) driving clearly excites the CLS mode.
In the case of the stub chain, the overlap between the compact localized states does not allow the same phenomena, on the other hand.

\section{Possible future research}
As an extension to Ref.~\cite{lee2023criticalA, lee2023criticalB} and Chapter~\ref{chpt2}, the interesting problem is to identify other quasiperiodic potentials that might give extended states in projected models, for example, by mapping the ABF models for weak perturbation onto the extended Harper model in the ergodic-delocalized part of the phase diagram.
This could be achieved by solving an inverse problem of reconstructing the full potential from the projected effective model.
Moreover, considering an interacting extended Harper model could also be an interesting problem. Then, the correlation of multifractal eigenstates plays an important role, and we might be able to observe novel physical phenomena.
An alternative approach to defining a flatband system involves the presence of a Dirac-delta-shaped density of states.
This criterion allows exploring non-trivial lattice configurations, such as Cayley trees and complete graphs.
I believe investigating the behavior of these systems to perturbations and interactions could offer a new physical phenomenon.

%now enable appendix numbering format and include any appendices
\appendix
%\include{appendix2_revised} % Appendix for Multivalley
%\include{appendix6_revised} % Appendix for Hybird - Graphene
%\include{appendix7_revised2} % Appendix for Hybrid - 2DGE

%next line adds the Bibliography to the contents page
\addcontentsline{toc}{chapter}{Bibliography}

%%uncomment next line to change bibliography name to references
\renewcommand{\bibname}{References}
\bibliographystyle{unsrt}
\bibliography{flatband, frustration, mbl, general, local}        %use a bibtex bibliography file refs.bib

\begin{thebibliography}{100}

\bibitem{anderson1958absence}
P.W. Anderson.
\newblock Absence of diffusion in certain random lattices.
\newblock {\em Phys. Rev.}, 109:1492--1505, March 1958.

\bibitem{doi:10.1142/7663}
Elihu Abrahams.
\newblock {\em 50 Years of Anderson Localization}.
\newblock WORLD SCIENTIFIC, 2010.

\bibitem{ruelle1969remark}
D.~Ruelle.
\newblock A remark on bound states in potential-scattering theory.
\newblock {\em Il Nuovo Cimento A (1965-1970)}, 61:655--662, 1969.

\bibitem{amrein1973characterization}
W~O Amrein and V~Georgescu.
\newblock Characterization of bound states and scattering states in quantum
  mechanics.
\newblock {\em Helv. Phys. Acta, v. 46, no. 5, pp. 635-658}, 1 1973.

\bibitem{enss1978asymptotic}
Volker Enss.
\newblock Asymptotic completeness for quantum mechanical potential scattering.
\newblock {\em Communications in Mathematical Physics}, 61:285--291, 1978.

\bibitem{hunziker2000quantum}
W.~Hunziker and I.~M. Sigal.
\newblock {The quantum N-body problem}.
\newblock {\em Journal of Mathematical Physics}, 41(6):3448--3510, 06 2000.

\bibitem{simon1990absence}
B.~Simon.
\newblock Absence of ballistic motion.
\newblock {\em Comm. Math. Phys}, 134:209--214, 1990.

\bibitem{del1995localization}
R.~del Rio, S.~Jitomirskaya, Y.~Last, and B.~Simon.
\newblock What is localization?
\newblock {\em Phys. Rev. Lett.}, 75:117--119, Jul 1995.

\bibitem{del1996operators}
R.~Del~Rio, S.~Jitomirskaya, Y.~Last, and B.~Simon.
\newblock Operators with singular continuous spectrum, iv. hausdorff
  dimensions, rank one perturbations, and localization.
\newblock {\em Journal d’Analyse Math{\'e}matique}, 69:153--200, 1996.

\bibitem{damanik2001multi}
D.~Damanik and P.~Stollmann.
\newblock Multi-scale analysis implies strong dynamical localization.
\newblock {\em Geometric \& Functional Analysis}, 11:11--29, 2001.

\bibitem{sutherland1986localization}
Bill Sutherland.
\newblock Localization of electronic wave functions due to local topology.
\newblock {\em Phys. Rev. B}, 34:5208--5211, October 1986.

\bibitem{lieb1989two}
Elliott~H. Lieb.
\newblock Two theorems on the hubbard model.
\newblock {\em Phys. Rev. Lett.}, 62:1201--1204, March 1989.

\bibitem{mielke1992exactgs}
A~Mielke.
\newblock Exact ground states for the hubbard model on the kagome lattice.
\newblock {\em Journal of Physics A: Mathematical and General},
  25(16):4335--4345, August 1992.

\bibitem{creutz1999end}
Michael Creutz.
\newblock End states, ladder compounds, and domain-wall fermions.
\newblock {\em Phys. Rev. Lett.}, 83:2636--2639, September 1999.

\bibitem{santos2004atomic}
L.~Santos, M.A. Baranov, J.I. Cirac, H.-U. Everts, H.~Fehrmann, and
  M.~Lewenstein.
\newblock Atomic quantum gases in kagom\'e lattices.
\newblock {\em Phys. Rev. Lett.}, 93:030601, July 2004.

\bibitem{wang2019highly}
Hanyu Wang, Biao Yang, Wei Xu, Yuancheng Fan, Qinghua Guo, Zhihong Zhu, and
  C.~T. Chan.
\newblock Highly degenerate photonic flat bands arising from complete graph
  configurations.
\newblock {\em Phys. Rev. A}, 100:043841, Oct 2019.

\bibitem{aoki1996hofstadter}
Hideo Aoki, Masato Ando, and Hajime Matsumura.
\newblock Hofstadter butterflies for flat bands.
\newblock {\em Phys. Rev. B}, 54:R17296--R17299, December 1996.

\bibitem{landau1930diamagnetismus}
L.~D. Landau.
\newblock Diamagnetismus der metalle.
\newblock {\em Zeitschrift f{\"u}r Physik}, 64:629--637, Sep 1930.

\bibitem{laia2011holographic}
Joao~N Laia and David Tong.
\newblock A holographic flat band.
\newblock {\em Journal of High Energy Physics}, 2011(11):1--19, Nov 2011.

\bibitem{rontgen2019quantum}
M.~R\"ontgen, C.V. Morfonios, I.~Brouzos, F.K. Diakonos, and P.~Schmelcher.
\newblock Quantum network transfer and storage with compact localized states
  induced by local symmetries.
\newblock {\em Phys. Rev. Lett.}, 123:080504, August 2019.

\bibitem{hsu2016bound}
Chia~Wei Hsu, Bo~Zhen, A.~Douglas Stone, John~D. Joannopoulos, and Marin
  Soljacic.
\newblock Bound states in the continuum.
\newblock {\em Nat. Rev. Mat.}, 1(2):16048, July 2016.

\bibitem{plotnik2011experimental}
Yonatan Plotnik, Or~Peleg, Felix Dreisow, Matthias Heinrich, Stefan Nolte,
  Alexander Szameit, and Mordechai Segev.
\newblock Experimental observation of optical bound states in the continuum.
\newblock {\em Phys. Rev. Lett.}, 107:183901, Oct 2011.

\bibitem{vicencio2013diffraction}
Rodrigo~A Vicencio and Cristian Mejía-Cortés.
\newblock Diffraction-free image transmission in kagome photonic lattices.
\newblock {\em Journal of Optics}, 16(1):015706, nov 2013.

\bibitem{xia2016demonstration}
Shiqiang Xia, Yi~Hu, Daohong Song, Yuanyuan Zong, Liqin Tang, and Zhigang Chen.
\newblock Demonstration of flat-band image transmission in optically induced
  lieb photonic lattices.
\newblock {\em Opt. Lett.}, 41(7):1435--1438, April 2016.

\bibitem{rojas-rojas2017quantum}
S.~Rojas-Rojas, L.~Morales-Inostroza, R.A. Vicencio, and A.~Delgado.
\newblock Quantum localized states in photonic flat-band lattices.
\newblock {\em Phys. Rev. A}, 96:043803, October 2017.

\bibitem{leykam2018perspective}
Daniel Leykam and Sergej Flach.
\newblock Perspective: Photonic flatbands.
\newblock {\em APL Photonics}, 3(7):070901, 2018.

\bibitem{vicencio2021photonic}
Rodrigo A.~Vicencio Poblete.
\newblock Photonic flat band dynamics.
\newblock {\em Advances in Physics: X}, 6(1):1878057, 2021.

\bibitem{maimaiti2017compact}
Wulayimu Maimaiti, Alexei Andreanov, Hee~Chul Park, Oleg Gendelman, and Sergej
  Flach.
\newblock Compact localized states and flat-band generators in one dimension.
\newblock {\em Phys. Rev. B}, 95(11):115135, March 2017.

\bibitem{maimaiti2019universal}
Wulayimu Maimaiti, Sergej Flach, and Alexei Andreanov.
\newblock Universal $d=1$ flat band generator from compact localized states.
\newblock {\em Phys. Rev. B}, 99:125129, March 2019.

\bibitem{maimaiti2021flat}
Wulayimu Maimaiti, Alexei Andreanov, and Sergej Flach.
\newblock Flat-band generator in two dimensions.
\newblock {\em Phys. Rev. B}, 103:165116, April 2021.

\bibitem{rhim2019classification}
Jun-Won Rhim and Bohm-Jung Yang.
\newblock Classification of flat bands according to the band-crossing
  singularity of bloch wave functions.
\newblock {\em Phys. Rev. B}, 99:045107, January 2019.

\bibitem{hwang2021flat}
Yoonseok Hwang, Jun-Won Rhim, and Bohm-Jung Yang.
\newblock Flat bands with band crossings enforced by symmetry representation.
\newblock {\em Phys. Rev. B}, 104:L081104, Aug 2021.

\bibitem{hwang2021general}
Yoonseok Hwang, Jun-Won Rhim, and Bohm-Jung Yang.
\newblock General construction of flat bands with and without band crossings
  based on wave function singularity.
\newblock {\em Phys. Rev. B}, 104:085144, Aug 2021.

\bibitem{graf2021designing}
Ansgar Graf and Fr\'ed\'eric Pi\'echon.
\newblock Designing flat-band tight-binding models with tunable multifold band
  touching points.
\newblock {\em Phys. Rev. B}, 104:195128, Nov 2021.

\bibitem{rontgen2018compact}
M.~R\"ontgen, C.V. Morfonios, and P.~Schmelcher.
\newblock Compact localized states and flat bands from local symmetry
  partitioning.
\newblock {\em Phys. Rev. B}, 97:035161, January 2018.

\bibitem{morfonios2021flat}
C.V. Morfonios, M.~R\"ontgen, M.~Pyzh, and P.~Schmelcher.
\newblock Flat bands by latent symmetry.
\newblock {\em Phys. Rev. B}, 104:035105, July 2021.

\bibitem{rontgen2021latent}
M.~R\"ontgen, M.~Pyzh, C.V. Morfonios, N.E. Palaiodimopoulos, F.K. Diakonos,
  and P.~Schmelcher.
\newblock Latent symmetry induced degeneracies.
\newblock {\em Phys. Rev. Lett.}, 126:180601, May 2021.

\bibitem{smith2019hidden}
Dallas Smith and Benjamin Webb.
\newblock Hidden symmetries in real and theoretical networks.
\newblock {\em Physica A: Statistical Mechanics and its Applications},
  514:855--867, 2019.

\bibitem{chern2015pt}
Gia-Wei Chern and Avadh Saxena.
\newblock $\mathcal {PT}$-symmetric phase in kagome-based photonic lattices.
\newblock {\em Opt. Lett.}, 40(24):5806--5809, December 2015.

\bibitem{qi2018defect}
Bingkun Qi, Lingxuan Zhang, and Li~Ge.
\newblock Defect states emerging from a non-hermitian flatband of photonic zero
  modes.
\newblock {\em Phys. Rev. Lett.}, 120:093901, February 2018.

\bibitem{zyuzin2018flat}
A.A. Zyuzin and A.Yu. Zyuzin.
\newblock Flat band in disorder-driven non-hermitian weyl semimetals.
\newblock {\em Phys. Rev. B}, 97:041203(R), January 2018.

\bibitem{leykam2017flat}
Daniel Leykam, Sergej Flach, and Y.D. Chong.
\newblock Flat bands in lattices with non-hermitian coupling.
\newblock {\em Phys. Rev. B}, 96:064305, August 2017.

\bibitem{ramezani2017non}
Hamidreza Ramezani.
\newblock Non-hermiticity-induced flat band.
\newblock {\em Phys. Rev. A}, 96:011802(R), July 2017.

\bibitem{tobias2019experimental}
Tobias Biesenthal, Mark Kremer, Matthias Heinrich, and Alexander Szameit.
\newblock Experimental realization of $\mathcal {P}\mathcal {T}$-symmetric flat
  bands.
\newblock {\em Phys. Rev. Lett.}, 123:183601, October 2019.

\bibitem{zhang2020nonhermitian}
S.M. Zhang and L.~Jin.
\newblock Non-hermitian aharonov-bohm cage, 2020.

\bibitem{zhang2020flatband}
S.M. Zhang and L.~Jin.
\newblock Flat band in two-dimensional non-hermitian optical lattices.
\newblock {\em Phys. Rev. A}, 100:043808, October 2019.

\bibitem{talkington2022dissipation}
Spenser Talkington and Martin Claassen.
\newblock Dissipation-induced flat bands.
\newblock {\em Phys. Rev. B}, 106:L161109, Oct 2022.

\bibitem{maimaiti2021nonhermitian}
Wulayimu Maimaiti and Alexei Andreanov.
\newblock Non-hermitian flat-band generator in one dimension.
\newblock {\em Phys. Rev. B}, 104:035115, Jul 2021.

\bibitem{leykam2018artificial}
Daniel Leykam, Alexei Andreanov, and Sergej Flach.
\newblock Artificial flat band systems: from lattice models to experiments.
\newblock {\em Adv. Phys.: X}, 3(1):1473052, 2018.

\bibitem{vidal1998aharonov}
Julien Vidal, R\'emy Mosseri, and Benoit Dou\ifmmode~\mbox{\c{c}}\else
  \c{c}\fi{}ot.
\newblock {Aharonov-Bohm} cages in two-dimensional structures.
\newblock {\em Phys. Rev. Lett.}, 81:5888--5891, December 1998.

\bibitem{aharonov1959significance}
Y.~Aharonov and D.~Bohm.
\newblock Significance of electromagnetic potentials in the quantum theory.
\newblock {\em Phys. Rev.}, 115:485--491, Aug 1959.

\bibitem{abilio1999magnetic}
C.C. Abilio, P.~Butaud, Th. Fournier, B.~Pannetier, J.~Vidal, S.~Tedesco, and
  B.~Dalzotto.
\newblock Magnetic field induced localization in a two-dimensional
  superconducting wire network.
\newblock {\em Phys. Rev. Lett.}, 83:5102--5105, December 1999.

\bibitem{naud2001aharonov}
C\'ecile Naud, Giancarlo Faini, and Dominique Mailly.
\newblock {Aharonov-Bohm} cages in 2d normal metal networks.
\newblock {\em Phys. Rev. Lett.}, 86:5104--5107, May 2001.

\bibitem{tadjine2016from}
Athmane Tadjine, Guy Allan, and Christophe Delerue.
\newblock From lattice hamiltonians to tunable band structures by lithographic
  design.
\newblock {\em Phys. Rev. B}, 94:075441, August 2016.

\bibitem{qiu2016designing}
Wen-Xuan Qiu, Shuai Li, Jin-Hua Gao, Yi~Zhou, and Fu-Chun Zhang.
\newblock Designing an artificial lieb lattice on a metal surface.
\newblock {\em Phys. Rev. B}, 94:241409, December 2016.

\bibitem{drost2017topological}
Robert Drost, Teemu Ojanen, Ari Harju, and Peter Liljeroth.
\newblock Topological states in engineered atomic lattices.
\newblock {\em Nat. Phys.}, 13:668 EP --, March 2017.

\bibitem{slot2017experimental}
Marlou~R. Slot, Thomas~S. Gardenier, Peter~H. Jacobse, Guido~C.P. van Miert,
  Sander~N. Kempkes, Stephan~J.M. Zevenhuizen, Cristiane~Morais Smith, Daniel
  Vanmaekelbergh, and Ingmar Swart.
\newblock Experimental realization and characterization of an electronic lieb
  lattice.
\newblock {\em Nat. Phys.}, 13:672 EP --, April 2017.

\bibitem{bercioux2009massless}
D.~Bercioux, D.F. Urban, H.~Grabert, and W.~H\"ausler.
\newblock Massless dirac-weyl fermions in a ${\mathcal {t}}\_{3}$ optical
  lattice.
\newblock {\em Phys. Rev. A}, 80:063603, December 2009.

\bibitem{urban2011barrier}
Daniel~F. Urban, Dario Bercioux, Michael Wimmer, and Wolfgang H\"ausler.
\newblock Barrier transmission of dirac-like pseudospin-one particles.
\newblock {\em Phys. Rev. B}, 84:115136, Sep 2011.

\bibitem{ruostekoski2009optical}
J.~Ruostekoski.
\newblock Optical kagome lattice for ultracold atoms with nearest neighbor
  interactions.
\newblock {\em Phys. Rev. Lett.}, 103:080406, August 2009.

\bibitem{shen2010single}
R.~Shen, L.B. Shao, Baigeng Wang, and D.Y. Xing.
\newblock Single dirac cone with a flat band touching on line-centered-square
  optical lattices.
\newblock {\em Phys. Rev. B}, 81:041410, January 2010.

\bibitem{apaja2010flat}
V.~Apaja, M.~Hyrk\"as, and M.~Manninen.
\newblock Flat bands, {Dirac} cones, and atom dynamics in an optical lattice.
\newblock {\em Phys. Rev. A}, 82:041402, October 2010.

\bibitem{taie2015coherent}
Shintaro Taie, Hideki Ozawa, Tomohiro Ichinose, Takuei Nishio, Shuta Nakajima,
  and Yoshiro Takahashi.
\newblock Coherent driving and freezing of bosonic matter wave in an optical
  lieb lattice.
\newblock {\em Science Advances}, 1(10):e1500854, 2015.

\bibitem{ozawa2017interaction}
Hideki Ozawa, Shintaro Taie, Tomohiro Ichinose, and Yoshiro Takahashi.
\newblock Interaction-driven shift and distortion of a flat band in an optical
  {Lieb} lattice.
\newblock {\em Phys. Rev. Lett.}, 118:175301, April 2017.

\bibitem{baba2008slow}
Toshihiko Baba.
\newblock Slow light in photonic crystals.
\newblock {\em Nat. Phot.}, 2:465 EP --, August 2008.

\bibitem{letartre2001group}
X.~Letartre, C.~Seassal, C.~Grillet, P.~Rojo-Romeo, P.~Viktorovitch,
  M.~Le~Vassor~d’Yerville, D.~Cassagne, and C.~Jouanin.
\newblock {Group velocity and propagation losses measurement in a single-line
  photonic-crystal waveguide on InP membranes}.
\newblock {\em Applied Physics Letters}, 79(15):2312--2314, 10 2001.

\bibitem{notomi2001extremely}
M.~Notomi, K.~Yamada, A.~Shinya, J.~Takahashi, C.~Takahashi, and I.~Yokohama.
\newblock Extremely large group-velocity dispersion of line-defect waveguides
  in photonic crystal slabs.
\newblock {\em Phys. Rev. Lett.}, 87:253902, Nov 2001.

\bibitem{takeda2004flat}
Hiroyuki Takeda, Tetsuya Takashima, and Katsumi Yoshino.
\newblock Flat photonic bands in two-dimensional photonic crystals with kagome
  lattices.
\newblock {\em J Phys.: Cond. Mat.}, 16(34):6317, 2004.

\bibitem{schulz2017photonic}
Sebastian~A. Schulz, Jeremy Upham, Liam O'Faolain, and Robert~W. Boyd.
\newblock Photonic crystal slow light waveguides in a kagome lattice.
\newblock {\em Opt. Lett.}, 42(16):3243--3246, August 2017.

\bibitem{li2008systematic}
Juntao Li, Thomas~P. White, Liam O'Faolain, Alvaro Gomez-Iglesias, and
  Thomas~F. Krauss.
\newblock Systematic design of flat band slow light in photonic crystal
  waveguides.
\newblock {\em Opt. Express}, 16(9):6227--6232, Apr 2008.

\bibitem{xu2015design}
Changqing Xu, Gang Wang, Zhi~Hong Hang, Jie Luo, C~T Chan, and Yun Lai.
\newblock Design of full-k-space flat bands in photonic crystals beyond the
  tight-binding picture.
\newblock {\em Sci. Rep.}, 5:18181, 2015.

\bibitem{endo2010tight}
Shimpei Endo, Takashi Oka, and Hideo Aoki.
\newblock Tight-binding photonic bands in metallophotonic waveguide networks
  and flat bands in kagome lattices.
\newblock {\em Phys. Rev. B}, 81:113104, March 2010.

\bibitem{feigenbaum2010resonant}
Eyal Feigenbaum and Harry~A. Atwater.
\newblock Resonant guided wave networks.
\newblock {\em Phys. Rev. Lett.}, 104:147402, April 2010.

\bibitem{nakata2012observation}
Yosuke Nakata, Takanori Okada, Toshihiro Nakanishi, and Masao Kitano.
\newblock Observation of flat band for terahertz spoof plasmons in a metallic
  kagom\'e lattice.
\newblock {\em Phys. Rev. B}, 85:205128, May 2012.

\bibitem{nixon2013observing}
Micha Nixon, Eitan Ronen, Asher~A. Friesem, and Nir Davidson.
\newblock Observing geometric frustration with thousands of coupled lasers.
\newblock {\em Phys. Rev. Lett.}, 110:184102, May 2013.

\bibitem{kajiwara2016observation}
Sho Kajiwara, Yoshiro Urade, Yosuke Nakata, Toshihiro Nakanishi, and Masao
  Kitano.
\newblock Observation of a nonradiative flat band for spoof surface plasmons in
  a metallic {Lieb} lattice.
\newblock {\em Phys. Rev. B}, 93:075126, February 2016.

\bibitem{szameit2010discrete}
Alexander Szameit and Stefan Nolte.
\newblock Discrete optics in femtosecond-laser-written photonic structures.
\newblock {\em J Phys. B: At. Mol. Opt. Phys.}, 43(16):163001, 2010.

\bibitem{guzman2014experimental}
D~Guzm{\'a}n-Silva, C~Mej{\'\i}a-Cort{\'e}s, M~A Bandres, M~C Rechtsman,
  S~Weimann, S~Nolte, M~Segev, A~Szameit, and R~A Vicencio.
\newblock Experimental observation of bulk and edge transport in photonic
  {Lieb} lattices.
\newblock {\em New J. Phys.}, 16(6):063061, 2014.

\bibitem{mukherjee2015observation}
Sebabrata Mukherjee, Alexander Spracklen, Debaditya Choudhury, Nathan Goldman,
  Patrik \"Ohberg, Erika Andersson, and Robert~R. Thomson.
\newblock Observation of a localized flat-band state in a photonic {Lieb}
  lattice.
\newblock {\em Phys. Rev. Lett.}, 114:245504, June 2015.

\bibitem{vicencio2015observation}
Rodrigo~A. Vicencio, Camilo Cantillano, Luis Morales-Inostroza, Basti\'an Real,
  Cristian Mej\'{\i}a-Cort\'es, Steffen Weimann, Alexander Szameit, and
  Mario~I. Molina.
\newblock Observation of localized states in {Lieb} photonic lattices.
\newblock {\em Phys. Rev. Lett.}, 114:245503, June 2015.

\bibitem{schneider2016exciton}
C.~Schneider, K.~Winkler, M.~D. Fraser, M.~Kamp, Y.~Yamamoto, E.~A.
  Ostrovskaya, and S.~Höfling.
\newblock Exciton-polariton trapping and potential landscape engineering.
\newblock {\em Reports on Progress in Physics}, 80(1):016503, nov 2016.

\bibitem{jacqmin2014direct}
T.~Jacqmin, I.~Carusotto, I.~Sagnes, M.~Abbarchi, D.D. Solnyshkov, G.~Malpuech,
  E.~Galopin, A.~Lema\^{\i}tre, J.~Bloch, and A.~Amo.
\newblock Direct observation of dirac cones and a flatband in a honeycomb
  lattice for polaritons.
\newblock {\em Phys. Rev. Lett.}, 112:116402, March 2014.

\bibitem{baboux2016bosonic}
F.~Baboux, L.~Ge, T.~Jacqmin, M.~Biondi, E.~Galopin, A.~Lema\^{\i}tre,
  L.~Le~Gratiet, I.~Sagnes, S.~Schmidt, H.E. T\"ureci, A.~Amo, and J.~Bloch.
\newblock Bosonic condensation and disorder-induced localization in a flat
  band.
\newblock {\em Phys. Rev. Lett.}, 116:066402, February 2016.

\bibitem{klembt2017polariton}
S.~Klembt, T.H. Harder, O.A. Egorov, K.~Winkler, H.~Suchomel, J.~Beierlein,
  M.~Emmerling, C.~Schneider, and S.~H\"ofling.
\newblock Polariton condensation in {S}- and {P}-flatbands in a two-dimensional
  {Lieb} lattice.
\newblock {\em Applied Phys. Lett.}, 111(23):231102, 2017.

\bibitem{whittaker2018exciton}
C.~E. Whittaker, E.~Cancellieri, P.~M. Walker, D.~R. Gulevich, H.~Schomerus,
  D.~Vaitiekus, B.~Royall, D.~M. Whittaker, E.~Clarke, I.~V. Iorsh, I.~A.
  Shelykh, M.~S. Skolnick, and D.~N. Krizhanovskii.
\newblock Exciton polaritons in a two-dimensional lieb lattice with spin-orbit
  coupling.
\newblock {\em Phys. Rev. Lett.}, 120:097401, Mar 2018.

\bibitem{derzhko2010low}
O.~Derzhko, J.~Richter, A.~Honecker, M.~Maksymenko, and R.~Moessner.
\newblock Low-temperature properties of the hubbard model on highly frustrated
  one-dimensional lattices.
\newblock {\em Phys. Rev. B}, 81:014421, January 2010.

\bibitem{nita2013spectral}
M.~Ni\ifmmode \mbox{\c{t}}\else \c{t}\fi{}\ifmmode~\u{a}\else \u{a}\fi{},
  B.~Ostahie, and A.~Aldea.
\newblock Spectral and transport properties of the two-dimensional lieb
  lattice.
\newblock {\em Phys. Rev. B}, 87:125428, March 2013.

\bibitem{flach2014detangling}
Sergej Flach, Daniel Leykam, Joshua~D. Bodyfelt, Peter Matthies, and Anton~S.
  Desyatnikov.
\newblock Detangling flat bands into {Fano} lattices.
\newblock {\em Europhys. Lett.}, 105(3):30001, 2014.

\bibitem{goda2006inverse}
Masaki Goda, Shinya Nishino, and Hiroki Matsuda.
\newblock Inverse anderson transition caused by flatbands.
\newblock {\em Phys. Rev. Lett.}, 96:126401, March 2006.

\bibitem{nishino2007flat}
Shinya Nishino, Hiroki Matsuda, and Masaki Goda.
\newblock Flat-band localization in weakly disordered system.
\newblock {\em J Phys. Soc. Jap.}, 76(2):024709, 2007.

\bibitem{cadez2021metal}
Tilen \ifmmode \check{C}\else \v{C}\fi{}ade\ifmmode~\check{z}\else \v{z}\fi{},
  Yeongjun Kim, Alexei Andreanov, and Sergej Flach.
\newblock Metal-insulator transition in infinitesimally weakly disordered flat
  bands.
\newblock {\em Phys. Rev. B}, 104:L180201, Nov 2021.

\bibitem{longhi2021inverse}
Stefano Longhi.
\newblock Inverse anderson transition in photonic cages.
\newblock {\em Opt. Lett.}, 46(12):2872--2875, Jun 2021.

\bibitem{li2022aharonov}
Hang Li, Zhaoli Dong, Stefano Longhi, Qian Liang, Dizhou Xie, and Bo~Yan.
\newblock Aharonov-bohm caging and inverse anderson transition in ultracold
  atoms.
\newblock {\em Phys. Rev. Lett.}, 129:220403, Nov 2022.

\bibitem{tasaki1992ferromagnetism}
Hal Tasaki.
\newblock Ferromagnetism in the hubbard models with degenerate single-electron
  ground states.
\newblock {\em Phys. Rev. Lett.}, 69:1608--1611, September 1992.

\bibitem{mielke1999ferromagnetism}
Andreas Mielke.
\newblock Ferromagnetism in single-band hubbard models with a partially flat
  band.
\newblock {\em Phys. Rev. Lett.}, 82:4312--4315, May 1999.

\bibitem{ramirez1994strongly}
A~P Ramirez.
\newblock Strongly geometrically frustrated magnets.
\newblock {\em Ann. Rev. Mat. Sci.}, 24(1):453--480, 1994.

\bibitem{derzhko2015strongly}
Oleg Derzhko, Johannes Richter, and Mykola Maksymenko.
\newblock Strongly correlated flat-band systems: The route from heisenberg
  spins to hubbard electrons.
\newblock {\em Int. J. Mod. Phys. B}, 29(12):1530007, 2015.

\bibitem{kuno2020flat}
Yoshihito Kuno, Takahiro Orito, and Ikuo Ichinose.
\newblock Flat-band many-body localization and ergodicity breaking in the
  {Creutz} ladder.
\newblock {\em New J. Phys.}, 22(1):013032, January 2020.

\bibitem{danieli2020many}
Carlo Danieli, Alexei Andreanov, and Sergej Flach.
\newblock Many-body flatband localization.
\newblock {\em Phys. Rev. B}, 102:041116, Jul 2020.

\bibitem{vakulchyk2021heat}
Ihor Vakulchyk, Carlo Danieli, Alexei Andreanov, and Sergej Flach.
\newblock Heat percolation in many-body flat-band localizing systems.
\newblock {\em Phys. Rev. B}, 104:144207, Oct 2021.

\bibitem{orito2021nonthermalized}
Takahiro Orito, Yoshihito Kuno, and Ikuo Ichinose.
\newblock Nonthermalized dynamics of flat-band many-body localization.
\newblock {\em Phys. Rev. B}, 103:L060301, February 2021.

\bibitem{danieli2022many}
Carlo Danieli, Alexei Andreanov, and Sergej Flach.
\newblock Many-body localization transition from flat-band fine tuning.
\newblock {\em Phys. Rev. B}, 105:L041113, Jan 2022.

\bibitem{tovmasyan2013geometry}
Murad Tovmasyan, Evert~P.L. van Nieuwenburg, and Sebastian~D. Huber.
\newblock Geometry-induced pair condensation.
\newblock {\em Phys. Rev. B}, 88:220510, December 2013.

\bibitem{tovmasyan2018preformed}
Murad Tovmasyan, Sebastiano Peotta, Long Liang, P\"aivi T\"orm\"a, and
  Sebastian~D. Huber.
\newblock Preformed pairs in flat {Bloch} bands.
\newblock {\em Phys. Rev. B}, 98:134513, October 2018.

\bibitem{danieli2021nonlinear}
Carlo Danieli, Alexei Andreanov, Thudiyangal Mithun, and Sergej Flach.
\newblock Nonlinear caging in all-bands-flat lattices.
\newblock {\em Phys. Rev. B}, 104:085131, August 2021.

\bibitem{danieli2021quantum}
Carlo Danieli, Alexei Andreanov, Thudiyangal Mithun, and Sergej Flach.
\newblock Quantum caging in interacting many-body all-bands-flat lattices.
\newblock {\em Phys. Rev. B}, 104:085132, August 2021.

\bibitem{doucot2002pairing}
Benoit Dou\ifmmode~\mbox{\c{c}}\else \c{c}\fi{}ot and Julien Vidal.
\newblock Pairing of cooper pairs in a fully frustrated josephson-junction
  chain.
\newblock {\em Phys. Rev. Lett.}, 88:227005, May 2002.

\bibitem{takayoshi2013phase}
Shintaro Takayoshi, Hosho Katsura, Noriaki Watanabe, and Hideo Aoki.
\newblock {Phase diagram and pair Tomonaga-Luttinger liquid in a Bose-Hubbard
  model with flat bands}.
\newblock {\em Phys. Rev. A}, 88:063613, December 2013.

\bibitem{phillips2015low}
L.~G. Phillips, G.~De~Chiara, P.~\"Ohberg, and M.~Valiente.
\newblock Low-energy behavior of strongly interacting bosons on a flat-band
  lattice above the critical filling factor.
\newblock {\em Phys. Rev. B}, 91:054103, February 2015.

\bibitem{mielke2018pair}
Andreas Mielke.
\newblock Pair formation of hard core bosons in flat band systems.
\newblock {\em J. Stat. Phys.}, 171(4):679--695, May 2018.

\bibitem{peotta2015superfluidity}
Sebastiano Peotta and P{\"{a}}ivi T{\"{o}}rm{\"{a}}.
\newblock Superfluidity in topologically nontrivial flat bands.
\newblock {\em Nat. Comm.}, 6:8944, November 2015.

\bibitem{julku2016geometric}
Aleksi Julku, Sebastiano Peotta, Tuomas~I. Vanhala, Dong-Hee Kim, and P\"aivi
  T\"orm\"a.
\newblock Geometric origin of superfluidity in the lieb-lattice flat band.
\newblock {\em Phys. Rev. Lett.}, 117:045303, July 2016.

\bibitem{tovmasyan2016effective}
Murad Tovmasyan, Sebastiano Peotta, P\"aivi T\"orm\"a, and Sebastian~D. Huber.
\newblock Effective theory and emergent $\text {SU}(2)$ symmetry in the flat
  bands of attractive hubbard models.
\newblock {\em Phys. Rev. B}, 94:245149, December 2016.

\bibitem{liang2017band}
Long Liang, Tuomas~I. Vanhala, Sebastiano Peotta, Topi Siro, Ari Harju, and
  P\"aivi T\"orm\"a.
\newblock Band geometry, berry curvature, and superfluid weight.
\newblock {\em Phys. Rev. B}, 95:024515, January 2017.

\bibitem{cao2018unconventional}
Yuan Cao, Valla Fatemi, Shiang Fang, Kenji Watanabe, Takashi Taniguchi,
  Efthimios Kaxiras, and Pablo Jarillo-Herrero.
\newblock Unconventional superconductivity in magic-angle graphene
  superlattices.
\newblock {\em Nature}, 556:43--50, 2018.

\bibitem{chan2022pairing}
Si~Min Chan, B.~Gr\'emaud, and G.~G. Batrouni.
\newblock Pairing and superconductivity in quasi-one-dimensional flat-band
  systems: Creutz and sawtooth lattices.
\newblock {\em Phys. Rev. B}, 105:024502, Jan 2022.

\bibitem{chan2022designer}
Si~Min Chan, B.~Gr\'emaud, and G.~G. Batrouni.
\newblock Designer flat bands: Topology and enhancement of superconductivity.
\newblock {\em Phys. Rev. B}, 106:104514, Sep 2022.

\bibitem{chan2023superconductivity}
Si~Min Chan, Alexei Andreanov, Sergej Flach, and G.~George Batrouni.
\newblock Superconductivity with wannier-stark flat bands, 2023.

\bibitem{lee2023criticalA}
Sanghoon Lee, Alexei Andreanov, and Sergej Flach.
\newblock Critical-to-insulator transitions and fractality edges in perturbed
  flat bands.
\newblock {\em Phys. Rev. B}, 107:014204, Jan 2023.

\bibitem{lee2023criticalB}
S.~Lee, S.~Flach, and Alexei Andreanov.
\newblock {Critical state generators from perturbed flatbands}.
\newblock {\em Chaos: An Interdisciplinary Journal of Nonlinear Science},
  33(7):073125, 07 2023.

\bibitem{chase2023compact}
Carys Chase-Mayoral, LQ~English, Yeongjun Kim, Sanghoon Lee, Noah Lape, Alexei
  Andreanov, PG~Kevrekidis, and Sergej Flach.
\newblock Compact localized states in electric circuit flatband lattices.
\newblock {\em arXiv preprint arXiv:2307.15319}, 2023.

\bibitem{kramer1993localization}
B.~Kramer and A.~MacKinnon.
\newblock Localization: theory and experiment.
\newblock {\em Rep. Prog. Phys.}, 56(12):1469--1564, December 1993.

\bibitem{aubry1980analyticity}
Serge Aubry and Gilles Andr{\'e}.
\newblock Analyticity breaking and {Anderson} localization in incommensurate
  lattices.
\newblock {\em Ann. Israel Phys. Soc}, 3(133):18, 1980.

\bibitem{avila2017spectral}
A.~Avila, S.~Jitomirskaya, and C.~A. Marx.
\newblock Spectral theory of extended harper’s model and a question by
  {Erdős} and {Szekeres}.
\newblock {\em Inventiones mathematicae}, 210:283--339, 2017.

\bibitem{xiao2021observation}
Teng Xiao, Dizhou Xie, Zhaoli Dong, Tao Chen, Wei Yi, and Bo~Yan.
\newblock Observation of topological phase with critical localization in a
  quasi-periodic lattice.
\newblock {\em Science Bulletin}, 66(21):2175--2180, 2021.

\bibitem{liu2022anomalous}
Tong Liu, Xu~Xia, Stefano Longhi, and Laurent Sanchez-Palencia.
\newblock {Anomalous mobility edges in one-dimensional quasiperiodic models}.
\newblock {\em SciPost Phys.}, 12:027, 2022.

\bibitem{zhang2022lyapunov}
Yi-Cai Zhang and Yan-Yang Zhang.
\newblock Lyapunov exponent, mobility edges, and critical region in the
  generalized aubry-andr\'e model with an unbounded quasiperiodic potential.
\newblock {\em Phys. Rev. B}, 105:174206, May 2022.

\bibitem{wang2022quantum}
Yucheng Wang, Long Zhang, Wei Sun, Ting-Fung~Jeffrey Poon, and Xiong-Jun Liu.
\newblock Quantum phase with coexisting localized, extended, and critical
  zones.
\newblock {\em Phys. Rev. B}, 106:L140203, Oct 2022.

\bibitem{shimasaki2022anomalous}
Toshihiko Shimasaki, Max Prichard, H.~Esat Kondakci, Jared Pagett, Yifei Bai,
  Peter Dotti, Alec Cao, Tsung-Cheng Lu, Tarun Grover, and David~M. Weld.
\newblock Anomalous localization and multifractality in a kicked quasicrystal,
  2022.

\bibitem{lin2023general}
Xiaoshui Lin, Xiaoman Chen, Guang-Can Guo, and Ming Gong.
\newblock The general approach to the critical phase with coupled quasiperiodic
  chains, 2023.

\bibitem{evers2008anderson}
Ferdinand Evers and Alexander~D. Mirlin.
\newblock Anderson transitions.
\newblock {\em Rev. Mod. Phys.}, 80:1355--1417, October 2008.

\bibitem{vicsek1992fractal}
Tamas Vicsek.
\newblock {\em Fractal growth phenomena}.
\newblock World scientific, 1992.

\bibitem{steel1960principles}
Robert George~Douglas Steel and James~Hiram Torrie.
\newblock {\em Principles and procedures of statistics}.
\newblock McGraw-Hill Book Company, Inc., New York, Toronto, London, 1960.

\bibitem{teschl2000jacobi}
G.~Teschl.
\newblock {\em Jacobi operators and completely integrable nonlinear lattices}.
\newblock Mathematical surveys and monographs, v. 72. American Mathematical
  Society, Providence, RI, 2000.

\bibitem{chang1997multifractal}
I.~Chang, K.~Ikezawa, and M.~Kohmoto.
\newblock Multifractal properties of the wave functions of the square-lattice
  tight-binding model with next-nearest-neighbor hopping in a magnetic field.
\newblock {\em Phys. Rev. B}, 55:12971--12975, May 1997.

\bibitem{han1994critical}
J.~H. Han, D.~J. Thouless, H.~Hiramoto, and M.~Kohmoto.
\newblock Critical and bicritical properties of harper's equation with
  next-nearest-neighbor coupling.
\newblock {\em Phys. Rev. B}, 50:11365--11380, Oct 1994.

\bibitem{kraus2012topological}
Yaacov~E. Kraus and Oded Zilberberg.
\newblock Topological equivalence between the fibonacci quasicrystal and the
  harper model.
\newblock {\em Phys. Rev. Lett.}, 109:116404, Sep 2012.

\bibitem{herbert1971localized}
D~C Herbert and R~Jones.
\newblock Localized states in disordered systems.
\newblock {\em J. Phys. C: Solid State Phys.}, 4(10):1145, jul 1971.

\bibitem{thouless1972relation}
D~J Thouless.
\newblock A relation between the density of states and range of localization
  for one dimensional random systems.
\newblock {\em J. Phys. C: Solid State Phys.}, 5(1):77, jan 1972.

\bibitem{avila2015global}
A.~Avila.
\newblock {Global theory of one-frequency Schrödinger operators}.
\newblock {\em Acta Mathematica}, 215(1):1 -- 54, 2015.

\bibitem{jitomirskaya2012analytic}
Svetlana Jitomirskaya and C.~A. Marx.
\newblock Analytic quasi-perodic cocycles with singularities and the lyapunov
  exponent of extended harper’s model.
\newblock {\em Commun. Math. Phys.}, 316(1):237--267, 2012.

\bibitem{jitomirskaya2019critical}
Svetlana Jitomirskaya and Igor Krasovsky.
\newblock Critical almost mathieu operator: hidden singularity, gap continuity,
  and the hausdorff dimension of the spectrum, 2019.

\bibitem{hiramoto1988dynamics}
Hisashi Hiramoto and Shuji Abe.
\newblock Dynamics of an electron in quasiperiodic systems. ii. harper's model.
\newblock {\em Journal of the Physical Society of Japan}, 57(4):1365--1371,
  1988.

\bibitem{geisel1991new}
T.~Geisel, R.~Ketzmerick, and G.~Petschel.
\newblock New class of level statistics in quantum systems with unbounded
  diffusion.
\newblock {\em Phys. Rev. Lett.}, 66:1651--1654, Apr 1991.

\bibitem{passaro1992anomalous}
B.~Passaro, C.~Sire, and V.~G. Benza.
\newblock Anomalous diffusion and conductivity in octagonal tiling models.
\newblock {\em Phys. Rev. B}, 46:13751--13755, Dec 1992.

\bibitem{guarneri1989spectral}
I.~Guarneri.
\newblock Spectral properties of quantum diffusion on discrete lattices.
\newblock {\em Europhysics Letters}, 10(2):95, sep 1989.

\bibitem{combes1993connections}
Jean-Michel Combes.
\newblock Connections between quantum dynamics and spectral properties of
  time-evolution operators.
\newblock In W.F. Ames, E.M. Harrell, and J.V. Herod, editors, {\em
  Differential Equations with Applications to Mathematical Physics}, volume 192
  of {\em Mathematics in Science and Engineering}, pages 59--68. Elsevier,
  1993.

\bibitem{last1996quantum}
Yoram Last.
\newblock Quantum dynamics and decompositions of singular continuous spectra.
\newblock {\em Journal of Functional Analysis}, 142(2):406--445, 1996.

\bibitem{wilkinson1994spectral}
Michael Wilkinson and Elizabeth~J. Austin.
\newblock Spectral dimension and dynamics for harper's equation.
\newblock {\em Phys. Rev. B}, 50:1420--1429, Jul 1994.

\bibitem{bellissard1995noncommutative}
Jean Bellissard.
\newblock Noncommutative geometry and quantum hall effect.
\newblock In S.~D. Chatterji, editor, {\em Proceedings of the International
  Congress of Mathematicians}, pages 1238--1246, Basel, 1995. Birkh{\"a}user
  Basel.

\bibitem{schulz1998anomalous}
H.~Schulz-Baldes and J.~Bellissard.
\newblock Anomalous transport: A mathematical framework.
\newblock {\em Reviews in Mathematical Physics}, 10(01):1--46, 1998.

\bibitem{dominguez2019aubry}
G~A Dom{\'{\i}}nguez-Castro and R~Paredes.
\newblock The aubry{\textendash}andr{\'{e}} model as a hobbyhorse for
  understanding the localization phenomenon.
\newblock {\em European Journal of Physics}, 40(4):045403, jun 2019.

\bibitem{amini2017spread}
M.~Amini.
\newblock Spread of wave packets in disordered hierarchical lattices.
\newblock {\em Europhysics Letters}, 117(3):30003, mar 2017.

\bibitem{detomasi2019survival}
Giuseppe~De Tomasi, Mohsen Amini, Soumya Bera, Ivan~M. Khaymovich, and
  Vladimir~E. Kravtsov.
\newblock {Survival probability in Generalized Rosenzweig-Porter random matrix
  ensemble}.
\newblock {\em SciPost Phys.}, 6:014, 2019.

\bibitem{khaymovich2021dynamical}
Ivan~M. Khaymovich and Vladimir~E. Kravtsov.
\newblock {Dynamical phases in a ``multifractal'' Rosenzweig-Porter model}.
\newblock {\em SciPost Phys.}, 11:045, 2021.

\bibitem{wang2016phase}
Jun Wang, Xia-Ji Liu, Gao Xianlong, and Hui Hu.
\newblock Phase diagram of a non-abelian aubry-andr\'e-harper model with
  $p$-wave superfluidity.
\newblock {\em Phys. Rev. B}, 93:104504, Mar 2016.

\bibitem{degottardi2013majorana}
Wade DeGottardi, Diptiman Sen, and Smitha Vishveshwara.
\newblock Majorana fermions in superconducting 1d systems having periodic,
  quasiperiodic, and disordered potentials.
\newblock {\em Phys. Rev. Lett.}, 110:146404, Apr 2013.

\bibitem{kang2020creutz}
Jin~Hyoun Kang, Jeong~Ho Han, and Y~Shin.
\newblock Creutz ladder in a resonantly shaken 1d optical lattice.
\newblock {\em New Journal of Physics}, 22(1):013023, jan 2020.

\bibitem{he2021flat}
Yanyan He, Ruosong Mao, Han Cai, Jun-Xiang Zhang, Yongqiang Li, Luqi Yuan,
  Shi-Yao Zhu, and Da-Wei Wang.
\newblock Flat-band localization in creutz superradiance lattices.
\newblock {\em Phys. Rev. Lett.}, 126:103601, Mar 2021.

\bibitem{martinez2023interaction}
Jeronimo G.~C. Martinez, Christie~S. Chiu, Basil~M. Smitham, and Andrew~A.
  Houck.
\newblock Interaction-induced escape from an aharonov-bohm cage, 2023.

\bibitem{caceres2022controlled}
Gabriel C\'aceres-Aravena, Diego Guzm\'an-Silva, Ignacio Salinas, and
  Rodrigo~A. Vicencio.
\newblock Controlled transport based on multiorbital aharonov-bohm photonic
  caging.
\newblock {\em Phys. Rev. Lett.}, 128:256602, Jun 2022.

\bibitem{zhao2018topological}
Erhai Zhao.
\newblock Topological circuits of inductors and capacitors.
\newblock {\em Annals of Physics}, 399:289--313, 2018.

\bibitem{maksymenko2012flatband}
M.~Maksymenko, A.~Honecker, R.~Moessner, J.~Richter, and O.~Derzhko.
\newblock Flat-band ferromagnetism as a pauli-correlated percolation problem.
\newblock {\em Phys. Rev. Lett.}, 109:096404, August 2012.

\bibitem{rhim2021singular}
Jun-Won Rhim and Bohm-Jung Yang.
\newblock Singular flat bands.
\newblock {\em Advances in Physics: X}, 6(1):1901606, 2021.

\bibitem{srednicki1994chaos}
Mark Srednicki.
\newblock Chaos and quantum thermalization.
\newblock {\em Phys. Rev. E}, 50:888--901, Aug 1994.

\bibitem{deutsch2018eigenstate}
Joshua~M Deutsch.
\newblock Eigenstate thermalization hypothesis.
\newblock {\em Reports on Progress in Physics}, 81(8):082001, jul 2018.

\bibitem{biroli2010effect}
Giulio Biroli, Corinna Kollath, and Andreas~M. L\"auchli.
\newblock Effect of rare fluctuations on the thermalization of isolated quantum
  systems.
\newblock {\em Phys. Rev. Lett.}, 105:250401, Dec 2010.

\bibitem{alba2015eigenstate}
Vincenzo Alba.
\newblock Eigenstate thermalization hypothesis and integrability in quantum
  spin chains.
\newblock {\em Phys. Rev. B}, 91:155123, Apr 2015.

\bibitem{ikeda2013finite}
Tatsuhiko~N. Ikeda, Yu~Watanabe, and Masahito Ueda.
\newblock Finite-size scaling analysis of the eigenstate thermalization
  hypothesis in a one-dimensional interacting bose gas.
\newblock {\em Phys. Rev. E}, 87:012125, Jan 2013.

\bibitem{imbrie2016many}
John~Z Imbrie.
\newblock On many-body localization for quantum spin chains.
\newblock {\em Journal of Statistical Physics}, 163:998--1048, Apr 2016.

\bibitem{olshanii1998atomic}
M.~Olshanii.
\newblock Atomic scattering in the presence of an external confinement and a
  gas of impenetrable bosons.
\newblock {\em Phys. Rev. Lett.}, 81:938--941, Aug 1998.

\bibitem{petrov2000regimes}
D.~S. Petrov, G.~V. Shlyapnikov, and J.~T.~M. Walraven.
\newblock Regimes of quantum degeneracy in trapped 1d gases.
\newblock {\em Phys. Rev. Lett.}, 85:3745--3749, Oct 2000.

\bibitem{dunjko2001bosons}
V.~Dunjko, V.~Lorent, and M.~Olshanii.
\newblock Bosons in cigar-shaped traps: Thomas-fermi regime, tonks-girardeau
  regime, and in between.
\newblock {\em Phys. Rev. Lett.}, 86:5413--5416, Jun 2001.

\bibitem{lenard1964momentum}
A.~Lenard.
\newblock {Momentum Distribution in the Ground State of the One‐Dimensional
  System of Impenetrable Bosons}.
\newblock {\em Journal of Mathematical Physics}, 5(7):930--943, 12 2004.

\bibitem{lenard1966one}
A.~Lenard.
\newblock {One‐Dimensional Impenetrable Bosons in Thermal Equilibrium}.
\newblock {\em Journal of Mathematical Physics}, 7(7):1268--1272, 12 2004.

\bibitem{vaidya1979one}
H.~G. Vaidya and C.~A. Tracy.
\newblock {One particle reduced density matrix of impenetrable bosons in one
  dimension at zero temperature}.
\newblock {\em Journal of Mathematical Physics}, 20(11):2291--2312, 07 2008.

\bibitem{rigol2004emergence}
Marcos Rigol and Alejandro Muramatsu.
\newblock Emergence of quasicondensates of hard-core bosons at finite momentum.
\newblock {\em Phys. Rev. Lett.}, 93:230404, Dec 2004.

\bibitem{rigol2005free}
Marcos Rigol and Alejandro Muramatsu.
\newblock Free expansion of impenetrable bosons on one-dimensional optical
  lattices.
\newblock {\em Mod. Phys. Lett. B}, 19(18):861--881, 2005.

\bibitem{aizenman2004bose}
Michael Aizenman, Elliott~H. Lieb, Robert Seiringer, Jan~Philip Solovej, and
  Jakob Yngvason.
\newblock Bose-einstein quantum phase transition in an optical lattice model.
\newblock {\em Phys. Rev. A}, 70:023612, Aug 2004.

\bibitem{drescher2017hard}
Moritz Drescher and Andreas Mielke.
\newblock Hard-core bosons in flat band systems above the critical density.
\newblock {\em The European Physical Journal B}, 90:1--8, 2017.

\bibitem{moudgalya2022quantum}
Sanjay Moudgalya, B~Andrei Bernevig, and Nicolas Regnault.
\newblock Quantum many-body scars and hilbert space fragmentation: a review of
  exact results.
\newblock {\em Reports on Progress in Physics}, 85(8):086501, jul 2022.

\bibitem{buvca2022out}
Berislav Bu\ifmmode~\check{c}\else \v{c}\fi{}a.
\newblock Out-of-time-ordered crystals and fragmentation.
\newblock {\em Phys. Rev. Lett.}, 128:100601, Mar 2022.

\bibitem{schecter2019weak}
Michael Schecter and Thomas Iadecola.
\newblock Weak ergodicity breaking and quantum many-body scars in spin-1 $xy$
  magnets.
\newblock {\em Phys. Rev. Lett.}, 123:147201, Oct 2019.

\bibitem{buvca2023unified}
Berislav Bu\ifmmode~\check{c}\else \v{c}\fi{}a.
\newblock Unified theory of local quantum many-body dynamics: Eigenoperator
  thermalization theorems.
\newblock {\em Phys. Rev. X}, 13:031013, Aug 2023.

\bibitem{nguyen2018symmetry}
H.S. Nguyen, F.~Dubois, T.~Deschamps, S.~Cueff, A.~Pardon, J.-L. Leclercq,
  C.~Seassal, X.~Letartre, and P.~Viktorovitch.
\newblock Symmetry breaking in photonic crystals: On-demand dispersion from
  flatband to dirac cones.
\newblock {\em Phys. Rev. Lett.}, 120:066102, February 2018.

\bibitem{ma2020direct}
Jina Ma, Jun-Won Rhim, Liqin Tang, Shiqi Xia, Haiping Wang, Xiuyan Zheng,
  Shiqiang Xia, Daohong Song, Yi~Hu, Yigang Li, et~al.
\newblock Direct observation of flatband loop states arising from nontrivial
  real-space topology.
\newblock {\em Physical Review Letters}, 124(18):183901, 2020.

\bibitem{masumoto2012exciton}
Naoyuki Masumoto, Na~Young Kim, Tim Byrnes, Kenichiro Kusudo, Andreas
  L{\"o}ffler, Sven H{\"o}fling, Alfred Forchel, and Yoshihisa Yamamoto.
\newblock Exciton--polariton condensates with flat bands in a two-dimensional
  kagome lattice.
\newblock {\em New J. Phys.}, 14(6):065002, 2012.

\bibitem{zhang2023non}
Weixuan Zhang, Haiteng Wang, Houjun Sun, and Xiangdong Zhang.
\newblock Non-abelian inverse anderson transitions.
\newblock {\em Physical Review Letters}, 130(20):206401, 2023.

\bibitem{wang2022observation}
Haiteng Wang, Weixuan Zhang, Houjun Sun, and Xiangdong Zhang.
\newblock Observation of inverse anderson transitions in aharonov-bohm
  topolectrical circuits.
\newblock {\em Physical Review B}, 106(10):104203, 2022.

\bibitem{kang2020}
Mingu Kang, Shiang Fang, Linda Ye, Hoi~Chun Po, Jonathan Denlinger, Chris
  Jozwiak, Aaron Bostwick, Eli Rotenberg, Efthimios Kaxiras, Joseph~G.
  Checkelsky, and Riccardo Comin.
\newblock Topological flat bands in frustrated kagome lattice cosn.
\newblock {\em Nature Communications}, 11(1):4004, 2020.

\bibitem{magnonic}
Silvia Tacchi, Jorge Flores-Farías, Daniela Petti, Felipe Brevis, Andrea
  Cattoni, Giuseppe Scaramuzzi, Davide Girardi, David Cortés-Ortuño,
  Rodolfo~A. Gallardo, Edoardo Albisetti, Giovanni Carlotti, and Pedro
  Landeros.
\newblock Experimental observation of flat bands in one-dimensional chiral
  magnonic crystals.
\newblock {\em Nano Letters}, 0(0):null, 0.

\bibitem{danieli2018compact}
C.~Danieli, A.~Maluckov, and S.~Flach.
\newblock Compact discrete breathers on flat-band networks.
\newblock {\em Low Temp. Phys.}, 44(7):678--687, 2018.

\bibitem{morales2016simple}
Luis Morales-Inostroza and Rodrigo~A. Vicencio.
\newblock Simple method to construct flat-band lattices.
\newblock {\em Phys. Rev. A}, 94:043831, October 2016.

\end{thebibliography}

\end{document}